\documentclass{JFM-FLM_Au}

\usepackage{enumitem}

\usepackage{longtable}

\hypersetup{hypertexnames=false}

\newcommand{\diff}{\mathrm{d}}

\lefttitle{E. Kanin, A. Möri, D. Garagash and B. Lecampion}
\righttitle{A semi-infinite hydraulic fracture propagating in a poroelastic medium}

\title{A semi-infinite hydraulic fracture propagating in a poroelastic medium}

\author{Evgenii Kanin\aff{1}, Andreas Möri\aff{1}, Dmitry Garagash\aff{2} \and Brice Lecampion\aff{1}}

\affiliation{
\aff{1}Geo-Energy Laboratory Institute of Civil Engineering, École Polytechnique Fédérale de Lausanne (EPFL), Lausanne, Switzerland
\aff{2}Department of Civil and Resource Engineering, Dalhousie University, Halifax, Canada}

\corresau{Brice Lecampion, brice.lecampion@epfl.ch}

\makeatletter
\def\ps@arxiv{%
    \ps@headings
    \def\@oddfoot{\hfill{\cppagefont\thepage}}%
    \def\@evenfoot{{\cppagefont\thepage}\hfill}%
}
\makeatother

\AtBeginDocument{\pagestyle{arxiv}}

\begin{document}
\maketitle

\thispagestyle{arxiv}

\begin{center}
    \small\itshape
    Manuscript under consideration for publication
    in the Journal of Fluid Mechanics.
\end{center}

\begin{abstract}
We solve the near-tip coupled hydromechanical problem of a hydraulic fracture propagating in a permeable poroelastic medium, represented locally as a semi-infinite crack moving at constant velocity. In the moving tip frame, all fracture fields are steady. The porous solid obeys linear poroelasticity, fracture propagation follows LEFM, fluid flow within the tensile fracture is governed by lubrication theory, and the fracturing and pore fluids are treated as identical Newtonian liquids. The primary unknowns are the fracture opening and the fluid displacement function along the fracture, determined from the resulting non-linear system of equations. In a viscosity--toughness scaling, the problem depends on four dimensionless parameters: a dimensionless permeability $\varrho$ (associated with the intensity of fluid leak-in), a dimensionless effective confining stress $\Sigma_0'$ (measuring the relative importance of fluid leak-off and leak-in), the drained--undrained Poisson's ratio contrast $\beta$, and the poroelastic stress coefficient $\eta$. The solution evolves from a drained, toughness-dominated near-tip asymptote with uniform net pressure to an undrained, storage-viscosity-dominated far-field asymptote. Numerical solutions are obtained over parameter ranges representative of various rock types. The fully coupled model is compared with simplified models, namely uncoupled and one-way coupled formulations, over distances from the fracture front spanning laboratory-to-field scales. The resulting relative difference maps in the $(\varrho,\Sigma_0')$ parameter space identify four characteristic regimes: negligible poroelastic effects, dominance of the direct mechanical poroelastic response, importance of the poroelastic backstress, and a fully coupled regime. The latter occurs mainly at large $\varrho$ and small $\Sigma_0'$, corresponding to strong leak-in relative to leak-off.
\end{abstract}



\section{Introduction}

Hydraulic fractures (HFs) are fluid-driven tensile cracks that propagate in solid media under internal fluid pressurization. Natural examples include magma-driven dykes in the Earth's crust \citep{spence1985magma, lister1991fluid, rubin1993tensile, rivalta2015review, dontsov2016propagation}, tensile jointing in overpressured saturated rock formations \citep{secor1965role, engelder1990natural}, and fluid-filled fractures in glacier beds \citep{van2007fracture, tsai2010model}. Engineered HFs are relevant to hydrocarbon reservoirs \citep{economides2000reservoir}, enhanced geothermal systems \citep{moska2021hydraulic, jia2022hydraulic}, and the assessment of fracture growth and containment integrity in geological formations used for subsurface CO$_2$ storage \citep{raziperchikolaee2013effect, huerta2020hydraulic}. From an industrial perspective, HFs are used to enhance hydraulic connectivity between low-permeability formations and wellbores by creating highly conductive fractures that provide preferential flow pathways. Numerical modelling of HFs is important for treatment design and optimization, mitigation of induced seismicity, and assessment of containment integrity in subsurface formations.

Hydraulic fractures commonly propagate in fluid-saturated porous formations, in which deformation of the solid skeleton and pore pressure diffusion are intrinsically coupled. Fluid exchange between the fracture and the surrounding formation redistributes pore pressure in the porous medium. Through poroelastic coupling, this pressure redistribution induces deformation of the solid skeleton and an additional normal stress acting on the fracture surfaces, commonly referred to as poroelastic backstress \citep{DeCh89, VaDe89, DeCh90, DeCh91, cheng1993review, Kovalyshen10, GaDe20}. Conversely, the deformation field associated with fracture opening generates pore pressure perturbations in the surrounding rock, which in turn modify the fluid exchange. HF propagation in permeable formations is therefore influenced not only by the contribution of fluid exchange to the fracture fluid balance, but also by this two-way coupling between solid deformation and pore pressure diffusion. Because hydraulic diffusion competes with fracture propagation, the poroelastic response depends on both the propagation velocity and the distance from the fracture front, giving rise to drained, partially drained, and undrained regions \citep{Clea78,Ruin78}.

HF propagation is governed by the interaction of rock failure, viscous flow of the fracturing fluid within the fracture, fluid exchange with the surrounding permeable formation, and the associated poroelastic response. These processes are particularly strongly coupled in the vicinity of the fracture front, where the fracture opening and the fracturing fluid pressure vary rapidly with distance from the tip. These variations produce large pressure gradients within the fracture channel and, through fluid exchange and solid deformation, steep pore pressure and stress gradients in the surrounding rock. The near-tip region therefore provides the natural setting for determining how poroelastic coupling modifies the local fracture fields and the physical mechanisms controlling fracture propagation.

The characteristic length scales associated with near-tip processes are often much smaller than the spatial resolution of finite fracture simulators \citep{detournay2016mechanics}. For this reason, several HF simulators determine the advancement of the fracture front by directly incorporating a near-tip asymptotic solution that depends on the local fracture propagation velocity. Examples include simulators based on the Implicit Level Set Algorithm (ILSA) \citep{peirce2008implicit, peirce2015modeling, dontsov2017multiscale, zia2020pyfrac} and the Implicit Moving Mesh Algorithm (IMMA) \citep{dontsov2016approximate, dontsov2017approximate, kanin2020radial}. In such approaches, the local near-tip region is represented by a steadily propagating semi-infinite crack, whose velocity is interpreted as the instantaneous local velocity of the parent fracture front. In hydraulic fracture simulators, the near-tip solution is effectively sampled over a finite spatial window measured from the fracture front, whose size is set by the spatial resolution of the numerical model. Since the near-tip solution is usually expressed in dimensionless variables, the portion sampled in finite fracture simulations depends on the material parameters and on the local propagation velocity of the parent fracture. This sampled portion determines the effective propagation regime of the finite fracture and, therefore, which physical processes control fracture advancement during the simulation.

A broad range of near-tip models has been developed to resolve the multiscale structure of the fracture fields in the vicinity of the fracture front. The tip region of a fully fluid-filled HF propagating in an impermeable elastic medium was analysed by \cite{spence1985self, lister1990buoyancy, desroches1994crack}. The role of a fluid lag in the near-tip structure of hydraulic fractures was investigated by \cite{rubin1993tensile, garagash2000tip}. For permeable formations, \cite{lenoach1995crack} and \cite{garagash2011multiscale} studied the influence of fluid exchange between the fracture and the surrounding rock using the pressure-independent Carter's leak-off law \citep{howard1957optimum}. An efficient approximation of the tip solution of \cite{garagash2011multiscale} was later developed by \cite{dontsov2015non}, enabling its integration into ILSA-based HF simulators \citep{dontsov2017multiscale, zia2020pyfrac}. More elaborate descriptions of fluid exchange were proposed by \cite{kanin2020near}, who considered a one-dimensional pressure-dependent exchange process, while formulations with two-dimensional pressure-dependent fluid exchange were examined by \cite{detournay2003near, Kovalyshen10}. In addition, \cite{Kovalyshen10} accounted for the poroelastic backstress generated by fluid exchange with the surrounding medium, which acts against fracture opening.

Other works have incorporated additional physical mechanisms into the problem of a steadily propagating semi-infinite hydraulic fracture. Non-Newtonian rheology of the fracturing fluid was considered by \cite{dontsov2018semi, bessmertnykh2021semi} for power-law fluids, by \cite{moukhtari2018semi} for Carreau fluids, and by \cite{bessmertnykh2019semi} for Herschel--Bulkley fluids. The influence of a cohesive zone and the limitations of linear elastic fracture mechanics (LEFM) as a propagation criterion in HF modelling were analysed by \cite{garagash2019cohesive}. In the aforementioned models, the flow inside the fracture channel is assumed to be laminar. The transition from laminar flow near the fracture front to turbulent flow farther behind the tip was examined by \cite{dontsov2016tip, lecampion2019slickwater, kanin2020turbulent}. Finally, a near-tip HF model accounting for proppant particles was developed by \cite{bessmertnykh2020effects}.

Among the models reviewed above, the formulations of \cite{detournay2003near} and \cite{Kovalyshen10} are the most closely related to the present work. \cite{detournay2003near} considered the near-tip cavity of a fluid-driven fracture that supports pore fluid circulation. In their formulation, the fracture opening was constrained by the LEFM square-root asymptote \citep{irwin1957analysis}, while fluid exchange with the surrounding permeable formation was treated as a two-dimensional pressure-dependent diffusion process. Their analysis showed that pore fluid flows into the fracture near the front and leaks back into the surrounding rock farther behind the tip, forming a near-tip circulation zone. This mechanism arises from the pressure-dependent nature of the fluid exchange and is absent in Carter-type leak-off models. A similar circulation zone was also discussed by \cite{kanin2020near} in the context of a one-dimensional pressure-dependent fluid exchange formulation.

The formulation of \cite{detournay2003near} was later generalized by \cite{Kovalyshen10}, who incorporated poroelastic effects associated with fluid exchange between the fracture and the surrounding poroelastic medium. In particular, the formulation of \cite{Kovalyshen10} accounted for the hydraulic-to-mechanical cross-coupling contribution to the normal stress on the fracture surfaces, i.e. the poroelastic backstress. However, the formulation was not fully coupled in the sense adopted in the present paper. The direct mechanical contribution to the normal stress was treated using the purely elastic crack response rather than the poroelastic one, and the mechanical-to-hydraulic cross-coupling contribution to the pore fluid pressure on the fracture surfaces, corresponding to deformation-induced pore pressure perturbations, was not retained.

Earlier related studies \citep{Clea78, Ruin78, atkinson1991plane} analysed steadily moving semi-infinite cracks in poroelastic media under prescribed mechanical and hydraulic boundary conditions, showing that poroelastic coupling produces a drained crack-tip zone embedded in a partially drained or undrained outer region and modifies the resistance to fracture propagation.

The objective of the present work is to develop a fully coupled model for the near-tip region of a hydraulic fracture propagating in a permeable poroelastic rock. To the best of our knowledge, a fully coupled solution for a steadily propagating semi-infinite hydraulic fracture in a poroelastic medium has not yet been obtained. For such a fracture, the plane strain poroelastic problem can be reduced to two boundary integral equations for the normal stress and pore fluid pressure on the fracture surfaces. These equations were derived by \cite{kanin2026steadily} using the fundamental solutions of plane strain poroelasticity for an instantaneous normal edge dislocation and an instantaneous fluid source \citep{detournay1987poroelastic, cheng2016poroelasticity}, followed by temporal and spatial superpositions. The formulation was validated against available analytical and semi-analytical solutions for reduced poroelastic fracture problems, including a tensile fracture subjected to exponentially distributed normal loading and a stress-free tensile fracture with an imposed exponential pore fluid pressure \citep{atkinson1991plane}. The resulting formulation contains both direct and cross-coupling contributions to the normal stress and pore fluid pressure. The direct contributions correspond to the mechanical poroelastic response in the normal stress equation and the hydraulic response in the pore fluid pressure equation. The cross-coupling terms represent the hydraulic-to-mechanical contribution, or poroelastic backstress, and the mechanical-to-hydraulic contribution responsible for deformation-induced pore pressure perturbations. \cite{kanin2026steadily} did not solve the hydraulic fracture problem, in which the fracture opening, fracturing fluid pressure and fluid exchange are determined self-consistently. In the present work, these boundary integral equations, supplemented by the closure relations on the fracture surfaces, are coupled with the LEFM propagation condition and lubrication flow inside the fracture. This yields a non-linear system of governing equations for the fracture opening, net pressure, and fluid displacement function.

Using this formulation, we quantify the influence of poroelastic coupling on the near-tip fracture fields and identify the mechanisms controlling this influence. By comparing the fully coupled formulation with several reduced formulations, we determine when poroelastic effects are negligible, when the direct mechanical poroelastic response is sufficient, when the poroelastic backstress provides the dominant coupling mechanism, and when deformation-induced pore pressure perturbations make the fully coupled formulation necessary.

The paper is organised as follows. Section~\ref{sec:model_formulation} presents the physical model, governing equations, and reduced formulations. Section~\ref{sec:mk_scaling} gives the normalized governing equations for the fully coupled and reduced models and introduces the dimensionless governing parameters. Section~\ref{sec:asymptotes_and_numerical_solution} derives the near-field and far-field asymptotic solutions and outlines the numerical procedure used to obtain the general solution of the fully coupled fracture tip problem. Section~\ref{sec:results_discussion} presents and discusses the numerical results, including the general solution structure, relative difference and regime maps, and detailed analyses of representative points associated with the identified characteristic coupling regimes. Finally, Section~\ref{sec:conclusions} summarizes the main conclusions and outlines possible extensions of the model.

\section{Model formulation}
\label{sec:model_formulation}

\subsection{Problem definition}

\begin{figure}[h!]
    \centering
    \includegraphics[width=1\textwidth]{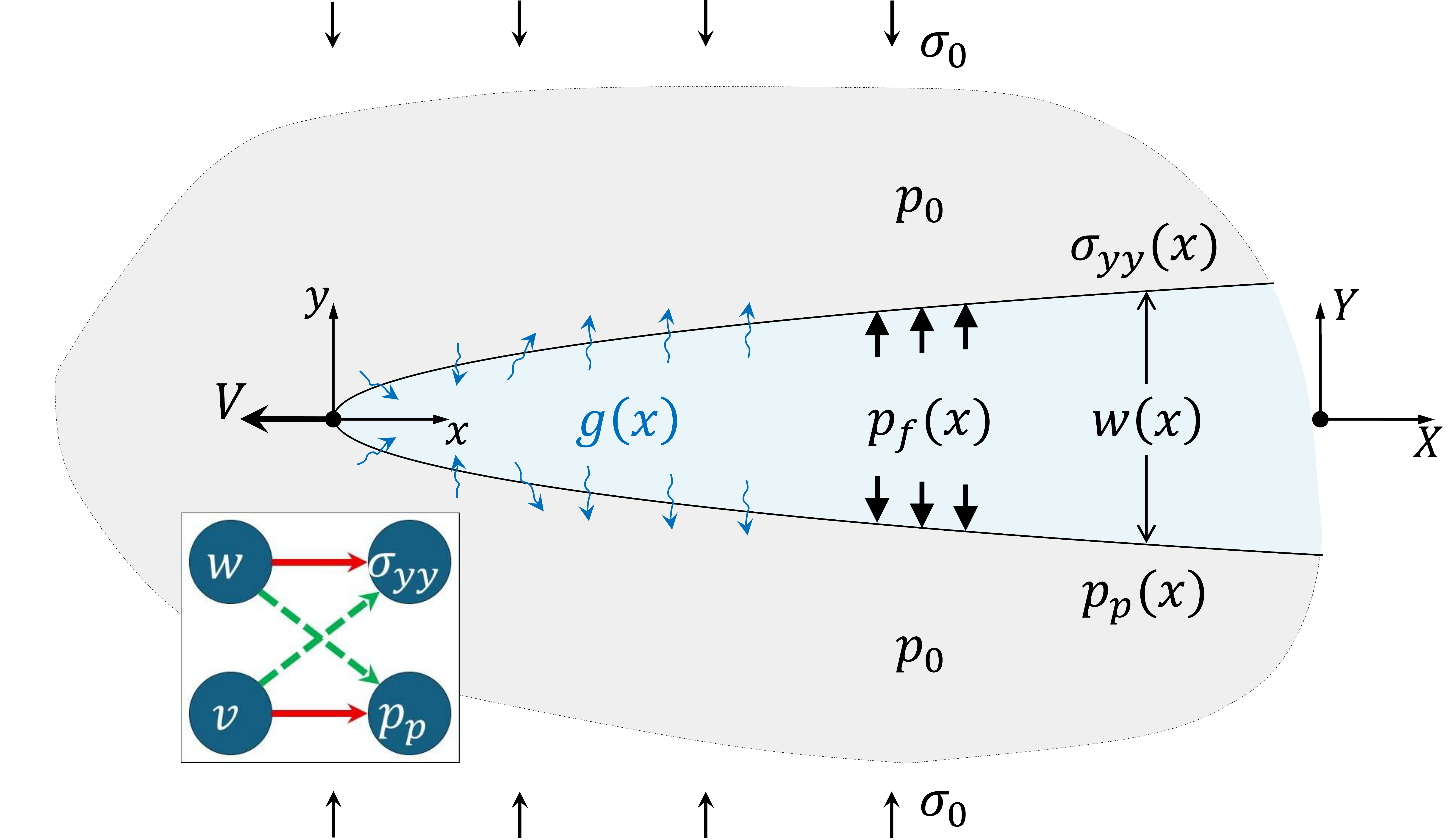}
    \caption{
    Schematic of a semi-infinite hydraulic fracture propagating at constant velocity $V$ along the negative $X-$direction. The fixed coordinate system $(X,Y)$ and the moving coordinate system $(x,y)$ attached to the fracture tip are both shown. The fracture opening and the fracturing fluid pressure are denoted by $w(x)$ and $p_f(x)$, respectively. The surrounding formation is initially subjected to a uniform pore pressure $p_0$ and a uniform compressive stress $\sigma_0$. Fluid exchange between the fracture and the porous medium is described by the local exchange rate $g(x)$ and the associated fluid displacement function $v(x)$. The blue arrows indicate the direction of fluid exchange across the fracture surfaces: leak-off from the fracture into the formation, or leak-in from the formation into the fracture. The inset shows the direct and cross-coupling contributions to the normal stress $\sigma_{yy}(x)$ and pore fluid pressure $p_p(x)$ on the fracture surfaces. Solid red arrows indicate the direct contributions $w\mapsto\sigma_{yy}^{M}$ and $v\mapsto p^{H}$; dashed green arrows indicate the cross-coupling contributions $v\mapsto\sigma_{yy}^{HM}$ and $w\mapsto p^{MH}$. The corresponding decompositions of $\sigma_{yy}(x)$ and $p_p(x)$ are given in Section~\ref{sec:governing_equations}.
    }
    \label{fig:semi_infinite_crack_sketch}
\end{figure}

The near-tip region of a hydraulic fracture can be analysed by considering a semi-infinite plane strain crack propagating at constant velocity $V$, representing the instantaneous local front velocity of the parent fracture. A schematic of the model is shown in figure~\ref{fig:semi_infinite_crack_sketch}. Owing to the constant propagation velocity, a coordinate system moving with the fracture tip can be introduced, in which the problem becomes steady. In this frame, the fracture fields depend only on the distance from the tip $x$, the propagation velocity, and the material parameters.

The surrounding rock is modelled as an isotropic linear poroelastic medium. Its response is characterized by a minimal set of independent parameters, chosen here as the drained Young's modulus $E$, the drained Poisson's ratio $\nu$, the Biot coefficient $\alpha$, and the Biot modulus $M$. The latter depends on the rock porosity $\phi$, the bulk moduli of the pore fluid $K_f$ and the solid grains $K_s$, and the Biot coefficient. We also use the derived elastic and poroelastic properties, namely the shear modulus $G$, the drained bulk modulus $K$, and the undrained Poisson's ratio $\nu_u$. Fracture propagation is described within the framework of LEFM, assuming quasi-static conditions and a damage/yielding zone that remains small relative to the characteristic length scales of the problem. The rock is characterized by the fracture toughness $K_{Ic}$.

The fracture surfaces are subjected to internal loading from the fracturing fluid pressure $p_f(x)$, which acts to open the crack against the in-situ confining stress $\sigma_0$. The fracturing fluid is assumed to be Newtonian, with viscosity $\mu$. Flow within the fracture is described using lubrication theory \citep{batchelor2000introduction}.

The surrounding rock is permeable, with intrinsic permeability $k$, and is saturated with a pore fluid initially at pressure $p_0$, whose properties are identical to those of the fracturing fluid. We assume that the fracture surfaces remain hydraulically open, with no filter cake, damaged zone, or other interfacial resistance limiting hydraulic communication with the surrounding porous medium. The associated fluid exchange is therefore modelled as a two-dimensional process \citep{detournay2003near, Kovalyshen10}. The local exchange rate is denoted by $g(x)$. We further introduce the fluid displacement function $v(x)$, which is related to $g(x)$ by \citep{cheng1998singular, Kovalyshen10}:
\begin{equation*}
    v(x)=\frac{1}{V}\int_0^x g(s)\diff s, \quad g(x)=V \frac{\diff v}{\diff x}.
\end{equation*}
Since $v(x)$ is proportional to the cumulative fluid exchange volume over the interval $s \in [0,x)$, it may also be interpreted as a normalized cumulative fluid exchange.

\subsection{Governing equations}
\label{sec:governing_equations}

We introduce a moving coordinate system $(x,y)$ attached to the tip of a semi-infinite fracture propagating at constant velocity $V$. The transformation between the fixed $(X,Y)$ and moving $(x,y)$ frames is given by $x = X + Vt$ and $y = Y$. In this moving frame, the problem becomes stationary. The governing equations are reduced to the fracture plane, $y = 0$, and formulated in terms of the fracture opening $w(x)$ and the fluid displacement function $v(x)$, from which the net pressure $p(x) = p_f(x) - \sigma_0$ is determined. All fracture characteristics are expressed as functions of the distance $x$ from the tip. Once the fracture tip problem is solved, the pore fluid pressure and stress fields in the surrounding poroelastic medium can be recovered from the corresponding full-space boundary integral representations \citep{kanin2026steadily}.

\subsubsection{Poroelastic response of the surrounding solid}

The response of the surrounding poroelastic solid is expressed through two boundary integral equations evaluated on the fracture surfaces. The first equation gives the normal stress $\sigma_{yy}(x)$, whereas the second determines the pore fluid pressure $p_p(x)$. Both equations contain direct and cross-coupling contributions and are written in terms of the fracture opening $w(x)$ and the fluid displacement function $v(x)$. The boundary integral equations were derived by \cite{kanin2026steadily} from the fundamental solutions of poroelasticity for an instantaneous normal edge dislocation and an instantaneous point source, followed by temporal and spatial superposition.

The boundary integral equation governing the normal stress distribution on the fracture surfaces (positive in compression) can be written as:
\begin{flalign}
    & \sigma_{yy}(x) = \sigma_0 + \sigma_{yy}^M(x) + \sigma_{yy}^{HM}(x), \nonumber \\
    & \sigma_{yy}^M(x) = \int_0^{\infty} \frac{\diff w}{\diff s} \sigma_{yy}^e(x-s) \diff s, \quad \sigma_{yy}^{HM}(x) =  - V \int_0^{\infty} \frac{\diff v}{\diff s} \sigma_{yy}^s(x-s) \diff s,
    \label{eq:normal_stress}
\end{flalign}
where $\sigma_{yy}^M(x)$ is the direct mechanical (poroelastic) contribution, and $\sigma_{yy}^{HM}(x)$ is the cross-coupling hydraulic-to-mechanical contribution (see the inset of figure~\ref{fig:semi_infinite_crack_sketch}). The cross-coupling term taken without the preceding minus sign is commonly referred to as the backstress \citep{cheng1993review, Kovalyshen10}.

Equation~\eqref{eq:normal_stress} involves the functions $\sigma_{yy}^e(x)$ and $\sigma_{yy}^s(x)$, which represent the normal stress on the fracture surfaces induced by a unit steadily moving normal edge dislocation and a unit steadily moving fluid source, respectively \citep{kanin2026steadily}:
\begin{flalign}
    & \sigma_{yy}^e(x) = \frac{E'}{4 \pi (1 - \beta)} \frac{1}{x} \mathcal{S}^e \left(\frac{V x}{2c}\right), \nonumber \\
    & \mathcal{S}^e (\xi) = 1 - \beta \left(\frac{1}{\xi} \mp e^\xi \mathrm{K}_1(|\xi|) - 2 \xi e^\xi\bigg[\mathrm{K}_0(|\xi|) \mp  \mathrm{K}_1(|\xi|)\bigg]\right); \label{eq:normal_stress_dislocation}\\
    & \sigma_{yy}^s(x) = \frac{\eta}{2 \pi \kappa} \mathcal{S}^s\left(\frac{V x}{2c}\right), \quad \mathcal{S}^s(\xi) = -\frac{1}{\xi} - e^\xi \bigg[\mathrm{K}_0(|\xi|) \mp \mathrm{K}_1(|\xi|)\bigg],
     \label{eq:normal_stress_source}
\end{flalign}
where the upper signs correspond to $\xi > 0$ and the lower signs to $\xi < 0$, and $\mathrm{K}_0$ and $\mathrm{K}_1$ denote modified Bessel functions of the second kind of orders zero and one, respectively. As shown by \citet{kanin2026steadily}, the function $\mathcal{S}^e(\xi)$ is regular at $\xi=0$. Consequently, the prefactor $1/x$ in $\sigma_{yy}^e(x)$ gives rise to the classical Cauchy-type singularity, accompanied by a weaker logarithmic contribution at the next order. By contrast, $\mathcal{S}^s(\xi)$ has a weak logarithmic singularity at $\xi=0$, implying that $\sigma_{yy}^s(x)$ is also logarithmically singular at $x=0$. 

In equations~\eqref{eq:normal_stress_dislocation} and~\eqref{eq:normal_stress_source}, $E' = E/(1-\nu^2)$ is the drained plane strain modulus, $c = \kappa / S$ is the hydraulic diffusivity (also referred to as the generalized consolidation coefficient \citep{rice1976some}), and $\kappa = k/\mu$ is the fluid mobility. The parameter $S$ is the uniaxial strain / constant stress storage coefficient \citep{cheng2016poroelasticity}, given by:
\begin{equation}
    S = \frac{1}{M} + \frac{\alpha^2}{K + 4G/3}.
    \label{eq:storage_coefficient}
\end{equation}

The dimensionless parameter $\beta$, which measures the contrast between the undrained and drained Poisson's ratios, is defined as:
\begin{equation}
    \beta = \frac{\nu_u - \nu}{1 - \nu},
    \label{eq:beta_coefficient}
\end{equation}
where the undrained Poisson's ratio is given by: 
\begin{equation}
    \nu_u = \frac{1}{2}\left[1 - \frac{G(1-2\nu)}{G + \alpha^2 M (1-2 \nu )}\right].
    \label{eq:undrained_poisson_ratio}
\end{equation}
The parameter $\beta$ lies in the range $0 \leq \beta \leq 0.5$, with the lower bound corresponding to the uncoupled limit ($\alpha = 0$, for which $\nu_u = \nu$ and $\beta = 0$) and the upper bound to the maximum difference, attained for $\nu = 0$ and $\nu_u = 0.5$.

The dimensionless parameter $\eta$, referred to as the poroelastic stress coefficient \citep{detournay1993fundamentals}, is defined as:
\begin{equation}
    \eta = \frac{\alpha (1-2\nu)}{2(1-\nu)}.
    \label{eq:eta_coefficient}
\end{equation}
It takes values in the range $0 \leq \eta \leq 0.5$. The lower bound corresponds to vanishing cross-coupling poroelastic effects, occurring either for $\alpha = 0$ (uncoupled case) or in the limit of an incompressible solid, $\nu = 0.5$, while the upper bound is attained for a poroelastic medium with $\alpha = 1$ and $\nu = 0$.

The boundary integral equation governing the pore fluid pressure on the fracture surfaces is given by:
\begin{flalign}
    & p_p(x) = p_0 + p^H(x) + p^{MH}(x), \nonumber \\
    & p^H(x) = V \int_0^{\infty} \frac{\diff v}{\diff s} p^s(x-s) \diff s, \quad p^{MH}(x) =  - \int_0^{\infty} \frac{\diff w}{\diff s} p^e(x-s) \diff s,
    \label{eq:pore_fluid_pressure}
\end{flalign}
where $p^H(x)$ is the direct hydraulic contribution, and $p^{MH}(x)$ is the cross-coupling mechanical-to-hydraulic contribution (see the inset of figure~\ref{fig:semi_infinite_crack_sketch}).

The functions $p^s(x)$ and $p^e(x)$ in equation~\eqref{eq:pore_fluid_pressure} represent the pore fluid pressure on the fracture surfaces induced by a unit steadily moving fluid source and a unit steadily moving normal edge dislocation, respectively \citep{kanin2026steadily}:
\begin{align}
    & p^s(x) = \frac{1}{2\pi\kappa} \mathcal{P}^s\left(\frac{V x}{2c}\right), \quad \mathcal{P}^s(\xi) = e^\xi \mathrm{K}_0 (|\xi|); \label{eq:fluid_pressure_source} \\
    & p^e(x) = \frac{\eta V}{2 \pi \kappa} \mathcal{P}^e\left(\frac{V x}{2c}\right), \quad \mathcal{P}^e(\xi) = \mathcal{S}^s(\xi).
     \label{eq:fluid_pressure_dislocation}
\end{align}
As shown by \cite{kanin2026steadily}, both $\mathcal{P}^s(\xi)$ and $\mathcal{P}^e(\xi)$ have weak logarithmic singularities at $\xi=0$. Consequently, both $p^s(x)$ and $p^e(x)$ are logarithmically singular at $x=0$.

\subsubsection{Closure relations}

The boundary integral equations \eqref{eq:normal_stress} and \eqref{eq:pore_fluid_pressure} are supplemented by the following closure relations on the fracture surfaces:
\begin{equation}
    \sigma_{yy}(x) = p_f(x) = p(x) + \sigma_0, \quad p_p(x) = p_f(x) = p(x) + \sigma_0,
    \label{eq:closure_relations}
\end{equation}
where the first relation expresses the balance between the normal stress and the fracturing fluid pressure, while the second imposes equality between the pore fluid pressure and the fracturing fluid pressure. The latter represents an idealized hydraulic boundary condition corresponding to perfect hydraulic communication between the fracture channel and the surrounding porous medium. This assumption is consistent with the assumed identical properties of the fracturing fluid and the pore fluid, as well as with the absence of additional hydraulic resistance at the interface, such as a filter cake or damaged zone \citep{Kovalyshen10, chekhonin2012hydraulic}.

Using the closure relations~\eqref{eq:closure_relations}, equations~\eqref{eq:normal_stress} and \eqref{eq:pore_fluid_pressure} can be rewritten as:
\begin{equation}
    p(x) = \sigma_{yy}^M(x) + \sigma_{yy}^{HM}(x), \quad p(x) + \sigma_0' = p^H(x) + p^{MH}(x),
\end{equation}
where $\sigma_0' = \sigma_0 - p_0$ is the effective confining stress.

\subsubsection{Fluid flow inside fracture}

Flow within the fracture is governed by lubrication theory \citep{batchelor2000introduction}, specifically by the mass conservation equation and the momentum balance (Poiseuille's law), both averaged over the fracture aperture:
\begin{flalign}
    & V \frac{\diff w}{\diff x} - \frac{\diff q}{\diff x} + V \frac{\diff v }{\diff x}  = 0, \label{eq:mass_conservation} \\
    & q(x) = \frac{w^3}{\mu'}\frac{\diff p}{\diff x}, \label{eq:momentum_conservation}
\end{flalign}
where $q(x)$ is the fluid flux within the fracture, and $\mu' = 12 \mu$ is the viscosity parameter.

Integrating equation~\eqref{eq:mass_conservation} from the fracture tip to a point $x > 0$, and substituting equation~\eqref{eq:momentum_conservation}, yields the lubrication equation:
\begin{equation}
    \frac{w^3}{\mu'} \frac{\diff p}{\diff x} = V \left(w + v\right).
    \label{eq:lubrication_equation}
\end{equation}

\subsubsection{Fracture propagation}

Within the framework of LEFM, fracture propagation occurs when the stress intensity factor at the tip equals the fracture toughness, i.e. $K_I = K_{Ic}$. This condition is enforced through the near-tip asymptotic behaviour of the fracture opening \citep{irwin1957analysis}:
\begin{equation}
    w(x) = \sqrt{\frac{32}{\pi}}\frac{K_{Ic}}{E'} \sqrt{x}, \quad x \to 0.
    \label{eq:propagation_criterion}
\end{equation}
The use of the drained plane strain modulus $E'$ in the propagation condition reflects the fact that the region immediately adjacent to the fracture tip is asymptotically drained \citep{rice1976some, Clea78, Ruin78}. Indeed, at a distance $r$ from the fracture tip, the characteristic loading time is $t_{\mathrm{load}} \sim r/V$, whereas the diffusion time over the same length scale is $t_{\mathrm{diff}} \sim r^2/c$. Their ratio is $t_{\mathrm{diff}}/t_{\mathrm{load}}\sim Vr/c \to 0$ as $r \to 0$. Thus, pore fluid pressure perturbations equilibrate locally in the vicinity of the fracture tip and do not enter the leading-order LEFM singular stress field. In this sense, the near-tip LEFM asymptote is governed by the drained elastic response of the surrounding rock.

\subsection{Overview of the reduced models}
\label{sec:reduced_models}

In this section, we outline several reduced formulations of the fully coupled model proposed in this work. The objective is to place the fully coupled formulation within the hierarchy of existing fracture tip models that involve different approximations of poroelastic coupling and fluid exchange. Moreover, comparison with the reduced models in terms of fracture opening and net pressure profiles enables the identification of regions in parameter space where different poroelastic mechanisms significantly influence fracture propagation.

We distinguish three principal classes of reduced models: one-way coupled models, elastic-response models with poroelastic backstress, and uncoupled models. These classes are presented in order of decreasing coupling complexity. In addition, we introduce an auxiliary model without any cross-coupling terms. For each principal class, we outline three representations of fluid exchange between the fracture and the surrounding porous medium. The most general is the two-dimensional pressure-dependent fluid exchange model (2D PDL, where L denotes leak-off/leak-in). The one-dimensional pressure-dependent fluid exchange model (1D PDL) and Carter's leak-off law are then considered as reduced descriptions. The corresponding formulations are described below in relation to the fully coupled model.

\subsubsection{One-way coupled models}
\label{sec:one-way_coupled_models}

In one-way coupled models, the mechanical response does not induce pore pressure perturbations. Consequently, the pore fluid pressure on the fracture surfaces is governed solely by hydraulic diffusion and remains independent of mechanical deformation. 

In the 2D PDL formulation, the normal stress on the fracture surfaces is governed by equation~\eqref{eq:normal_stress}, whereas the pore fluid pressure is given by \eqref{eq:pore_fluid_pressure} without the cross-coupling term $p^{MH}(x)$ \citep{detournay2003near, Kovalyshen10}:
\begin{equation*}
    \sigma_{yy}(x) = \sigma_0 + \sigma_{yy}^M(x) + \sigma_{yy}^{HM}(x), \quad p_p(x) = p_0 + p^H(x).
\end{equation*}


In the 1D PDL formulation, the normal stress is governed by \eqref{eq:normal_stress}, while the pore fluid pressure is described by a one-dimensional linear diffusion model normal to the fracture plane \citep{detournay2003near}. This approximation neglects hydraulic diffusion in the formation along the fracture, or equivalently assumes that fluid exchange with the surrounding porous medium is locally controlled by diffusion normal to the fracture surfaces. The one-dimensional pore fluid pressure equation can be eliminated in favour of the corresponding fluid displacement function, obtained as its inverse \citep{kanin2020near}:
\begin{flalign*}
    \sigma_{yy}(x) = \sigma_0 + \sigma_{yy}^M(x) + \sigma_{yy}^{HM}(x), \quad p_p(x) = p_0 + p^H_\mathrm{1D}(x) \quad \text{or} \quad v(x) = v_\mathrm{1D}(x),
\end{flalign*}
where
\begin{flalign}
    & p^H_\mathrm{1D}(x) = \frac{1}{2\sqrt{\pi}}\sqrt{\frac{V}{\kappa S}} \int_{0}^{x} \frac{\diff v}{\diff s}\frac{\diff s}{\sqrt{x-s}},     \label{eq:pore_fluid_pressure_uncoupled_1D} \\
    & v_\mathrm{1D}(x) = \frac{2}{\sqrt{\pi}}\sqrt{\frac{\kappa S}{V}}\int_{0}^{x}\frac{p_p\left(s\right)-p_0}{\sqrt{x-s}}\diff s.
    \label{eq:fluid_displacement_function_1D}
\end{flalign}

Carter's leak-off formulation corresponds to one-dimensional pressure-independent fluid exchange. In this approximation, the pressure used to calculate leak-off is obtained by setting the fluid pressure in the fracture channel equal to the in-situ confining stress, $p_f(x) = \sigma_0$. Through the closure relation~\eqref{eq:closure_relations}, this also sets the pore fluid pressure on the fracture surfaces to $p_p(x)=\sigma_0$. In this case, the pore fluid pressure equation need not be solved. Instead, the fluid displacement function is prescribed directly and is obtained from equation~\eqref{eq:fluid_displacement_function_1D} by setting $p_p(x)=\sigma_0$ \citep{garagash2011multiscale}:
\begin{equation}
    v(x) = v_\mathrm{C}(x) = \frac{4}{\sqrt{\pi}}\sqrt{\frac{\kappa S}{V}}\sigma_0'\sqrt{x}.
    \label{eq:fluid_displacement_function_Carter}
\end{equation}
Substitution of $v(x) = v_\mathrm{C}(x)$ into the hydraulic-to-mechanical cross-coupling term gives $\sigma_{yy}^{HM}(x)=0$ \citep{Kovalyshen10}. Thus, Carter's leak-off formulation does not generate backstress. The normal stress on the fracture surfaces is therefore governed by \eqref{eq:normal_stress} without the backstress term, while retaining the direct mechanical poroelastic contribution: 
\begin{equation*}
    \sigma_{yy}(x) = \sigma_0 + \sigma_{yy}^M(x).
\end{equation*}

\subsubsection{Auxiliary model without cross-coupling terms}
\label{sec:auxiliary_no_cross_coupling}

To facilitate the subsequent assessment of the influence of different poroelastic coupling mechanisms on the fracture fields, we introduce an auxiliary formulation in which both cross-coupling terms are omitted. Fluid exchange between the fracture and the surrounding porous medium is retained in its two-dimensional pressure-dependent form (2D PDL). Accordingly, the normal stress and pore fluid pressure on the fracture surfaces are governed by equations~\eqref{eq:normal_stress} and~\eqref{eq:pore_fluid_pressure}, respectively, with the hydraulic-to-mechanical contribution to the normal stress $\sigma_{yy}^{HM}(x)$ and the mechanical-to-hydraulic contribution to the pore fluid pressure $p^{MH}(x)$ omitted:
\begin{equation*}
    \sigma_{yy}(x) = \sigma_0+\sigma_{yy}^{M}(x),
    \quad p_p(x) = p_0+p^{H}(x).
\end{equation*}
This auxiliary formulation serves solely as a diagnostic model for isolating the effects of different poroelastic coupling mechanisms and is not intended to represent a distinct physical approximation of the fully coupled model.

\subsubsection{Elastic-response models with backstress}

Elastic-response models with backstress retain the hydraulic-to-mechanical cross-coupling contribution to the normal stress, but approximate the direct mechanical contribution  $\sigma_{yy}^M(x)$ by the purely elastic response, which is recovered by setting $\beta = 0$ \citep{BiEs68}:
\begin{equation}
    \sigma_{yy}^{M, e}(x) = \frac{E'}{4\pi} \int_0^{\infty} \frac{\diff w}{\diff s}\frac{\diff s}{x - s},
    \label{eq:sigma_MM_elastic}
\end{equation}
The use of $\sigma_{yy}^{M,e}(x)$ may be justified if the poroelastic contribution has a negligible influence on fracture characteristics in the near- and intermediate-field. However, employing $\sigma_{yy}^{M,e}(x)$ does not recover the correct far-field asymptotic behaviour of the fracture opening and net pressure. Indeed, the far-field asymptotic analysis in Section~\ref{sec:far_field} shows that the opening--pressure relation is governed by the undrained, rather than drained, mechanical response.

The 2D PDL formulation was presented by \cite{Kovalyshen10}, with governing equations:
\begin{equation*}
    \sigma_{yy}(x) = \sigma_0 + \sigma_{yy}^{M, e}(x) + \sigma_{yy}^{HM}(x), \quad p_p(x) = p_0 + p^H(x).
\end{equation*}

The 1D PDL formulation can be written as:
\begin{equation*}
    \sigma_{yy}(x) = \sigma_0 + \sigma_{yy}^{M, e}(x) + \sigma_{yy}^{HM}(x), \quad p_p(x) = p_0 + p^H_\mathrm{1D}(x) \quad \mathrm{or} \quad v(x) = v_\mathrm{1D}(x).
\end{equation*}

The Carter's leak-off formulation coincides with that of the uncoupled case and is therefore presented in Section~\ref{sec:uncoupled_models}.

\subsubsection{Uncoupled models}
\label{sec:uncoupled_models}

In uncoupled models, pore pressure diffusion does not induce deformation and, conversely, deformation does not generate pore pressure perturbations. These models correspond to the limit $\alpha = 0$. The formulations for the 2D PDL, 1D PDL, and Carter's leak-off cases are obtained from the corresponding one-way coupled models (Section~\ref{sec:one-way_coupled_models}) by setting $\beta = \eta = 0$.

The 2D PDL formulation was presented by \cite{detournay2003near, Kovalyshen10}, with governing equations:
\begin{equation*}
    \sigma_{yy}(x) = \sigma_0 + \sigma_{yy}^{M, e}(x), \quad p_p(x) = p_0 + p^H(x).
\end{equation*}
It should be noted that \cite{detournay2003near} did not explicitly consider the normal stress equation, instead assuming that the fracture opening is governed by the LEFM asymptote \eqref{eq:propagation_criterion}.

The 1D PDL formulation was considered by \cite{kanin2020near}, and can be written as:
\begin{equation*}
    \sigma_{yy}(x) = \sigma_0 + \sigma_{yy}^{M, e}(x), \quad p_p(x) = p_0 + p^H_\mathrm{1D}(x) \quad \mathrm{or} \quad v(x) = v_\mathrm{1D}(x).
\end{equation*}

Finally, \cite{garagash2011multiscale} analysed the Carter's leak-off formulation, for which:
\begin{equation*}
    \sigma_{yy}(x) = \sigma_0 + \sigma_{yy}^{M, e}(x),  \quad v(x) = v_\mathrm{C}(x).
\end{equation*}

\section{Normalization of the governing equations}
\label{sec:mk_scaling}

This section presents the governing equations in dimensionless form. We adopt the $mk-$scaling \citep{garagash2011multiscale}, which is widely used in the analysis of fracture tip problems. This scaling originates from the classical hydraulic fracture tip solution in an impermeable elastic medium, where the near-field behaviour is described by the toughness-dominated $k-$asymptote, and the far-field behaviour by the viscosity-dominated $m-$asymptote. The complete tip solution provides a smooth transition between these two limiting regimes. The characteristic length $\ell_{mk}$ is defined as the distance from the fracture tip at which the formally extended $k-$ and $m-$asymptotes for the fracture opening intersect, and therefore sets the transition length scale between the two regimes. The corresponding opening and pressure scales, $w_{mk}$ and $p_{mk}$, characterize the fracture opening and net pressure at this distance from the tip. The scale of the fluid displacement function, $v_{mk}$, is chosen equal to the opening scale.

\subsection{Governing equations in the $mk-$scaling}
\label{sec:normalized_equations_mk_scaling}

Following the $mk-$scaling, the characteristic length $\ell_{mk}$, opening $w_{mk}$, fluid displacement function $v_{mk}$, and pressure $p_{mk}$ are defined as:
\begin{equation}
    \ell_{mk} = \frac{K_{Ic}^6}{E'^4 V^2 \mu'^2}, \quad w_{mk} = v_{mk} = \frac{K_{Ic}^4}{E'^3 V \mu'}, \quad p_{mk} = \frac{E'^2V\mu'}{K_{Ic}^2}.
    \label{eq:mk_scaling}
\end{equation}

The normalized distance from the fracture tip $\xi$, the fracture opening $\Omega$, the net pressure $\Pi$, the fluid displacement function $\Upsilon$, the normal stress perturbation $\Sigma$, and the pore pressure perturbation $\Pi_p$ are defined as follows:
\begin{equation}
    \xi = \frac{x}{\ell_{mk}}, \quad \Omega = \frac{w}{w_{mk}}, \quad \Pi = \frac{p}{p_{mk}}, \quad \Upsilon = \frac{v}{v_{mk}}, \quad \Sigma = \frac{\sigma_{yy} - \sigma_0}{p_{mk}}, \quad \Pi_p = \frac{p_p - p_0}{p_{mk}}.
    \label{eq:dimensionless_variables_mk_scaling}
\end{equation}

Substituting these dimensionless variables into equations~\eqref{eq:normal_stress}, \eqref{eq:pore_fluid_pressure}, \eqref{eq:closure_relations}, \eqref{eq:lubrication_equation}, and \eqref{eq:propagation_criterion} yields their dimensionless form in the $mk-$scaling:

\begin{itemize}    
    \item poroelastic boundary integral equations:
    \begin{itemize}[label=$\circ$]
        \item normal stress equation:
        \begin{flalign}
            & \Sigma(\xi) = \Sigma^M(\xi) + \Sigma^{HM}(\xi), \text{ with} \nonumber \\
            & \Sigma^M(\xi) = \frac{1}{4\pi (1-\beta)} \int_0^{\infty} \frac{\diff \Omega}{\diff s}  \mathcal{S}^e \left(\frac{\mathcal{S}(\xi-s)}{\varrho}\right) \frac{\diff s}{\xi-s}, \nonumber \\ 
            & \Sigma^{HM}(\xi) = - \frac{\eta}{2\pi \varrho} \int_0^{\infty} \frac{\diff \Upsilon}{\diff s} \mathcal{S}^s \left(\frac{\mathcal{S}(\xi-s)}{\varrho}\right) \diff s;
            \label{eq:normal_stress_dimensionless_mk_scaling}
        \end{flalign}

        \item pore fluid pressure equation:
        \begin{flalign}
            & \Pi_p(\xi) = \Pi^H(\xi) + \Pi^{MH}(\xi), \text{ with}  \nonumber \\
            & \Pi^H(\xi) = \frac{1}{2\pi \varrho} \int_{0}^{\infty} \frac{\diff \Upsilon}{\diff s} \mathcal{P}^s \left(\frac{\mathcal{S}(\xi-s)}{\varrho}\right)\diff s, \nonumber \\
            & \Pi^{MH}(\xi) = - \frac{\eta}{2\pi \varrho}\int_{0}^{\infty} \frac{\diff \Omega}{\diff s} \mathcal{P}^e \left(\frac{\mathcal{S}(\xi-s)}{\varrho}\right) \diff s;
            \label{eq:pore_fluid_pressure_dimensionless_mk_scaling}
        \end{flalign}

    \end{itemize}
    \item closure relations:
    \begin{equation}
        \Sigma(\xi) = \Pi(\xi), \quad \Pi_p(\xi) = \Pi(\xi) + \Sigma_0';
        \label{eq:closure_relations_mk_scaling}
    \end{equation}

    \item lubrication equation:
    \begin{equation}
        \Omega^3 \frac{\diff \Pi}{\diff \xi} = \Omega + \Upsilon;
        \label{eq:lubrication_equation_dimensionless_mk_scaling}
    \end{equation}

    \item propagation criterion:
    \begin{equation}
        \Omega(\xi) = \sqrt{\frac{32}{\pi}}\sqrt{\xi}, \quad \xi \to 0.
        \label{eq:propagation_criterion_dimensionless_mk_scaling}
    \end{equation}
\end{itemize}
In the normalized system, we introduced the following dimensionless parameters:
\begin{equation}
    \varrho = \frac{\kappa}{V} \frac{p_{mk}}{v_{mk}} = \frac{12 k E'^5 V \mu'}{K_{Ic}^6}, \quad \Sigma_0' = \frac{\sigma'_0}{p_{mk}} = \frac{\sigma_0' K_{Ic}^2}{E'^2 V \mu'}, \quad \mathcal{S} = \frac{S E'}{2} = \frac{G S}{1-\nu},
    \label{eq:dimensionless_governing_parameters_mk_scaling}
\end{equation}
where $\varrho$ is the dimensionless permeability, $\Sigma_0'$ is the dimensionless effective confining stress, and $\mathcal{S}$ is the dimensionless storage coefficient \citep{Kovalyshen10}.  

Under the $mk-$scaling, the fully coupled solution is governed by four independent dimensionless parameters, $\varrho$, $\Sigma_0'$, $\beta$, and $\eta$, while the dimensionless storage coefficient $\mathcal{S}$ can be expressed in terms of the dimensionless poroelastic parameters as:
\begin{equation}
    \mathcal{S} = \frac{G S}{1-\nu} = \frac{2 (1 - \beta)\eta^2}{\beta}.
    \label{eq:dimensionless_storage}
\end{equation}

The four independent dimensionless governing parameters can be separated into two groups. The parameters $\beta$ and $\eta$ characterize the poroelastic coupling, whereas, for fixed $\beta$ and $\eta$ (and hence fixed $\mathcal{S}$), the parameters $\varrho$ and $\Sigma_0'$ govern the fluid exchange regime. The parameter combination appearing in the arguments of the integral kernels in the boundary integral equations~\eqref{eq:normal_stress_dimensionless_mk_scaling} and \eqref{eq:pore_fluid_pressure_dimensionless_mk_scaling} can be written as:
\begin{equation*}
    \frac{\mathcal{S}}{\varrho}
    =
    \frac{\ell_{mk}}{\ell_{dif}},
    \quad
    \ell_{dif}=\frac{2c}{V},
\end{equation*}
where $\ell_{dif}$ is the diffusion length scale \citep{detournay2003near, Kovalyshen10}. Consequently, the kernel argument $\mathcal{S}(\xi-s)/\varrho$ represents the separation between the observation and source points along the fracture normalized by the diffusion length scale. The parameter $\Sigma_0'$ is the effective confining stress normalized by the pressure scale $p_{mk}$ and enters directly into the closure relation~\eqref{eq:closure_relations_mk_scaling}, thereby controlling the pressure contrast that drives fluid exchange between the fracture and the surrounding formation. These features make $(\varrho, \Sigma_0')$ a natural parameterization of the fully coupled near-tip problem with two-dimensional fluid exchange. 

In the uncoupled limit $\alpha=0$, the dimensionless poroelastic parameters vanish: $\beta=\eta=0$. However, the dimensionless storage coefficient $\mathcal{S}$ remains finite, as follows from its original definition in equation~\eqref{eq:dimensionless_governing_parameters_mk_scaling}, provided that the storage coefficient $S$ is finite. Therefore, the uncoupled model depends on three independent dimensionless parameters, $\varrho$, $\Sigma_0'$, and $\mathcal{S}$, the latter being treated as an independent governing parameter.

\subsection{Reduced models in dimensionless form}
\label{sec:reduced_models_normalized}

In this section, we present the normalized form of the governing equations for the reduced models introduced in Section~\ref{sec:reduced_models}. We also identify the number of independent dimensionless governing parameters in each formulation.

\subsubsection{One-way coupled models}

In the 2D PDL formulation, the normal stress on the fracture surfaces is governed by equation~\eqref{eq:normal_stress_dimensionless_mk_scaling}, while the pore fluid pressure is given by equation \eqref{eq:pore_fluid_pressure_dimensionless_mk_scaling} without cross-coupling term $\Pi^{MH}(\xi)$:
\begin{equation*}
    \Sigma(\xi) = \Sigma^M(\xi) + \Sigma^{HM}(\xi), \quad \Pi_p(\xi) = \Pi^H(\xi).
\end{equation*}
The solution is governed by four dimensionless parameters, $\varrho$, $\Sigma_0'$, $\beta$, and $\eta$.

In the 1D PDL formulation, the normal stress is governed by \eqref{eq:normal_stress_dimensionless_mk_scaling}, while fluid exchange is described by the reduced pore fluid pressure equation or, alternatively, by the fluid displacement function obtained as its inverse:
\begin{equation*}
    \Sigma(\xi) = \Sigma^M(\xi) + \Sigma^{HM}(\xi), \quad \Pi_p(\xi) = \Pi^H_\mathrm{1D}(\xi) \quad \text{or} \quad \Upsilon\left(\xi\right)=\Upsilon_\mathrm{1D}\left(\xi\right),
\end{equation*}
where
\begin{flalign}
    & \Pi^H_\mathrm{1D}(\xi) = \frac{1}{\pi\zeta^3} \int_{0}^{\xi} \frac{\diff \Upsilon}{\diff s} \frac{\diff s}{\sqrt{\xi - s}},
    \label{eq:pore_fluid_pressure_uncoupled_1D_dimensionless_mk_scaling} \\
    & \Upsilon\left(\xi\right)=\Upsilon_\mathrm{1D}\left(\xi\right)=\zeta^3\int_{0}^{\xi}{\frac{\Pi_p(s)}{\sqrt{\xi-s}}\diff s} = \chi\sqrt{\xi} + \zeta^3\int_{0}^{\xi}{\frac{\Pi(s)}{\sqrt{\xi-s}}\diff s}, \label{eq:fluid_displacement_function_1D_dimensionless}
\end{flalign}
Here, we introduce the dimensionless leak-in and leak-off coefficients,
\begin{equation}
    \zeta = \left(2\sqrt{\frac{2\varrho\mathcal{S}}{\pi}}\right)^{1/3},
    \quad
    \chi = 2\Sigma_0'\zeta^3,
    \label{eq:zeta_chi}
\end{equation}
following \citep{kanin2020near} and \citep{garagash2011multiscale}, respectively. The solution is governed by four dimensionless parameters: $\varrho$, $\Sigma_0'$, $\beta$, and $\eta$.

The parameter $\chi$ is referred to as the dimensionless leak-off coefficient, since it determines the amplitude of the Carter-type contribution, $\chi\sqrt{\xi}$, in \eqref{eq:fluid_displacement_function_1D_dimensionless} \citep{garagash2011multiscale}. The parameter $\zeta$ is referred to as the dimensionless leak-in coefficient \citep{kanin2020near}, while its cube, $\zeta^3$, quantifies the pressure-dependent contribution to the fluid displacement function $\Upsilon_{\mathrm{1D}}(\xi)$. Since $\Pi(\xi)<0$ in the semi-infinite fracture problem, this contribution is negative and therefore represents cumulative leak-in from the surrounding formation into the fracture.

For fixed dimensionless poroelastic parameters $\beta$ and $\eta$, and hence fixed $\mathcal{S}$, the pair $(\chi,\zeta)$ provides a parameterization of the fluid exchange regime equivalent to $(\varrho,\Sigma_0')$. Indeed, equations~\eqref{eq:zeta_chi} can be inverted to give:
\begin{equation*}
    \varrho = \frac{\pi \zeta^6}{8 \mathcal{S}}, \quad \Sigma_0' = \frac{\chi}{2\zeta^3}.
\end{equation*}
Consequently, $\varrho\propto\zeta^6$ characterizes the amount of leak-in, whereas $\Sigma_0'$ characterizes the ratio of leak-off to leak-in contributions through $\chi/\zeta^3=2\Sigma_0'$.

Although the two parameterisations are equivalent when $\beta$ and $\eta$ are fixed, we retain $(\varrho,\Sigma_0')$ in the analysis of the fully coupled problem with two-dimensional fluid exchange. The parameters $\chi$ and $\zeta$ arise naturally in the reduced one-dimensional fluid exchange model, with $\chi$ and $\zeta^3$ appearing as the coefficients of the leak-off and pressure-dependent contributions, respectively. In contrast, $\varrho$ and $\Sigma_0'$ enter directly into the governing equations of the full two-dimensional formulation. Moreover, both $\chi$ and $\zeta$ depend on $\mathcal{S}$ and would therefore mix the fluid exchange parameterisation with the poroelastic coupling parameters when $\beta$ and $\eta$ are varied.

In the Carter's leak-off formulation, the normal stress is governed by \eqref{eq:normal_stress_dimensionless_mk_scaling} without the cross-coupling term $\Sigma^{HM}(x)$, while the fluid displacement function is prescribed and obtained from \eqref{eq:fluid_displacement_function_1D_dimensionless} by setting $\Pi(\xi)=0$:
\begin{equation*}
    \Sigma(\xi) = \Sigma^M(\xi), \quad \Upsilon\left(\xi\right)=\Upsilon_\mathrm{C}\left(\xi\right),
\end{equation*}
where 
\begin{equation}
    \Upsilon_\mathrm{C}\left(\xi\right)=\chi\sqrt{\xi}.
    \label{eq:fluid_displacement_function_Carter_dimensionless}
\end{equation}
In this case, the problem is governed by three dimensionless parameters: $\mathcal{S}/\varrho$, $\beta$, and $\chi$.

\subsubsection{Auxiliary model without cross-coupling terms}

In the auxiliary model without cross-coupling terms, the normal stress and pore fluid pressure on the fracture surfaces are governed by equations~\eqref{eq:normal_stress_dimensionless_mk_scaling} and~\eqref{eq:pore_fluid_pressure_dimensionless_mk_scaling}, respectively, with the terms $\Sigma^{HM}(\xi)$ and $\Pi^{MH}(\xi)$ omitted:
\begin{equation*}
    \Sigma(\xi)=\Sigma^M(\xi), \quad \Pi_p(\xi) = \Pi^H(\xi).
\end{equation*}
Its solution is governed by four independent dimensionless parameters, $\varrho$, $\Sigma_0'$, $\beta$, and $\mathcal{S}$.

\subsubsection{Elastic-response models with backstress}

The 2D PDL formulation \citep{Kovalyshen10} is governed by:
\begin{equation*}
    \Sigma(\xi)=\Sigma^{M,e}(\xi)+\Sigma^{HM}(\xi), \quad \Pi_p(\xi) = \Pi^H(\xi),
\end{equation*}
where the normalized direct mechanical contribution to the normal stress on the fracture surfaces is given by:
\begin{equation}
    \Sigma^{M,e}(\xi) = \frac{1}{4\pi} \int_0^{\infty} \frac{\diff \Omega}{\diff s}  \frac{\diff s}{\xi-s},  \label{eq:normal_stress_uncoupled_dimensionless_mk_scaling} 
\end{equation}
The solution depends on four dimensionless parameters, $\varrho$, $\Sigma_0'$, $\mathcal{S}$, and $\eta$.

The 1D PDL formulation can be written as:
\begin{equation*}
    \Sigma(\xi)=\Sigma^{M,e}(\xi)+\Sigma^{HM}(\xi), \quad \Pi_p(\xi) = \Pi^H_\mathrm{1D}(\xi) \quad \mathrm{or} \quad \Upsilon(\xi) = \Upsilon_\mathrm{1D}(\xi).
\end{equation*}
The solution is governed by four dimensionless parameters, $\varrho$, $\Sigma_0'$, $\mathcal{S}$, and $\eta$.

The Carter's leak-off formulation coincides with the uncoupled case and is presented in Section~\ref{sec:uncoupled_models_dimensionless}.

\subsubsection{Uncoupled models}
\label{sec:uncoupled_models_dimensionless}

In the 2D PDL formulation \citep{detournay2003near, Kovalyshen10}, the normalized governing equations are given by:
\begin{equation*}
    \Sigma(\xi)=\Sigma^{M,e}(\xi), \quad \Pi_p(\xi) = \Pi^H(\xi).
\end{equation*}
The solution is controlled by three dimensionless parameters, $\varrho$, $\Sigma_0'$, $\mathcal{S}$.

The 1D PDL formulation \citep{kanin2020near} can be written as: 
\begin{equation*}
    \Sigma(\xi)=\Sigma^{M,e}(\xi), \quad \Pi_p(\xi) = \Pi^H_\mathrm{1D}(\xi) \quad \mathrm{or} \quad \Upsilon(\xi) = \Upsilon_\mathrm{1D}(\xi),
\end{equation*}
with two governing parameters, $\chi$ and $\zeta$.

Finally, the dimensionless governing equations for the Carter's leak-off formulation \citep{garagash2011multiscale} read:
\begin{equation*}
    \Sigma(\xi)=\Sigma^{M,e}(\xi), \quad \Upsilon(\xi) = \Upsilon_\mathrm{C}(\xi),
\end{equation*}
and the solution is governed by the dimensionless leak-off coefficient $\chi$.

Table~\ref{tab:reduced_models} summarizes the contributions to the normal stress and pore fluid pressure on the fracture surfaces, together with the governing parameters for the fully coupled and all reduced models.

\begin{table}
    \begin{center}
        \def~{\hphantom{0}}
        \begin{tabular}{lccc}
        Model &  Normal stress & Pore fluid pressure & Governing parameters \\ 
        \hline \hline
        Fully coupled & $\Sigma^M$, $\Sigma^{HM}$ & $\Pi^H$, $\Pi^{MH}$ & $\varrho$, $\Sigma_0'$, $\beta$, $\eta$ \\
        \hline
        One-way coupled (2D PDL) & $\Sigma^M$, $\Sigma^{HM}$ & $\Pi^H$ & $\varrho$, $\Sigma_0'$, $\beta$, $\eta$ \\
        One-way coupled (1D PDL) & $\Sigma^M$, $\Sigma^{HM}$ & $\Pi^H_\mathrm{1D}$ or $\Upsilon_\mathrm{1D}$ & $\varrho$, $\Sigma_0'$, $\beta$, $\eta$ \\
        One-way coupled (Carter's leak-off) & $\Sigma^M$ & $\Upsilon_\mathrm{C}$ & $\mathcal{S}/\varrho$, $\beta$, $\chi$ \\
        \hline
        Auxiliary, no cross-coupling (2D PDL)
        & $\Sigma^M$
        & $\Pi^H$
        & $\varrho$, $\Sigma_0'$, $\beta$, $\mathcal{S}$ \\
        \hline
        Elastic response with backstress (2D PDL) & $\Sigma^{M, e}$, $\Sigma^{HM}$ & $\Pi^H$ & $\varrho$, $\Sigma_0'$, $\mathcal{S}$, $\eta$ \\
        Elastic response with backstress (1D PDL) & $\Sigma^{M, e}$, $\Sigma^{HM}$ & $\Pi^H_\mathrm{1D}$ or $\Upsilon_\mathrm{1D}$ & $\varrho$, $\Sigma_0'$, $\mathcal{S}$, $\eta$ \\
        \hline
        Uncoupled (2D PDL) & $\Sigma^{M, e}$ & $\Pi^H$ & $\varrho$, $\Sigma_0'$, $\mathcal{S}$\\
        Uncoupled (1D PDL) & $\Sigma^{M, e}$ & $\Pi^H_\mathrm{1D}$ or $\Upsilon_\mathrm{1D}$ & $\chi$, $\zeta$ \\
        Uncoupled (Carter's leak-off) & $\Sigma^{M, e}$ & $\Upsilon_\mathrm{C}$ & $\chi$\\
        \end{tabular}
    \caption{Summary of governing equations and dimensionless parameters for the fully coupled and reduced models. The parameters $\chi$ and $\zeta$ are related to $\varrho$, $\Sigma_0'$, and $\mathcal{S}$ through \eqref{eq:zeta_chi}.}
    \label{tab:reduced_models}
    \end{center}
\end{table}

\section{Asymptotic limits and numerical solution}
\label{sec:asymptotes_and_numerical_solution}

In this section, we first examine the behaviour of the fracture opening $\Omega(\xi)$, the net pressure $\Pi(\xi)$, and the fluid displacement function $\Upsilon(\xi)$ in the near-field ($\xi \to 0$) and far-field ($\xi \to \infty$) regions of a semi-infinite hydraulic fracture in a poroelastic material. We then briefly describe the numerical solver used to compute the general fully coupled solution.

\subsection{Near-field region}
\label{sec:near_field}

The near-tip behaviour of the fracture opening is described by the LEFM asymptote \eqref{eq:propagation_criterion_dimensionless_mk_scaling}, $\Omega_0(\xi) = \sqrt{32/\pi}\sqrt{\xi}$. Assuming that the fluid pressure at the fracture tip remains non-singular, the lubrication equation~\eqref{eq:lubrication_equation_dimensionless_mk_scaling} requires that the fluid displacement function balances the fracture opening, such that $\Upsilon_0(\xi) = -\Omega_0(\xi)$. This relation ensures a uniform fluid pressure in the vicinity of the fracture tip, $\Pi_0(\xi) = \mathrm{const}$, with its magnitude determined as part of the numerical solution. Since the net pressure is negative along the entire fracture, as a direct consequence of the semi-infinite geometry, this constant is also negative.

Accordingly, the near-field characteristics of the semi-infinite fracture are as follows:
\begin{equation}
    \Omega_0(\xi) = \sqrt{\frac{32}{\pi}}\sqrt{\xi}, \quad \Pi_0(\xi) = \mathrm{const}, \quad \Upsilon_0(\xi) = -\sqrt{\frac{32}{\pi}}\sqrt{\xi}.
    \label{eq:near-field_asymptote_mk_scaling}
\end{equation}

Thus, the near-field is governed by the classical toughness-dominated fracture tip asymptote with drained elastic moduli. In this region, pore fluid influx from the surrounding formation into the fracture locally compensates for the opening-induced increase in fracture volume, so that the fracturing fluid pressure remains uniform at leading order.

\subsection{Far-field region}
\label{sec:far_field}

We seek the far-field asymptotic behaviour of the fracture opening and the fluid displacement function in the form of power-law functions, $\Omega_\infty(\xi) = A_\Omega \xi^{a_\Omega}$ and $\Upsilon_\infty(\xi) = A_\Upsilon \xi^{a_\Upsilon}$.

We first consider equation~\eqref{eq:pore_fluid_pressure_dimensionless_mk_scaling} together with the corresponding closure relation~\eqref{eq:closure_relations_mk_scaling}. The leading-order term of the fluid displacement function is obtained by balancing $\Pi^H(\xi)$ with $\Sigma_0'$, while assuming that the remaining terms, namely the net pressure $\Pi(\xi)$ and the deformation-induced pore pressure perturbation $\Pi^{MH}(\xi)$, are negligible for large $\xi$. Following the methodology of \cite{detournay2003near, Kovalyshen10}, this yields:
\begin{equation*}
    \Sigma_0' = \Pi^H_\infty(\xi) = \frac{a_\Upsilon A_\Upsilon }{2^{3/2}\sqrt{\varrho \mathcal{S}} }\frac{\Gamma (a_\Upsilon)}{\Gamma \left(a_\Upsilon + \frac{1}{2}\right)}\xi^{a_\Upsilon-1/2},
\end{equation*}
where $\Gamma(\cdot)$ denotes the Euler gamma function, giving:
\begin{equation*}
    \Upsilon_\infty(\xi) = 4 \sqrt{\frac{2}{\pi}} \sqrt{\varrho \mathcal{S}}\Sigma_0'\sqrt{\xi} = \chi\sqrt{\xi},
\end{equation*}
where equation~\eqref{eq:zeta_chi} has been used in the second equality. Thus, the dimensionless leak-off coefficient $\chi$ determines the amplitude of the leading-order far-field fluid displacement function in the fully coupled model. This expression coincides with the fluid displacement function prescribed by Carter's leak-off law~\eqref{eq:fluid_displacement_function_Carter_dimensionless} \citep{garagash2011multiscale}. Hence, although fluid exchange in the fully coupled model is two-dimensional and pressure-dependent, the leading-order far-field fluid displacement recovers the classical Carter-type cumulative leak-off response.

We now turn to equation~\eqref{eq:normal_stress_dimensionless_mk_scaling} supplemented by the corresponding closure relation~\eqref{eq:closure_relations_mk_scaling}. Analysis of the term $\Sigma^M(\xi)$ shows that, in the far-field, its contribution is asymptotically equivalent to:
\begin{equation*}
    \Sigma^M_\infty(\xi) = \frac{1}{4\pi (1-\beta)} \int_0^{\infty} \frac{\diff \Omega}{\diff s} \frac{\diff s}{\xi-s},
\end{equation*}
while the term $\Sigma^{HM}(\xi)$ vanishes owing to the leading-order term of the fluid displacement function, with the second-order correction remaining asymptotically small compared to $\Sigma^M_\infty(\xi)$. This limiting form therefore corresponds to the mechanical response of the surrounding rock as an elastic medium characterized by undrained moduli.

Using the fact that, in the far field, the leading-order contribution to the Cauchy-type integral is governed by the leading-order opening asymptote $\Omega_\infty(\xi)$, which reflects the localization property of the Cauchy kernel, we obtain:
\begin{equation*}
    \Pi_\infty(\xi) = \frac{a_\Omega A_\Omega}{4 (1-\beta)}  \cot{(\pi a_\Omega)}\xi^{a_\Omega - 1} 
\end{equation*}

To determine the leading-order far-field opening and pressure, we use the storage-viscosity balance in the lubrication equation~\eqref{eq:lubrication_equation_dimensionless_mk_scaling}, analogous to the $m-$vertex solution of \cite{garagash2011multiscale}. In this balance, the fluid displacement function is assumed to be subdominant relative to the fracture opening on the right-hand side of \eqref{eq:lubrication_equation_dimensionless_mk_scaling}. Upon substituting $\Omega_\infty(\xi)$ and $\Pi_\infty(\xi)$ into \eqref{eq:lubrication_equation_dimensionless_mk_scaling} and neglecting the fluid displacement function relative to the fracture opening, we obtain:
\begin{equation*}
    \Omega_\infty\left(\xi\right)=2^{1/3}3^{5/6}\left(1-\beta\right)^{1/3}\xi^{2/3}, \quad \Pi_\infty(\xi)=-6^{-2/3}\left(1-\beta\right)^{-2/3}\xi^{-1/3}.
\end{equation*}
The far-field net pressure is therefore negative, consistent with the sign of the net pressure along the entire fracture, which follows from the semi-infinite geometry. The assumption that the fluid displacement function is subdominant in the lubrication equation is self-consistent, since $\Upsilon_\infty(\xi) \propto \xi^{1/2}$, whereas $\Omega_\infty(\xi) \propto \xi^{2/3}$; hence $|\Upsilon_\infty|/\Omega_\infty = O(\xi^{-1/6}) \to 0$ as $\xi \to \infty$. Moreover, the obtained pressure asymptote satisfies $|\Pi_\infty(\xi)|/\Sigma_0' = O(\xi^{-1/3}) \to 0$ as $\xi \to \infty$, confirming that the net pressure is negligible compared with the dimensionless effective confining stress in the far field.

To verify the remaining assumptions made during the derivation, we next determine the far-field behaviour of the cross-coupling terms $\Sigma^{HM}(\xi)$ (backstress) and $\Pi^{MH}(\xi)$ (mechanical-to-hydraulic contribution) that were originally neglected. Examination of the term $\Pi^{MH}(\xi)$ indicates that, in the far-field, its contribution is asymptotically equivalent to:
\begin{equation*}
    \Pi^{MH}_\infty(\xi) = \frac{\eta}{2\pi \mathcal{S}}\int_{0}^{\infty} \frac{\diff \Omega}{\diff s} \frac{\diff s}{\xi-s}.
\end{equation*}
Evaluating this Cauchy-type integral using the far-field opening asymptote $\Omega_\infty(\xi)$ gives:
\begin{equation*}
    \Pi^{MH}_\infty(\xi) = -6^{-2/3}\left(1-\beta\right)^{-2/3}\frac{\beta}{\eta}\xi^{-1/3}.
\end{equation*}

The derivation of $\Sigma^{HM}_\infty(\xi)$ requires the second-order correction to the fluid displacement function. Similarly to the leading-order behaviour, we seek the next-order contribution in power-law form, $\Upsilon_\infty^{II}(\xi) = B_{\Upsilon} \xi^{b_\Upsilon}$. Balancing $\Pi_\infty(\xi)$, $\Pi^{MH}_\infty(\xi)$, and the second-order correction to $\Pi^H_\infty(\xi)$ in equation~\eqref{eq:pore_fluid_pressure_dimensionless_mk_scaling}, supplemented by the corresponding closure relation~\eqref{eq:closure_relations_mk_scaling}, yields:
\begin{equation*}
    -6^{-2/3}\left(1-\beta\right)^{-2/3}\xi^{-1/3} = \frac{b_\Upsilon B_\Upsilon }{2^{3/2}\sqrt{\varrho \mathcal{S}} }\frac{\Gamma (b_\Upsilon)}{\Gamma \left(b_\Upsilon + \frac{1}{2}\right)}\xi^{b_\Upsilon-1/2}-6^{-2/3}\left(1-\beta\right)^{-2/3}\frac{\beta}{\eta}\xi^{-1/3},
\end{equation*}
from which we obtain:
\begin{equation*}
    \Upsilon_\infty^{II}(\xi) = 2^{7/3} 3^{1/3}(1-\beta)^{-1/6}(\beta-\eta)\sqrt{\frac{\varrho}{\beta}}\frac{\Gamma\left(\frac{2}{3}\right)}{\Gamma\left(\frac{1}{6}\right)} \xi^{1/6}.
\end{equation*}
Consequently, the leading-order term and the second-order correction of $\Pi^H(\xi)$ are given by:
\begin{equation*}
    \Pi^H_\infty(\xi) = \Sigma_0' + 6^{-2/3}\left(1-\beta\right)^{-2/3} \frac{\beta - \eta}{\eta} \xi^{-1/3}.
\end{equation*}

To derive the far-field behaviour of $\Sigma^{HM}(\xi)$, we rewrite its expression following \cite{Kovalyshen10} as:
\begin{equation*}
    \Sigma^{HM}(\xi) = \frac{\eta}{2\pi\mathcal{S}}\int_{0}^{\infty}{\frac{\diff \Upsilon}{\diff s}\frac{\diff s}{\xi-s}}+\frac{\eta}{2\pi\mathcal{S}}\frac{\diff}{\diff \xi}\int_{0}^{\infty}{\frac{\diff \Upsilon}{\diff s}\mathcal{P}^s\left(\frac{\mathcal{S}\left(\xi-s\right)}{\varrho}\right)\diff s}.
\end{equation*}
The first term is evaluated by substituting the second-order correction to the fluid displacement function, $\Upsilon^{II}_\infty(\xi)$, into the Cauchy-type integral, while the second term follows directly from the second-order correction to $\Pi^H_\infty(\xi)$:
\begin{equation*}
    \Sigma^{HM}_\infty(\xi) = 2^{-2/3}3^{-1/6}\left(1-\beta\right)^{-7/6}\left(\beta-\eta\right)\frac{\sqrt{\beta\varrho}}{\eta}\frac{\Gamma\left(\frac{2}{3}\right)}{\Gamma\left(\frac{1}{6}\right)}\xi^{-5/6}-6^{-5/3}\left(1-\beta\right)^{-5/3}\frac{\beta\varrho\left(\beta-\eta\right)}{\eta^2}\xi^{-4/3}.
\end{equation*}
The derived expressions confirm the self-consistency of the assumptions used in the derivation. The relevant ratios behave as  $|\Pi^{MH}_\infty(\xi)|/\Sigma_0' = O(\xi^{-1/3}) \to 0$, and $|\Sigma^{HM}_\infty(\xi)|/|\Sigma_\infty^M(\xi)| = O(\xi^{-1/2}) \to 0$ as $\xi \to \infty$. Thus, $\Sigma_0' \gg {|\Pi^{MH}_\infty(\xi)|}$ and $|\Sigma^M_\infty(\xi)| \gg |\Sigma^{HM}_\infty(\xi)|$ in the far field.

Collecting the results of this section, the far-field solution for the semi-infinite fracture is given by:
\begin{flalign}
    & \Omega_\infty(\xi) = 2^{1/3}3^{5/6}\left(1-\beta\right)^{1/3}\xi^{2/3}, \quad \Pi_\infty(\xi) = -6^{-2/3}\left(1-\beta\right)^{-2/3}\xi^{-1/3}, \nonumber \\ 
    &\Upsilon_\infty(\xi) = 4 \sqrt{\frac{2}{\pi}} \sqrt{\varrho \mathcal{S}}\Sigma_0'\sqrt{\xi} = \chi\sqrt{\xi}.
    \label{eq:far-field_asymptote_mk_scaling}
\end{flalign}
Thus, the far-field behaviour of a steadily moving semi-infinite hydraulic fracture in a poroelastic medium corresponds to a storage-viscosity-dominated regime, in which the opening-pressure relation is governed by the undrained mechanical response of the surrounding rock. In this limit, the rock therefore behaves as an elastic medium characterized by its undrained moduli.

\subsection{Computation of the general solution}

The general solution of the fully coupled fracture tip problem is obtained numerically. Although the governing equations are formulated above in the $mk-$scaling (Section~\ref{sec:mk_scaling}), the computations are performed in the diffusion scaling introduced in Appendix~\ref{sec:diffusion_scaling}. This choice removes the explicit dependence of the integral kernels on the parameter combination $\mathcal{S}/\varrho$. The resulting solution is then mapped back to the $mk-$scaling. A detailed description of the numerical approach is provided in Appendix~\ref{sec:modeling_approach}. In brief, the solver uses a Chebyshev spectral collocation framework on a mapped finite interval, with the near- and far-field asymptotic behaviours incorporated into the approximation of the gradients of the fracture opening and fluid displacement function.

For each prescribed set of dimensionless governing parameters, $\varrho$, $\Sigma_0'$, $\beta$, and $\eta$, the non-linear system obtained after discretization of the governing equations is solved using the \texttt{root} routine from the \texttt{scipy.optimize} module of the SciPy library \citep{2020SciPy-NMeth}. The computed solution consists of the profiles of the fracture opening $\Omega(\xi)$, net pressure $\Pi(\xi)$, and fluid displacement function $\Upsilon(\xi)$.   

The results presented in Section~\ref{sec:results_discussion} are computed with $N_p=300$ primary nodes. To assess numerical convergence, we repeat the fully coupled calculations with $N_p=100$, $200$, $300$, $400$, $500$, and $600$ for two representative parameter combinations. The corresponding analysis is presented in Appendix~\ref{sec:numerical_convergence}.

\section{Results and discussion}
\label{sec:results_discussion}

\subsection{Representative values of the governing parameters}
\label{sec:representative_values_governing_parameters}

As noted above, the solution in the $mk-$scaling (Section~\ref{sec:normalized_equations_mk_scaling}) depends on four dimensionless parameters: $\varrho$, $\Sigma_0'$, $\beta$ and $\eta$. These parameters are functions of the poroelastic and hydraulic properties of the rock, its fracture toughness, the effective confining stress, the fluid viscosity, and the fracture propagation velocity. To ground the subsequent discussion, we first estimate realistic ranges of these dimensionless parameters.

Three classes of rocks are considered: sandstones, granites, and shales. For each class, table~\ref{tab:properties_of_different_rocks} compiles experimentally measured poroelastic and hydraulic properties of different rocks reported in the literature. It is important to bear in mind that these properties may depend significantly on the applied effective confining stress. The experimental values listed here correspond to conditions typical of $\sigma_0' \in [10,  20]$ MPa. 

We report the drained Young's modulus $E$, the drained Poisson's ratio $\nu$, the Biot coefficient $\alpha$, the porosity $\phi$, the solid grain bulk modulus $K_s$, and the permeability $k$. From these parameters, we then compute the corresponding values of the undrained Poisson's ratio $\nu_u$, the dimensionless parameter $\beta$, and the poroelastic stress coefficient $\eta$. The Biot modulus $M$ can be expressed as \citep{detournay1993fundamentals, coussy2004poromechanics}:
\begin{equation*}
    \frac{1}{M} = \frac{\phi}{K_f} + \frac{\alpha - \phi}{K_s},
\end{equation*}
where the fluid bulk modulus is taken as $K_f = 1.5$ GPa, corresponding to a light oil. The Biot modulus, together with the drained elastic moduli and Biot coefficient, determines $\nu_u$ via \eqref{eq:undrained_poisson_ratio}. The dimensionless poroelastic parameters are evaluated from $\beta=(\nu_u-\nu)/(1-\nu)$ and $\eta=\alpha (1-2\nu)/[2 (1-\nu)]$.

\begin{table}
    \begin{center}
        \def~{\hphantom{0}}
        \begin{tabular}{lccccccccc}
        Rock & $E$, GPa & $\nu$ & $\alpha$ & $\phi$ & $K_s$, GPa & $\nu_u$ & $k$, mD & $\beta$ & $\eta$ \\ 
        \hline \hline
        Ruhr sandstone & 29.1 & 0.12 & 0.65	& 0.02 & 36 & 0.29 & 0.2 & 0.19 & 0.28 \\
        Berea sandstone & 14.4 & 0.2 & 0.79 & 0.19 & 36 & 0.29 & 190 & 0.11 & 0.3 \\
        Weber sandstone & 27.6 & 0.15 & 0.64 & 0.06 & 36 & 0.26 & 1 & 0.12 & 0.26 \\
        Ohio sandstone & 16.1 & 0.18 & 0.74 & 0.19 & 31 & 0.26 & 5.6 & 0.1 & 0.3 \\
        Pecos sandstone & 13.7 & 0.16 & 0.83 & 0.2 & 39 & 0.28 & 0.8 & 0.14 & 0.34 \\
        Boise sandstone & 9.7 & 0.15 & 0.85 & 0.26 & 42 & 0.29 & 800 & 0.16 & 0.35 \\
        Kayenta sandstone & 15.3 & 0.22 & 0.76 & 0.2 & 37.9 & 0.29 & 700 & 0.09 & 0.27 \\
        Fontainebleau sandstone & 11.3 & 0.23 & 0.82 & 0.09 & 36 & 0.37 & 40.5 & 0.18 & 0.29 \\
        St. Peter sandstone & 21.6 & 0.2 & 0.63 & 0.1 & 35 & 0.27 & 25 & 0.09 & 0.24 \\
        Flechtinger sandstone & 9.3 & 0.22 & 0.87 & 0.13 & 41.5 & 0.37 & 0.9 & 0.19 & 0.31 \\
        Bentheimer sandstone & 12.9 & 0.22 & 0.75 & 0.24 & 27 & 0.29 & 870 & 0.09 & 0.27 \\
        Navajo sandstone & 13.4 & 0.22 & 0.75 & 0.2 & 34 & 0.3 & 76 & 0.1 & 0.27 \\
        \hline
        Charcoal granite & 48.3 & 0.27 & 0.27 & 0.02 & 45 & 0.29 & $4 \cdot 10^{-4}$ & 0.03 & 0.09 \\
        Westerly granite & 37.5 & 0.25 & 0.47 & 0.01 & 45 & 0.33 & $10^{-4}$ & 0.1 & 0.16 \\
        Sierra White granite & 43.7 & 0.22 & 0.56 & 0.008 & 50.4 & 0.33 & $7 \cdot 10^{-4}$ & 0.14 & 0.2 \\
        Lac du Bonnet granite & 70 & 0.21 & 0.3 & 0.003 & 58 & 0.27 & $10^{-4}$ & 0.08 & 0.11 \\
        Stanstead granite & 66 & 0.16 & 0.44 & 0.008 & 58 & 0.25 & 0.006 & 0.11 & 0.18 \\
        Grimsel granite & 40.5 & 0.25 & 0.54 & 0.005 & 59 & 0.36 & $6\cdot 10^{-4}$ & 0.14 & 0.18 \\
        Aue granite & 53.9 & 0.19 & 0.35 & 0.002 & 45 & 0.28 & 0.009 & 0.11 & 0.13 \\
        Barre granite & 30 & 0.25 & 0.65 & 0.007 & 55 & 0.38 & $2 \cdot 10^{-4}$ & 0.17 & 0.22 \\
        Chelmsford granite & 19.5 & 0.25 & 0.75 & 0.001 & 55 & 0.43 & $7 \cdot 10^{-4}$ & 0.24 & 0.25 \\
        \hline
        Opalinus clay & 3.2 & 0.3 & 0.7 & 0.13 & 8.9 & 0.4 & $7 \cdot 10^{-6}$ & 0.15 & 0.2 \\
        Eau Claire shale & 19.2 & 0.18 & 0.8 & 0.1 & 49.3 & 0.3 & 0.003 & 0.15 & 0.31 \\
        Callovo-Oxfordian clay & 3.6 & 0.3 & 0.86 & 0.14 & 21.7 & 0.43 & $10^{-5}$ & 0.18 & 0.25 \\
        Bakken shale & 14.5 & 0.26 & 0.78 & 0.007 & 42.6 & 0.43 & $10^{-6}$ & 0.22 & 0.25 \\
        Mancos shale & 24 & 0.25 & 0.6 & 0.007 & 40.3 & 0.37 & 0.02 & 0.16 & 0.2 \\
        Marcellus shale & 26.4 & 0.21 & 0.65 & 0.008 & 43.3 & 0.36 & $7 \cdot 10^{-4}$ & 0.19 & 0.24 \\
        \end{tabular}
    \caption{
        Poroelastic and hydraulic properties of various materials. Representative formations include the following sandstones: Ruhr, Berea, Weber, Ohio, Pecos, and Boise \citep{detournay1993fundamentals}, Kayenta \citep{wang2000theory}, Fontainebleau \citep{zamora1990experimental, song2008hydromechanical}, St. Peter \citep{ma2019volumetric}, Flechtinger and Bentheimer \citep{blocher2014direct, hassanzadegan2014effects}, and Navajo \citep{hasanov2019hydraulic}; the following granites: Charcoal and Westerly \citep{detournay1993fundamentals}, Sierra White \citep{miller1991laboratory, zhou2024poroelastic}, Lac du Bonnet and Stanstead \citep{selvadurai2021poroelastic}, Grimsel \citep{selvadurai2019multi}, Aue \citep{tan2015experimental}, Barre and Chelmsford \citep{coyner1984effects}; the following shales/claystones: Opalinus and Eau Claire \citep{kim2020coupling}, Callovo-Oxfordian \citep{belmokhtar2017poroelasticity}, Bakken \citep{he2016new, ling2016comparisons}, Mancos \citep{gutierrez2015determination}, and Marcellus \citep{el2019impact, yildirim2025new}.
    } 
    \label{tab:properties_of_different_rocks}
    \end{center}
\end{table}

We then determine representative ranges of the effective confining stress $\sigma_0'$ for each rock class. To this end, we assume a tectonically relaxed sedimentary basin under a normal-faulting stress regime and estimate the minimum horizontal stress using Eaton's relation \citep{eaton1969fracture}:
\begin{equation*}
    \sigma_0 = \frac{\nu}{1-\nu}\sigma_V + \alpha p_p \frac{1-2\nu}{1-\nu},
\end{equation*}
where $\sigma_V = \rho_s g H$ denotes the vertical stress, and $p_p = k_p \rho_f g H$ is the pore fluid pressure. Here, $\rho_s$ and $\rho_f$ are the overlying rock and fluid densities, $g$ is the gravitational acceleration, $H$ is the formation depth, and $k_p$ is a dimensionless coefficient characterizing depleted ($k_p < 1$), hydrostatic ($k_p = 1$), and overpressured ($k_p > 1$) conditions. 
We adopt different ranges for $\nu$, $\alpha$, $k_p$, and $\rho_s$ for each rock class (table~\ref{tab:properties_of_rock_classes}), while taking $\rho_s \in [2200, 2600]$ kg/m$^3$, $\rho_f = 10^3$ kg/m$^3$, and $H \in [1, 5]$ km as common values. The ranges of drained Poisson's ratio and Biot coefficient are consistent with the values reported for individual materials in table~\ref{tab:properties_of_different_rocks}. The resulting estimates yield ranges of $\sigma_0'$ reported in table~\ref{tab:properties_of_rock_classes}. We set a lower bound for $\sigma_0'$  to 1 MPa.

\begin{table}
    \begin{center}
        \def~{\hphantom{0}}
        \begin{tabular}{lcccccc}
        Rock class & $\nu$ & $\alpha$ & $k_p$ & $\sigma_0'$, MPa & $K_{Ic}$, MPa $\cdot$ m$^{1/2}$ & $k$, mD \\ 
        \hline \hline
        Sandstones & $[0.1, 0.25]$ & $[0.6, 0.9]$ & $[0.4, 2]$ & $[1, 35]$ & $[0.5, 2]$ & $[0.1, 10^3]$ \\
        Granites & $[0.15, 0.3]$ & $[0.3, 0.8]$ & $[0.5, 1.5]$ & $[1, 40]$ & $[1.5, 3]$ & $[10^{-5}, 10^{-2}]$ \\
        Shales/Clays & $[0.15, 0.35]$ & $[0.6, 1]$ & $[0.5, 2]$ & $[1, 55]$ & $[0.5, 2]$ & $[10^{-6}, 0.1]$ \\
        \end{tabular}
    \caption{
    Representative ranges of poroelastic, hydraulic, and fracture properties for different rock classes.
    } 
    \label{tab:properties_of_rock_classes}
    \end{center}
\end{table}

To estimate representative values of $\varrho$ and $\Sigma_0'$ \eqref{eq:dimensionless_governing_parameters_mk_scaling}, for each material listed in table~\ref{tab:properties_of_different_rocks}, we vary the fracture toughness $K_{Ic}$, fluid viscosity $\mu$, permeability $k$, effective confining stress $\sigma_0'$, and crack propagation velocity $V$. The ranges of $K_{Ic}$, $\sigma_0'$, and $k$ for sandstones, granites, and shales are given in table~\ref{tab:properties_of_rock_classes}. We adopt common ranges for the fluid viscosity $\mu \in [1, 5]$ cP and the fracture propagation velocity $V \in [10^{-3}, 1]$ m/s for all rock classes. The permeability values reported for individual materials in table~\ref{tab:properties_of_different_rocks} are not used directly, but instead serve to define a realistic range for each rock class listed in table~\ref{tab:properties_of_rock_classes}.

This procedure may be interpreted as follows. A rock formation, characterized by given poroelastic properties, may exhibit a range of hydraulic and fracture properties and be subject to different initial conditions, while the fracture propagates over a range of velocities. The resulting combinations yield a cloud of points in the $(\varrho, \Sigma_0')$ parameter space for each rock class, with boundaries that can be approximated by power-law functions.

\begin{figure}[h!]
    \centerline{
    \includegraphics[width=1\textwidth]{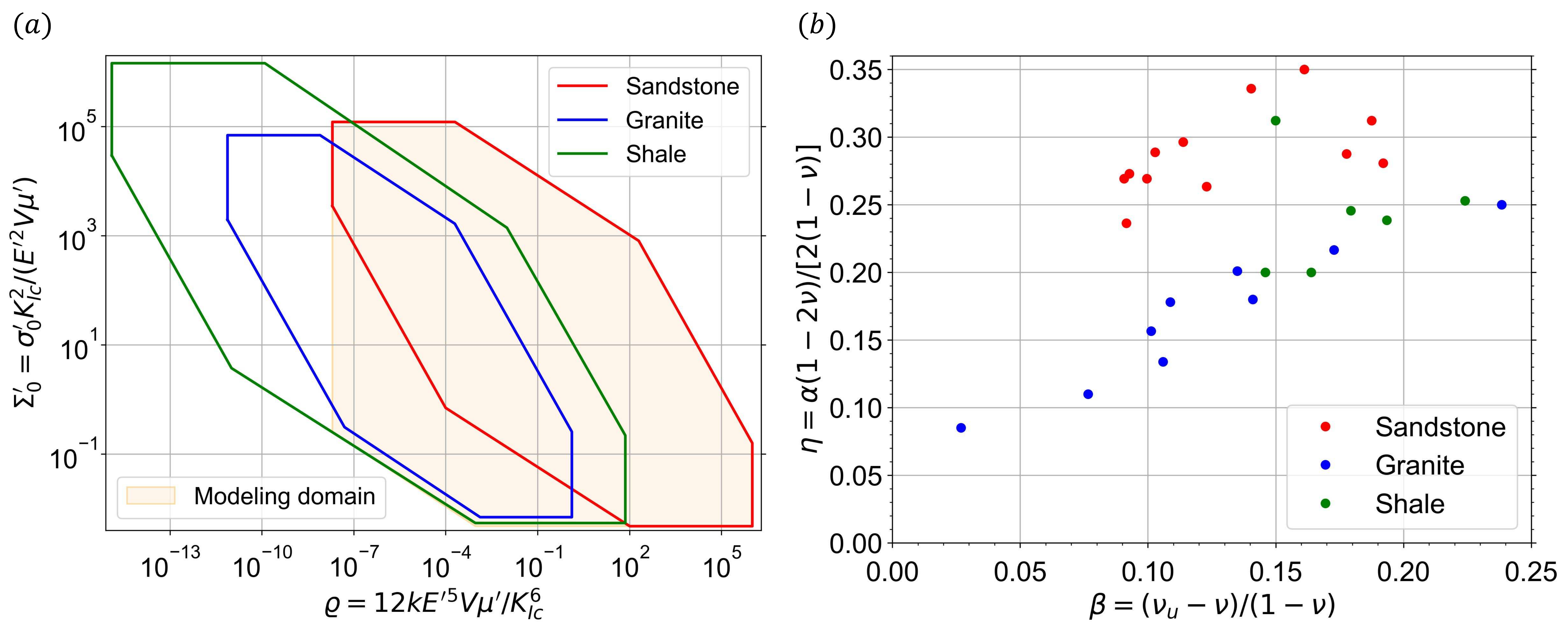}
    }
    \caption{
    Representative ranges of the governing parameters for sandstones (red), granites (blue), and shales (green). Panel (a) shows the admissible regions in the $(\varrho, \Sigma_0')$ parameter space, determined using the procedure described in the main text; the orange-shaded region indicates the modelling domain considered in this study. Panel (b) presents the values of the dimensionless parameters $\beta$ and $\eta$ for the materials listed in table~\ref{tab:properties_of_different_rocks}. 
    }
    \label{fig:representative_ranges_governing_parameters}
\end{figure}

The resulting admissible regions in the $(\varrho, \Sigma_0')$ parameter space are shown in figure~\ref{fig:representative_ranges_governing_parameters}a, together with the domain analysed in the present study. Modelling in the region to the left of this domain, which includes portions of the granite and shale zones, is not pursued because fluid exchange between the fracture and the surrounding formation is negligible. This will be confirmed by the results presented in Section~\ref{sec:variation_rho}: the solution in this low-$\varrho$ region approaches the effectively impermeable response, with poroelastic effects associated with fluid exchange having only a limited influence on the fracture fields.

Figure~\ref{fig:representative_ranges_governing_parameters}b shows the cloud of points in the $(\beta, \eta)$ parameter space for the materials listed in table~\ref{tab:properties_of_different_rocks}. Based on this figure, we identify the following admissible ranges: $\beta \in [0.03, 0.24]$ and $\eta \in [0.09, 0.35]$.

\subsection{Solution across the dimensionless parameter space}
\label{sec:general_solution}

As shown in figure~\ref{fig:representative_ranges_governing_parameters}, the dimensionless permeability $\varrho$ and dimensionless effective confining stress $\Sigma_0'$ span considerably broader ranges than the poroelastic coupling parameters $\beta$ and $\eta$. More importantly, the two parameter pairs govern distinct physical aspects of the problem. The parameters $\varrho$ and $\Sigma_0'$ characterize fluid exchange between the fracture and the surrounding porous medium: $\sqrt{\varrho} \propto \zeta^3$ scales the leak-in contribution, whereas $\Sigma_0' \propto \chi/\zeta^3$ measures the relative importance of leak-off and leak-in. By contrast, $\beta$ and $\eta$ describe the poroelastic coupling, with $\beta$ quantifying the contrast between the drained and undrained elastic responses of the rock and $\eta$ controlling the strength of the coupling between pore pressure and rock deformation.

We therefore fix the poroelastic coupling parameters at representative values, $\beta=0.15$ and $\eta=0.3$, and first explore the fully coupled solution across the $(\varrho,\Sigma_0')$ fluid exchange parameter space within the modelling domain highlighted in orange in figure~\ref{fig:representative_ranges_governing_parameters}a. We examine the influence of $\varrho$ and $\Sigma_0'$ on the fracture opening $\Omega(\xi)$, net pressure $\Pi(\xi)$, and fluid displacement function $\Upsilon(\xi)$ as functions of the dimensionless distance from the fracture tip, $\xi=x/\ell_{mk}$. The influence of variations in $\beta$ and $\eta$ within their admissible ranges is examined in Appendix~\ref{sec:beta_eta_variation} for the representative points analysed further in Section~\ref{sec:solution_structure_in_regimes}. 

We consider three types of variation in the $(\varrho,\Sigma_0')$ parameter space. First, we vary the dimensionless permeability $\varrho$ at fixed $\Sigma_0'$. When the remaining dimensionless governing parameters are fixed, this variation may be interpreted as a change in rock permeability, with larger values of $\varrho$ corresponding to fracture propagation in more permeable formations. In terms of the alternative fluid exchange parameterisation $(\chi, \zeta)$, both $\zeta$ and $\chi$ increase with $\varrho$. Second, we vary the dimensionless effective confining stress $\Sigma_0'$ at fixed $\varrho$. Increasing $\Sigma_0'$ under otherwise fixed conditions may be interpreted as an increase in the effective confining stress, corresponding, for example, to a transition of the initial formation state from overpressured towards depleted reservoir conditions. Along this path, $\zeta$/leak-in remains unchanged, whereas $\chi$/leak-off increases with $\Sigma_0'$. Finally, we vary $\varrho$ and $\Sigma_0'$ simultaneously while keeping their product constant. This path in the $(\varrho,\Sigma_0')$ parameter space is motivated by the progressive decrease in fracture propagation velocity during hydraulic fracture growth in practical applications, at both laboratory and field scales. Indeed, for fixed rock properties and in-situ effective confining stress, $\varrho \propto V$, whereas $\Sigma_0' \propto V^{-1}$. A decrease in propagation velocity therefore leads to a decrease in $\varrho$ and a simultaneous increase in $\Sigma_0'$, while their product, $\varrho \Sigma_0' = 12 k E'^3 \sigma_0'/K_{Ic}^4$, remains constant. Equivalently, along this path, $\zeta$ decreases and $\chi$ increases, while the product $\chi\zeta^3$ remains constant. 

\subsubsection{Effect of the dimensionless permeability $\varrho=\dfrac{12 k E'^5 V \mu'}{K_{Ic}^6}$}
\label{sec:variation_rho}

We first examine how the solution depends on the dimensionless permeability $\varrho$, which governs the strength of the leak-in contribution. To this end, numerical calculations are performed along horizontal transects of the modelling domain (figure~\ref{fig:representative_ranges_governing_parameters}a), at fixed values of the dimensionless effective confining stress, $\Sigma_0' = \{10^{-2}, 10, 10^4\}$, while varying $\varrho$. 

\begin{figure}[h!]
    \centerline{
    \includegraphics[width=1\textwidth]{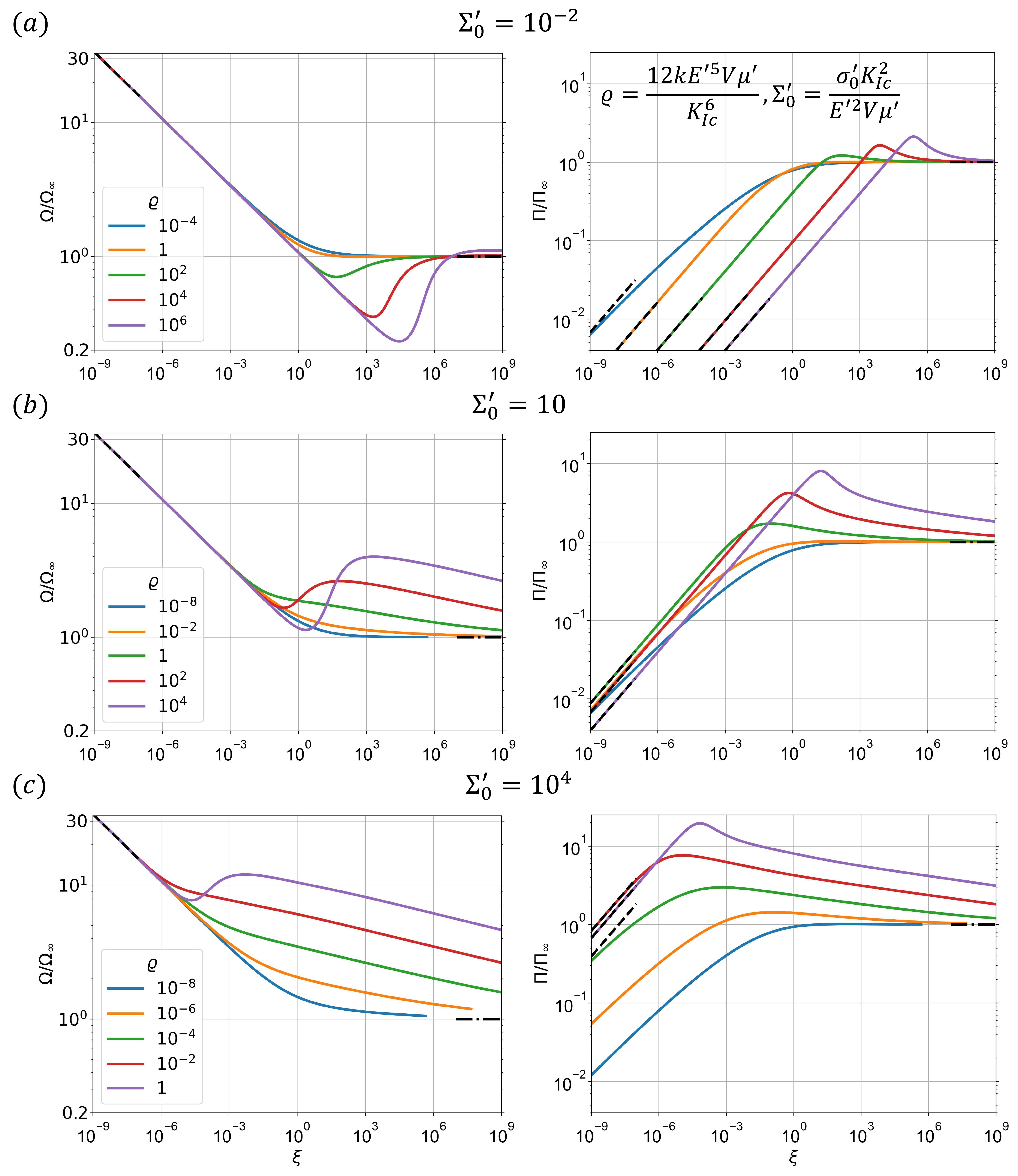}
    }
    \caption{
    Fracture opening $\Omega(\xi)$ (left column) and net pressure $\Pi(\xi)$ (right column) profiles for varying dimensionless permeability $\varrho$ along three horizontal transects of the modelling domain: $\Sigma_0' = 10^{-2}$ (a), $\Sigma_0' = 10$ (b), and $\Sigma_0' = 10^4$ (c). The dimensionless poroelastic parameters are fixed at $\beta = 0.15$ and $\eta = 0.3$. The profiles are shown in normalized form as $\Omega/\Omega_\infty$ and $\Pi/\Pi_\infty$, where $\Omega_\infty$ and $\Pi_\infty$ are the far-field asymptotes given by \eqref{eq:far-field_asymptote_mk_scaling}. The different extents of the profiles in $\xi$ result from the parameter-dependent transformation of the common computational interval from the diffusion scaling to the $mk-$scaling. The black dashed lines indicate the near-field asymptotic behaviour~\eqref{eq:near-field_asymptote_mk_scaling}, whereas the black dash-dotted lines denote the far-field behaviour~\eqref{eq:far-field_asymptote_mk_scaling}. For the net pressure, the near-field asymptote is shown only when it is attained within the plotted range of $\xi=x/\ell_{mk}$.
    }
    \label{fig:variation_rho}
\end{figure}

The fracture opening and net pressure profiles, normalized by their far-field asymptotes~\eqref{eq:far-field_asymptote_mk_scaling}, are shown in figure~\ref{fig:variation_rho}. This normalization emphasizes differences in the intermediate field, while the fracture opening itself always increases with distance from the tip. 

The results demonstrate that, for sufficiently small $\varrho$, the opening profiles collapse onto a common curve, while differences in the net pressure are confined mainly to the near-tip plateau, where the pressure remains uniform. This indicates that fluid exchange between the fracture and the surrounding permeable formation is negligible, such that the fracture fields are close to those for an effectively impermeable rock. The value of $\varrho$ at which the solution begins to depart from this limit depends on $\Sigma_0'$.

Increasing $\varrho$ broadens the transition between the near-field and far-field regimes and shifts it farther away from the fracture tip. The fracture fields reach their far-field asymptotes at larger distances as a result. For small $\varrho$, the opening and net pressure profiles remain monotonic. As $\varrho$ increases, both fields develop a non-monotonic behaviour within the intermediate region, reflecting the growing influence of fluid exchange and poroelastic effects. Thus, at fixed $\Sigma_0'$, the dimensionless permeability $\varrho$ controls the spatial scale over which the fully coupled solution departs from the limiting asymptotic behaviours. 

The spatial range over which the near-field asymptote remains valid depends on both $\varrho$ and $\Sigma_0'$. For $\Sigma_0'=10^{-2}$ and $10$, this range extends farther from the fracture tip as $\varrho$ increases, for both the opening and the net pressure. By contrast, for $\Sigma_0'=10^4$, the spatial extent of the near-field domain for the opening varies non-monotonically with $\varrho$, although it generally decreases as $\varrho$ increases, whereas the corresponding range for the net pressure continues to expand. 

The tip value of the dimensionless net pressure, before normalization by the far-field asymptote, is negative, as is the entire net pressure profile, owing to the semi-infinite fracture geometry. Its variation with $\varrho$ exhibits different trends depending on $\Sigma_0'$. For $\Sigma_0'=10^{-2}$, it increases monotonically with $\varrho$, meaning that it becomes less negative and its magnitude decreases. For $\Sigma_0'=10$ and $10^4$, it varies non-monotonically, with an overall tendency to become less negative in the former case and more negative in the latter. This complex behaviour reflects the competition between fluid exchange and poroelastic effects, whose relative importance is controlled jointly by $\varrho$ and $\Sigma_0'$.

The overall behaviour of the opening and net pressure profiles is \textit{qualitatively} consistent with that reported by \cite{kanin2020near} for an uncoupled fracture tip model with one-dimensional pressure-dependent fluid exchange. In that study, simultaneously increasing the leak-off and leak-in parameters, $\chi$ and $\zeta$, at fixed $\chi/\zeta^3$ produced a similar evolution of the fracture fields. For fixed poroelastic parameters, $\beta$ and $\eta$, this path corresponds to an equivalent trajectory in the $(\varrho,\Sigma_0')$ fluid exchange parameter space, along which $\varrho$ increases while $\Sigma_0'$ remains fixed.

\begin{figure}[h!]
    \centerline{
    \includegraphics[width=1\textwidth]{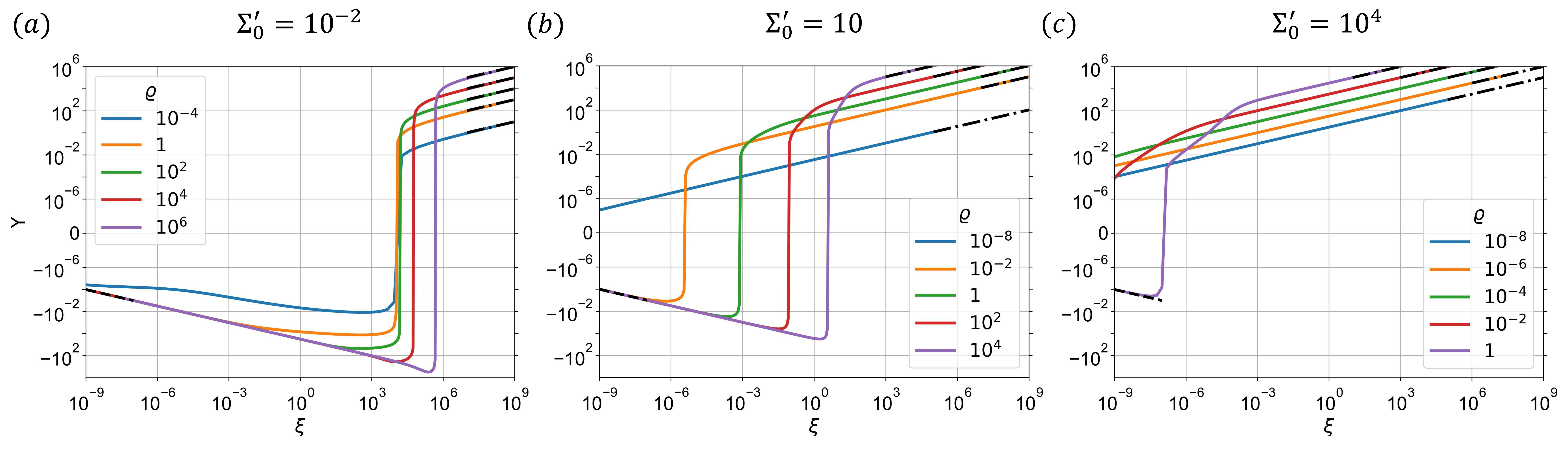}
    }
    \caption{
    Profiles of the fluid displacement function $\Upsilon(\xi)$ for varying dimensionless permeability $\varrho$ along three horizontal transects of the modelling domain: $\Sigma_0' = 10^{-2}$ (a), $\Sigma_0' = 10$ (b), and $\Sigma_0' = 10^4$ (c). The dimensionless poroelastic parameters are fixed at $\beta = 0.15$ and $\eta = 0.3$. The profiles span different ranges of $\xi$ owing to the parameter-dependent transformation from the diffusion scaling to the $mk-$scaling. The black dashed lines indicate the near-field asymptotic behaviour~\eqref{eq:near-field_asymptote_mk_scaling}, whereas the black dash-dotted lines denote the far-field behaviour~\eqref{eq:far-field_asymptote_mk_scaling}. A symmetric logarithmic scale is used on the vertical axis to show both negative and positive values of $\Upsilon$.
    }
    \label{fig:variation_rho_Y}
\end{figure}

The dimensionless fluid displacement function profiles are shown in figure~\ref{fig:variation_rho_Y}.
The signed profiles of $\Upsilon(\xi)$ reveal a transition from negative to positive values when the corresponding zero-crossing lies within the plotted range. In some cases, particularly in panels (b) and (c), the zero-crossing is not visible because it occurs closer to the fracture tip than the lower bound of the plotted range of $\xi$. Since $\Upsilon$ represents cumulative fluid exchange, its sign reflects the net balance of fluid exchange between the fracture tip and a given position. Near the fracture tip $\Upsilon<0$, indicating that cumulative leak-in from the surrounding formation into the fracture dominates. Farther away from the tip $\Upsilon>0$, indicating that cumulative leak-off from the fracture into the porous medium dominates. The interval over which $\Upsilon$ changes sign from negative to positive may therefore be interpreted as a local circulation zone, where fluid enters the fracture near the tip and returns to the surrounding rock farther from the tip, so that the cumulative fluid exchange vanishes at the zero-crossing of $\Upsilon$. The existence of such a circulation zone was previously discussed by \cite{detournay2003near} and \cite{kanin2020near}, who analysed fracture tip solutions with two-dimensional and one-dimensional pressure-dependent fluid exchange, respectively. The size of this circulation zone increases with $\varrho$: as permeability increases, hydraulic communication between the fracture and the surrounding rock becomes more efficient, shifting the sign change of $\Upsilon$ farther away from the fracture tip. At the same time, the far-field magnitude of the dimensionless fluid displacement function increases with $\varrho$ at fixed $\Sigma_0'$.

\subsubsection{Effect of the dimensionless effective confining stress $\Sigma_0'=\dfrac{\sigma_0' K_{Ic}^2}{E'^2 V \mu'}$}

We now consider vertical transects of the modelling domain (figure~\ref{fig:representative_ranges_governing_parameters}a), corresponding to fixed values of the dimensionless permeability, $\varrho = \{10^{-3}, 10, 10^5\}$, while varying $\Sigma_0'$, which measures the relative importance of leak-off and leak-in. 

\begin{figure}[h!]
    \centerline{
    \includegraphics[width=1\textwidth]{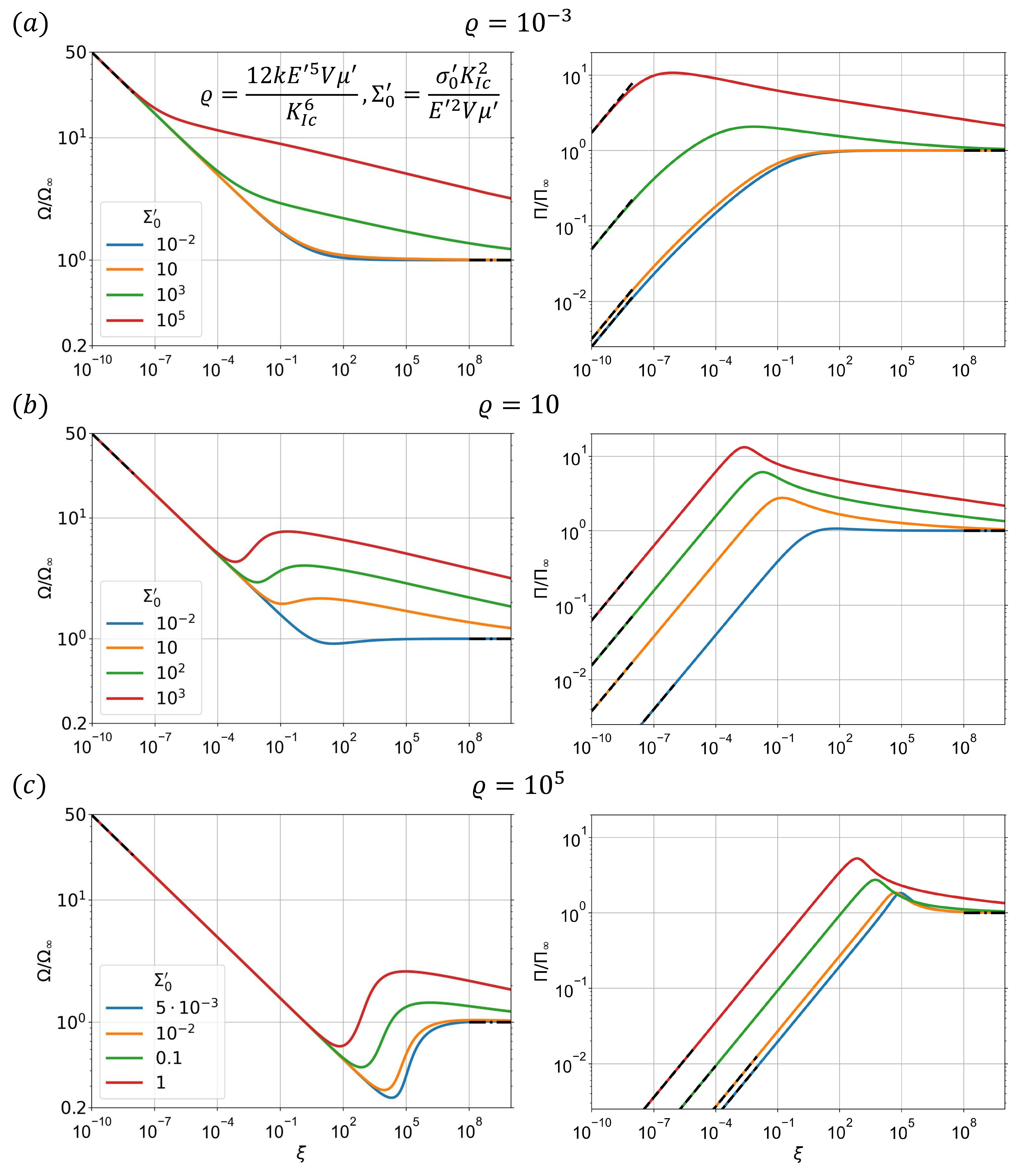}
    }
    \caption{
    Fracture opening $\Omega(\xi)$ (left column) and net pressure $\Pi(\xi)$ (right column) profiles for varying dimensionless effective confining stress $\Sigma_0'$ along three vertical transects of the modelling domain: $\varrho = 10^{-3}$ (a), $\varrho = 10$ (b), and $\varrho = 10^5$ (c). The dimensionless poroelastic parameters are fixed at $\beta = 0.15$ and $\eta = 0.3$. The profiles are shown in normalized form as $\Omega/\Omega_\infty$ and $\Pi/\Pi_\infty$, where $\Omega_\infty$ and $\Pi_\infty$ are the far-field asymptotes given by \eqref{eq:far-field_asymptote_mk_scaling}. The black dashed lines indicate the near-field asymptotic behaviour~\eqref{eq:near-field_asymptote_mk_scaling}, whereas the black dash-dotted lines denote the far-field behaviour~\eqref{eq:far-field_asymptote_mk_scaling}.
    }
    \label{fig:variation_sigma}
\end{figure}

The fracture opening and net pressure profiles, similarly normalized by their far-field asymptotes~\eqref{eq:far-field_asymptote_mk_scaling}, are shown in figure~\ref{fig:variation_sigma}. 
For $\varrho = 10^{-3}$ and sufficiently small $\Sigma_0'$, the opening profiles collapse onto a common curve, while the net pressure profiles differ only in the near-tip region. In this limit, corresponding to a low-permeability formation under overpressured conditions, fluid exchange between the fracture and the surrounding rock is weak, such that the solution approaches that for an effectively impermeable medium. As $\Sigma_0'$ increases, the profiles progressively depart from this limiting response. This indicates that increasing the effective confining stress, or equivalently decreasing the initial pore pressure under otherwise fixed conditions, enhances pressure-dependent fluid exchange even when the dimensionless permeability $\varrho$ is relatively small.

Increasing $\Sigma_0'$ reduces the range over which the near-field asymptote remains valid and shifts the approach to the far-field asymptote farther away from the fracture tip. As a result, the intermediate region expands, and both the fracture opening and the net pressure develop increasingly pronounced non-monotonic behaviour. An exception is observed for $\varrho=10^{-3}$ at sufficiently small $\Sigma_0'$, where the profiles remain monotonic. The tip value of the dimensionless net pressure, before normalization by the far-field asymptote, is negative and decreases monotonically with increasing $\Sigma_0'$ for all values of $\varrho$ considered; equivalently, its absolute magnitude increases.

\begin{figure}[h!]
    \centerline{
    \includegraphics[width=1\textwidth]{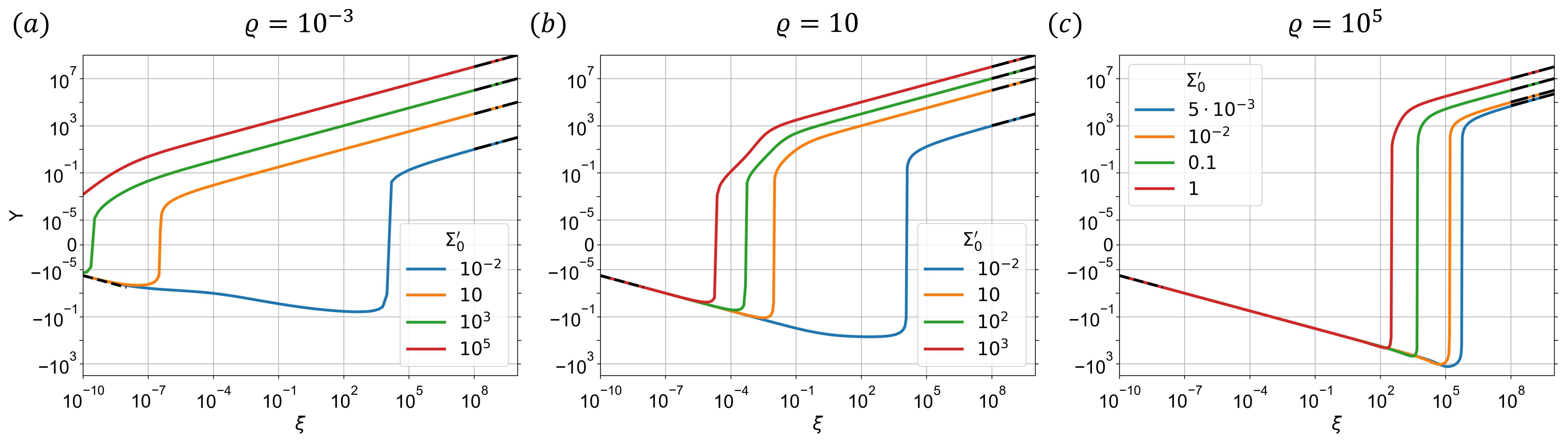}
    }
    \caption{
    Profiles of the fluid displacement function $\Upsilon(\xi)$ for varying dimensionless effective confining stress $\Sigma_0'$ along three vertical transects of the modelling domain: $\varrho = 10^{-3}$ (a), $\varrho = 10$ (b), and $\varrho = 10^5$ (c). The dimensionless poroelastic parameters are fixed at $\beta = 0.15$ and $\eta = 0.3$. The black dashed lines indicate the near-field asymptotic behaviour~\eqref{eq:near-field_asymptote_mk_scaling}, whereas the black dash-dotted lines denote the far-field behaviour~\eqref{eq:far-field_asymptote_mk_scaling}. A symmetric logarithmic scale is used on the vertical axis to show both negative and positive values of $\Upsilon$.
    }
    \label{fig:variation_sigma_Y}
\end{figure}

The dimensionless fluid displacement function profiles are shown in figure~\ref{fig:variation_sigma_Y}.
As $\Sigma_0'$ increases, the zero-crossing of the fluid displacement function, corresponding to the transition from $\Upsilon(\xi)<0$ to $\Upsilon(\xi)>0$, shifts closer to the fracture tip for all values of $\varrho$ considered. In other words, the spatial extent of the near-tip circulation zone decreases with increasing effective confining stress. At the same time, increasing $\Sigma_0'$ at fixed $\varrho$ increases the far-field magnitude of the dimensionless fluid displacement function.

\subsubsection{Effect of the fracture propagation velocity $V$}

We finally examine the solution along a trajectory in the $(\varrho,\Sigma_0')$ parameter space for which the product $\varrho\Sigma_0'$ is kept constant. As a representative example, we take $\varrho\Sigma_0'=100$. Numerical solutions are obtained for $\varrho=\{10^{-3},10^{-1},10,10^3\}$, corresponding to $\Sigma_0'=\{10^5,10^3,10,10^{-1}\}$, respectively. Along this trajectory, moving from larger $\varrho$ and smaller $\Sigma_0'$ to smaller $\varrho$ and larger $\Sigma_0'$ corresponds to a decrease in the fracture propagation velocity.

\begin{figure}[h!]
    \centerline{
    \includegraphics[width=1\textwidth]{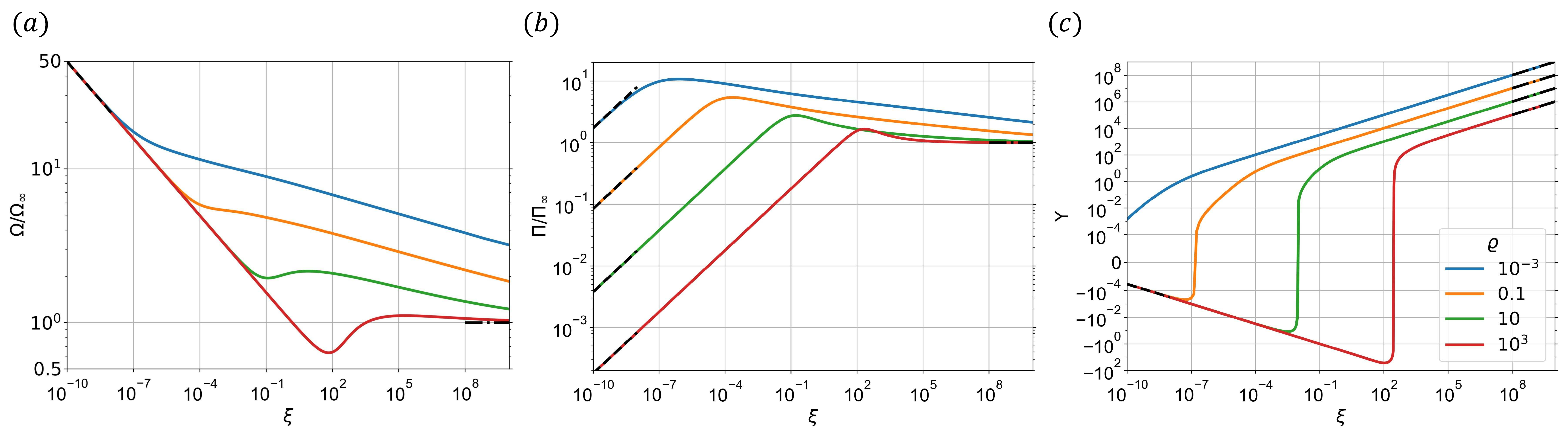}
    }
    \caption{
    Fracture opening $\Omega(\xi)$ (a), net pressure $\Pi(\xi)$ (b), and fluid displacement function $\Upsilon(\xi)$ (c) along the trajectory $\varrho\Sigma_0'=100$ in the $(\varrho,\Sigma_0')$ parameter space. The selected dimensionless permeabilities are $\varrho=\{10^{-3},10^{-1},10,10^3\}$, corresponding to the dimensionless effective confining stresses $\Sigma_0'=\{10^5,10^3,10,10^{-1}\}$, respectively. The dimensionless poroelastic parameters are fixed at $\beta = 0.15$ and $\eta = 0.3$. The opening and net pressure profiles are normalized by their far-field asymptotes, $\Omega_\infty$ and $\Pi_\infty$, given by~\eqref{eq:far-field_asymptote_mk_scaling}. The black dashed lines indicate the near-field asymptotic behaviour, whereas the black dash-dotted lines denote the far-field behaviour. A symmetric logarithmic scale is used for $\Upsilon(\xi)$ on the vertical axis to show both negative and positive values.
    }
    \label{fig:variation_rho_sigma}
\end{figure}

The resulting profiles are shown in figure~\ref{fig:variation_rho_sigma}: panels (a) and (b) present the fracture opening and net pressure, normalized by their far-field asymptotes~\eqref{eq:far-field_asymptote_mk_scaling}, while panel (c) shows the fluid displacement function.

The results show that, although $\varrho\Sigma_0' = \mathrm{const}$, the solution varies significantly along this trajectory in the $(\varrho,\Sigma_0')$ parameter space. As the fracture propagation velocity decreases, i.e. moving from large to small $\varrho$ and from small to large $\Sigma_0'$, the fracture opening and net pressure profiles exhibit several trends. The range of validity of the near-field asymptote shrinks, while the approach to the far-field asymptote is delayed. This leads to a broader intermediate region between the near- and far-field limits. The tip value of the net pressure, before normalization by the far-field asymptote, also becomes more negative, so that its absolute value increases. The zero-crossing of $\Upsilon(\xi)$ moves closer to the fracture tip, indicating a reduction of the near-tip circulation zone, whereas the far-field magnitude of $\Upsilon$ increases.

\subsection{Relative importance of the poroelastic coupling mechanisms}
\label{sec:maps_rel_dif_and_regime}

In this section, we quantify the relative importance of the different poroelastic coupling mechanisms over the $(\varrho, \Sigma_0')$ parameter space. The analysis proceeds in two steps. First, relative difference maps are constructed by comparing the fully coupled model with two reduced models introduced in Section~\ref{sec:reduced_models}, namely the uncoupled and one-way coupled. Their normalized formulations are given in Section~\ref{sec:reduced_models_normalized} and summarized in table~\ref{tab:reduced_models}. The fully coupled and one-way coupled models are evaluated at the representative values of the dimensionless poroelastic parameters $\beta=0.15$ (contrast between the undrained and drained Poisson's ratios) and $\eta=0.3$ (poroelastic stress coefficient), which yield $\mathcal{S} = 1.02$ (dimensionless storage coefficient) through equation~\eqref{eq:dimensionless_storage}. For consistency, the same value of $\mathcal{S}$ is prescribed in the uncoupled model, although $\mathcal{S}$ is an independent governing parameter in that formulation. The comparison is based on the dimensionless fracture opening $\Omega(\xi)$ and net pressure $\Pi(\xi)$ profiles. These maps identify regions where a given reduced formulation approximates the fully coupled solution within a prescribed accuracy. In the second step, the relative difference maps are used to define characteristic regimes in the $(\varrho, \Sigma_0')$ parameter space.

\subsubsection{Relative difference maps}
\label{sec:rel_dif_maps}

We construct relative difference maps by comparing the fully coupled poroelastic model with two reduced formulations: the uncoupled model and the one-way coupled model. For the present comparison, both reduced formulations employ a two-dimensional pressure-dependent fluid exchange description (2D PDL). The fully coupled and one-way coupled models are governed by the same set of dimensionless parameters, $\varrho$, $\Sigma_0'$, $\beta$, and $\eta$, with $\beta=0.15$ and $\eta=0.3$ adopted here. The one-way coupled model neglects the effect of rock deformation on the pore fluid pressure, such that $\Pi^{MH}(\xi) = 0$, while retaining both contributions to the normal stress on the fracture surfaces: the direct mechanical poroelastic contribution and the hydraulic-to-mechanical cross-coupling contribution, i.e. the backstress. The uncoupled model corresponds to $\alpha = 0$, and therefore to $\beta = \eta = 0$; its governing parameters are $\varrho$, $\Sigma_0'$, and $\mathcal{S}$. The dimensionless storage coefficient $\mathcal{S}$ remains finite in this limit and is defined independently through equation~\eqref{eq:dimensionless_governing_parameters_mk_scaling}. For consistency with the fully coupled and one-way coupled calculations, we prescribe $\mathcal{S} = 1.02$ in the uncoupled model, corresponding to the value obtained from equation~\eqref{eq:dimensionless_storage} at $\beta=0.15$ and $\eta=0.3$.

The discrepancy between the fully coupled and reduced solutions is quantified by:
\begin{equation}
    \Delta = \max \left(\Delta_\Omega, \Delta_\Pi\right),
    \label{eq:metrics_for_comparison}
\end{equation}
where
\begin{equation}
    \Delta_\Omega = \max_{0< \xi < \xi_0} \left|\frac{\Omega(\xi) - \Omega^*(\xi)}{\Omega(\xi)}\right|, \quad \Delta_\Pi = \max_{0 < \xi < \xi_0} \left|\frac{\Pi(\xi) - \Pi^*(\xi)}{\Pi(\xi)}\right|.
    \label{eq:metrics_for_comparison_2}
\end{equation}
Here, $\Omega(\xi)$ and $\Pi(\xi)$ denote the fracture opening and net pressure profiles of the fully coupled solution, whereas $\Omega^*(\xi)$ and $\Pi^*(\xi)$ are the corresponding fields obtained from a reduced model, either uncoupled or one-way coupled. The fluid displacement function is not included in the metric~\eqref{eq:metrics_for_comparison}, since the main differences in $\Upsilon$ are localized around its zero-crossing, which determines the spatial extent of the near-tip circulation zone. Moreover, small shifts in the zero-crossing position can lead to artificially large pointwise relative differences because $\Upsilon$ vanishes there. We therefore compare the fully coupled solution with the reduced formulations using only the fracture opening and net pressure profiles.

In evaluating $\Delta_\Omega$ and $\Delta_\Pi$, the maximum is taken over the finite dimensionless interval $\xi \in (0,\xi_0)$, where $\xi_0=\ell/\ell_{mk}$ and $\ell$ is the physical distance from the fracture tip over which the comparison is performed. For each point $(\varrho,\Sigma_0')$, an admissible range of the length scale $\ell_{mk}$ is evaluated using the dimensional model parameters discussed in Section~\ref{sec:representative_values_governing_parameters}. The distributions of the minimum, mean, and maximum values of $\ell_{mk}$ over the $(\varrho,\Sigma_0')$ parameter space are provided in figure~\ref{fig:l_mk_distribution} in the supplementary material. In the following analysis, the mean values of $\ell_{mk}$, shown in figure~\ref{fig:l_mk_distribution}b, are used as representative values for evaluating $\xi_0$. In addition to the limiting case $\xi_0 = \infty$, for which the maximum in equation~\eqref{eq:metrics_for_comparison_2} is evaluated along the entire semi-infinite fracture, we consider $\ell=10^{-2}$, $0.1$, $1$, $10$, and $10^2$ m. These values span laboratory-to-field scales and allow us to assess how the importance of poroelastic coupling depends on the physical extent of the near-tip region, which is assumed to scale with the characteristic size of a finite hydraulic fracture \citep{garagash2011multiscale}.

\begin{figure}[h!]
    \centerline{
    \includegraphics[width=1\textwidth]{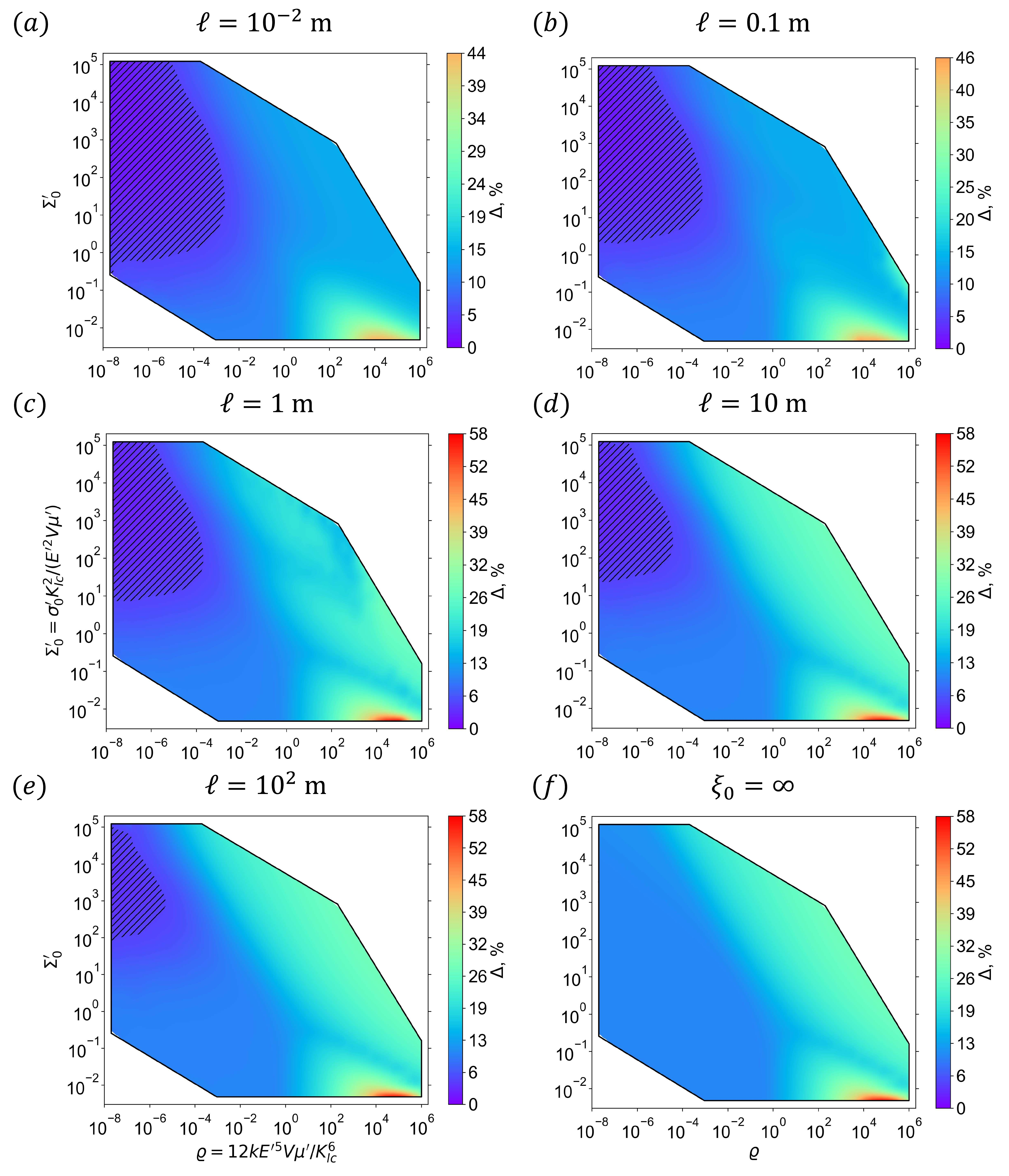}
    }
    \caption{
    Relative difference $\Delta$~\eqref{eq:metrics_for_comparison} between the fully coupled and uncoupled models, evaluated over the $(\varrho,\Sigma_0')$ parameter space. Panels (a)--(e) show the results for the finite physical comparison intervals $\ell=10^{-2}$ m (a), $\ell=0.1$ m (b), $\ell=1$ m (c), $\ell=10$ m (d), and $\ell=10^2$ m (e). Panel (f) corresponds to the limiting case $\xi_0=\infty$, in which the comparison is performed over the entire semi-infinite fracture. The colour scale is common to all panels, so that identical colours correspond to the same values of $\Delta$. The black contour outlines the modelling domain shown in figure~\ref{fig:representative_ranges_governing_parameters}a. Hatched regions indicate $\Delta<5\%$, where the uncoupled model approximates the fully coupled solution within the prescribed accuracy. The dimensionless poroelastic parameters are fixed at $\beta=0.15$ and $\eta=0.3$.
    }
    \label{fig:Delta_fully_coupled_vs_uncoupled}
\end{figure}

Figure~\ref{fig:Delta_fully_coupled_vs_uncoupled} shows the relative difference between the fully coupled and uncoupled solutions for five finite physical comparison intervals, $\ell=10^{-2},0.1,1,10$, and $10^2$ m, together with the limiting case $\xi_0=\infty$.

For finite comparison lengths, the uncoupled model approximates the fully coupled solution within the prescribed accuracy, chosen here as $5\%$, only in a limited part of the parameter space, mainly at small $\varrho$ and moderate-to-large $\Sigma_0'$ (hatched regions in figure~\ref{fig:Delta_fully_coupled_vs_uncoupled}). This domain identifies the part of the parameter space where the neglected poroelastic effects have a sufficiently small impact on the opening and net pressure profiles over the selected comparison interval. As the comparison length $\ell$ increases, the region where $\Delta<5\%$ progressively shrinks, indicating that poroelastic effects become more important when a larger portion of the fracture tip solution is included in the comparison. The hatched domains should be interpreted as representative, since their precise boundaries may shift for specific choices of the dimensional parameters. This dependence arises because, at each point in the $(\varrho,\Sigma_0')$ parameter space, the conversion from $\ell$ to $\xi_0$ is performed using a representative mean value of $\ell_{mk}$ through the relation $\xi_0=\ell/\ell_{mk}$.

When the metric~\eqref{eq:metrics_for_comparison} is evaluated along the entire semi-infinite fracture, the uncoupled approximation no longer satisfies the prescribed accuracy, $\Delta < 5\%$, for any value of $(\varrho,\Sigma'_0)$. This behaviour is expected because the fully coupled and uncoupled models have different far-field asymptotic behaviours~\eqref{eq:far-field_asymptote_mk_scaling}. The fully coupled solution, evaluated here for the representative poroelastic parameter $\beta=0.15$, approaches an undrained elastic response in the far field. By contrast, the uncoupled model corresponds to $\alpha=0$ and therefore $\beta=0$; in this limit, the drained and undrained elastic responses coincide. Consequently, although the uncoupled model can approximate the fracture fields over a finite interval near the tip, it cannot reproduce the full semi-infinite fully coupled solution within the prescribed accuracy. For smaller values of $\beta$, the difference between the far-field storage-viscosity asymptotes associated with the drained and undrained elastic responses would decrease. As a result, the uncoupled model could satisfy the $5\%$ criterion even in the limiting case $\xi_0=\infty$. The largest discrepancies, reaching approximately $58\%$, occur at large $\varrho$ and small $\Sigma_0'$, where poroelastic effects strongly influence the near-tip fracture fields.

A more detailed inspection of the relative difference map in the limiting case $\xi_0=\infty$ (figure~\ref{fig:Delta_fully_coupled_vs_uncoupled}f) reveals how the overall importance of poroelastic effects varies across the parameter space. Along horizontal transects of the modelling domain (at constant $\Sigma_0'$), the relative difference approaches a plateau at small $\varrho$, reflecting the mismatch between the drained and undrained far-field asymptotes of the uncoupled and fully coupled models. As $\varrho$ increases, the discrepancy generally increases for $\Sigma_0' > 0.3$, indicating that poroelastic effects become stronger with increasing formation permeability. For small values of the dimensionless effective confining stress, $\Sigma_0' < 10^{-2}$, the dependence is non-monotonic: the relative discrepancy $\Delta$ first increases and then decreases with $\varrho$. Intermediate values, $10^{-2} < \Sigma_0' < 0.3$, exhibit a more complex dependence on $\varrho$. A complementary reading follows from vertical transects of the modelling domain (at constant $\varrho$). For $\varrho < 0.1$, the relative difference starts from the far-field plateau and then increases with $\Sigma_0'$. For larger $\varrho$, the relative discrepancy $\Delta$ is already larger than the far-field plateau at small $\Sigma_0'$. As $\Sigma_0'$ increases, the relative difference either remains approximately constant for $0.1<\varrho<1$ or decreases for $\varrho > 1$ over an intermediate range, before increasing again at large $\Sigma_0'$. Overall, for the transects discussed above, the relative difference tends to increase with $\Sigma_0'$ once $\Sigma_0'$ becomes sufficiently large, which in dimensional terms corresponds to a transition from overpressured towards more depleted reservoir conditions.

\begin{figure}[h!]
    \centerline{
    \includegraphics[width=1\textwidth]{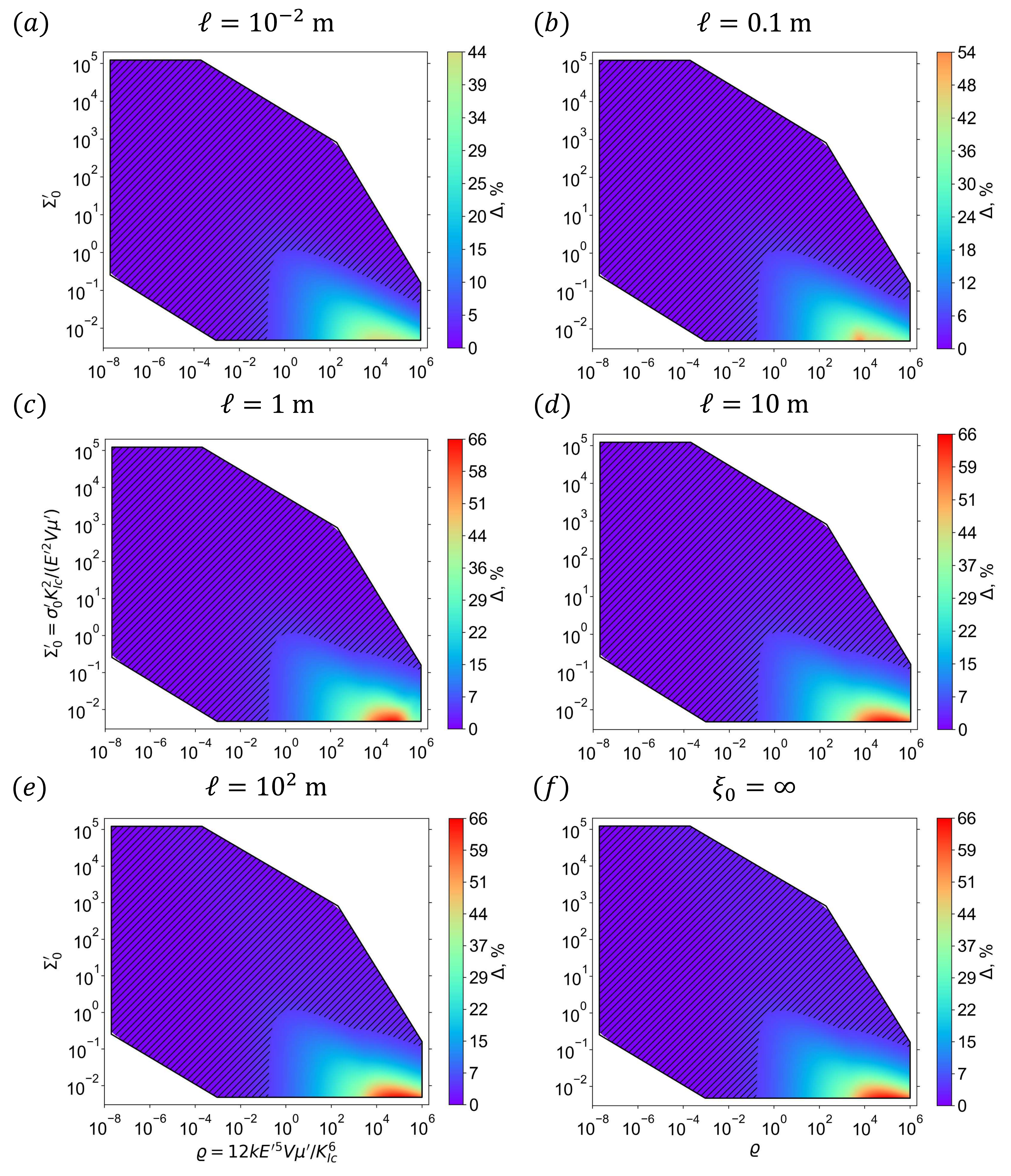}
    }
    \caption{
    Relative difference $\Delta$~\eqref{eq:metrics_for_comparison} between the fully coupled and one-way coupled models, evaluated over the $(\varrho,\Sigma_0')$ parameter space. Panels (a)--(e) show the results for the finite physical comparison intervals $\ell=10^{-2}$ m (a), $\ell=0.1$ m (b), $\ell=1$ m (c), $\ell=10$ m (d), and $\ell=10^2$ m (e). Panel (f) corresponds to the limiting case $\xi_0=\infty$, in which the comparison is performed over the entire semi-infinite fracture. The colour scale is common to all panels, so that identical colours correspond to the same values of $\Delta$. The black contour outlines the modelling domain shown in figure~\ref{fig:representative_ranges_governing_parameters}a. Hatched regions indicate $\Delta<5\%$, where the one-way coupled model approximates the fully coupled solution within the prescribed accuracy. The dimensionless poroelastic parameters are fixed at $\beta=0.15$ and $\eta=0.3$.
    }
    \label{fig:Delta_fully_coupled_vs_one_way_coupled}
\end{figure}

Figure~\ref{fig:Delta_fully_coupled_vs_one_way_coupled} presents the relative difference maps between the fully coupled and one-way coupled solutions for five finite physical comparison intervals, $\ell=10^{-2},0.1,1,10$, and $10^2$ m, together with the limiting case $\xi_0=\infty$.

Unlike the uncoupled case, the region where $\Delta<5\%$ occupies most of the modelling domain and changes only weakly with the comparison interval. This indicates that, over a wide part of the parameter space, the one-way coupled model captures the dominant poroelastic effects by retaining the direct mechanical poroelastic contribution to the normal stress and the hydraulic-to-mechanical coupling term, i.e. the poroelastic backstress. The deformation-induced pore pressure perturbation therefore has only a secondary influence in this region.

The main exception occurs at large $\varrho$ and small $\Sigma_0'$, where the relative difference between the fully coupled and one-way coupled solutions becomes substantial, reaching approximately $66\%$ in the limiting case $\xi_0=\infty$ (figure~\ref{fig:Delta_fully_coupled_vs_one_way_coupled}f). In this region, the mechanical-to-hydraulic coupling significantly modifies the pore pressure field, thereby affecting the fracture opening and net pressure profiles. Thus, the use of the fully coupled formulation is required in this lower-right part of the $(\varrho,\Sigma_0')$ parameter space.

\subsubsection{Regime maps}
\label{sec:regime_maps}

The relative difference maps introduced in Section~\ref{sec:rel_dif_maps} are now used to define characteristic regimes associated with different levels of poroelastic coupling and to determine their applicability domains in the $(\varrho,\Sigma_0')$ parameter space. We denote by $\mathcal{D}$ the entire modelling domain shown in figure~\ref{fig:representative_ranges_governing_parameters}a. For a given comparison interval, $\xi \in (0,\xi_0)$, where $\xi_0=\ell/\ell_{mk}$, and a prescribed accuracy threshold, here taken as $\Delta<5\%$, we define three subdomains of $\mathcal{D}$, denoted by $\mathcal{D}_0$, $\mathcal{D}_1$, and $\mathcal{D}_2$, within which the corresponding reduced models approximate the fully coupled solution with the prescribed accuracy.

For the adopted metric and threshold, these subdomains are uniquely determined using the representative value of $\ell_{mk}$ assigned to each point $(\varrho,\Sigma_0')$ to convert the physical length $\ell$ into $\xi_0$. Their boundaries are not universal, however, because different combinations of dimensional parameters may correspond to the same point $(\varrho,\Sigma_0')$ while yielding different values of $\ell_{mk}$. For a given physical length $\ell$, such variations modify $\xi_0$ and may therefore shift the domain boundaries.

The domain $\mathcal{D}_0$ corresponds to the region where the uncoupled model approximates the fully coupled solution within the prescribed accuracy. The domain $\mathcal{D}_1$ is defined analogously using the auxiliary model without cross-coupling terms introduced in Section~\ref{sec:auxiliary_no_cross_coupling}. Finally, $\mathcal{D}_2$ denotes the region where the one-way coupled model approximates the fully coupled solution within the same accuracy.

The domains $\mathcal{D}_0$ and $\mathcal{D}_2$ correspond to the hatched regions in the relative difference maps comparing the fully coupled solution with the uncoupled (figure~\ref{fig:Delta_fully_coupled_vs_uncoupled}) and one-way coupled (figure~\ref{fig:Delta_fully_coupled_vs_one_way_coupled}) models, respectively. The domain $\mathcal{D}_1$ is determined in the same manner, and the corresponding maps are provided in figure~\ref{fig:Delta_fully_coupled_vs_no_cross_couplings} in the supplementary material.

For the combined metric $\Delta=\max(\Delta_\Omega,\Delta_\Pi)$ and all comparison intervals considered, the computed domains are strictly nested: $\mathcal{D}_0 \subsetneq \mathcal{D}_1 \subsetneq \mathcal{D}_2 \subsetneq \mathcal{D}$, where $\subsetneq$ denotes strict set inclusion. We use these nested domains to define the applicability regions of the characteristic coupling regimes.

Accordingly, Regime 0 corresponds to a negligible influence of poroelastic effects on the fracture opening and net pressure profiles. In this regime, the uncoupled model approximates the fully coupled solution. The applicability region of Regime 0 is therefore defined as: $\mathcal{R}_0 = \mathcal{D}_0$.

Regime 1 is defined as: $\mathcal{R}_1 = \mathcal{D}_1  \setminus \mathcal{D}_0$, where $\setminus$ denotes set difference. In this regime, the uncoupled model is no longer sufficient, whereas the auxiliary model without cross-coupling terms approximates the fracture opening and net pressure profiles of the fully coupled solution. Thus, the direct mechanical poroelastic contribution to the normal stress must be retained, whereas the poroelastic backstress and the deformation-induced pore pressure perturbation do not significantly affect the fracture fields. 

Regime 2 is characterized by a significant influence of the poroelastic backstress on the fracture fields. Its applicability region is defined as: $\mathcal{R}_2 = \mathcal{D}_2 \setminus \mathcal{D}_1$. In this regime, the one-way coupled formulation is required to reproduce the fracture opening and net pressure profiles of the fully coupled solution. Therefore, both contributions to the normal stress must be retained: the direct mechanical poroelastic contribution and the hydraulic-to-mechanical cross-coupling contribution, i.e. the poroelastic backstress. By contrast, the deformation-induced pore pressure perturbation has only a secondary influence on the fracture fields. 

Finally, Regime 3 is defined as: $\mathcal{R}_3 = \mathcal{D} \setminus \mathcal{D}_2$. In this part of the parameter space, the one-way coupled formulation no longer approximates the fully coupled solution. Hence, all poroelastic coupling mechanisms must be retained. In other words, the fully coupled model is required to obtain the fracture opening and net pressure profiles.

Thus, Regimes 0--2 correspond to cases in which the mechanical-to-hydraulic coupling is sufficiently weak, so that deformation-induced pore pressure perturbations do not significantly impact the near-tip fracture fields. Consequently, in these regimes the semi-infinite steadily propagating fracture problem can be treated as at most one-way coupled. Only in Regime 3 must the fully coupled formulation be applied.

For the representative value $\beta=0.15$ considered here, Regime 0 is not observed when the metric $\Delta$~\eqref{eq:metrics_for_comparison} is evaluated over the entire semi-infinite fracture, i.e. for $\xi_0=\infty$. This is because the uncoupled and fully coupled models have different far-field asymptotic behaviours as already discussed. 

\begin{figure}[h!]
    \centerline{
    \includegraphics[width=1\textwidth]{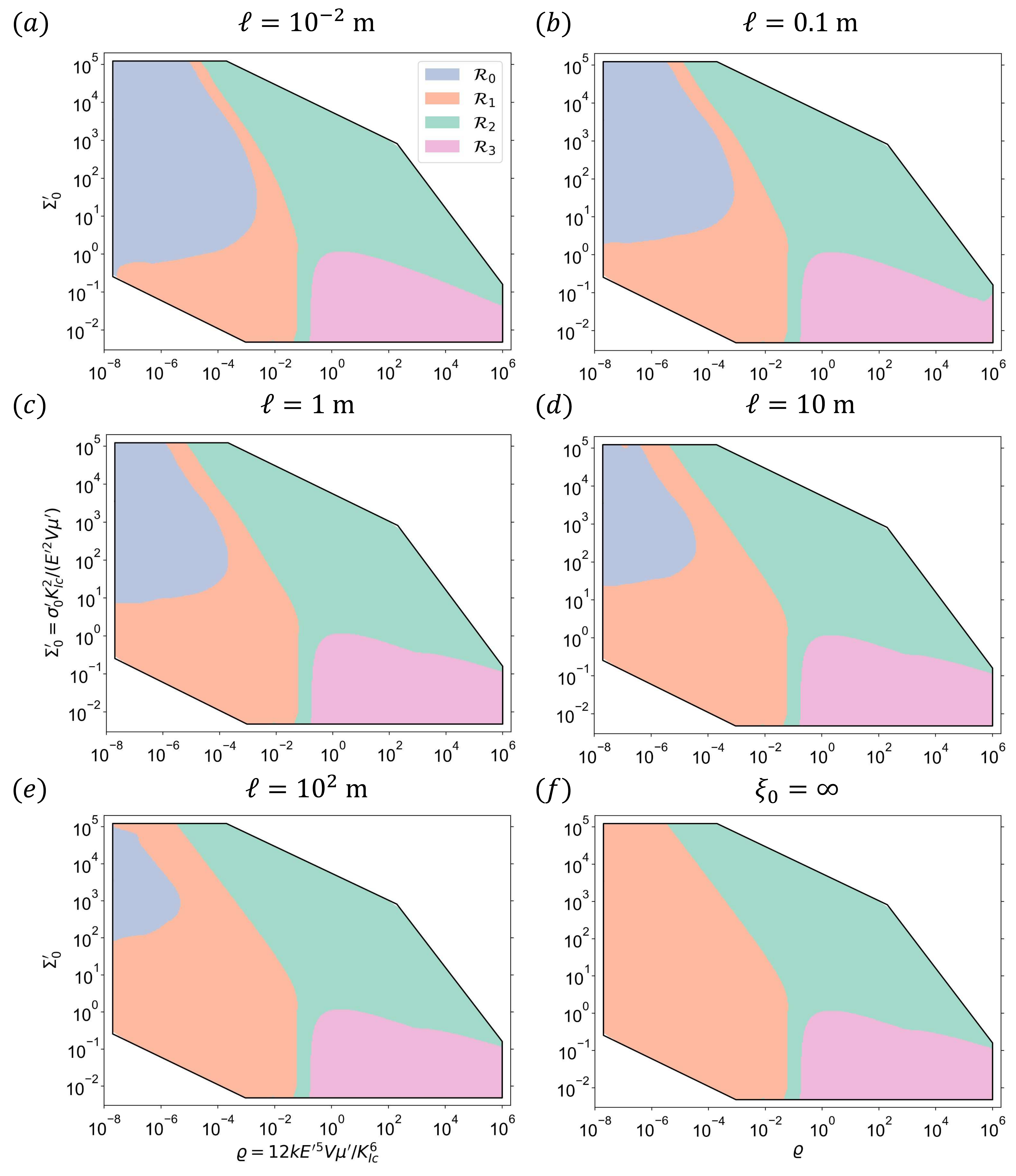}
    }
    \caption{
    Regime maps identifying the dominant poroelastic coupling mechanisms in the $(\varrho,\Sigma_0')$ parameter space. Panels (a)--(e) are constructed from relative difference maps comparing the fully coupled and reduced models over the finite physical comparison intervals: $\ell=10^{-2}$ m (a), $\ell=0.1$ m (b), $\ell=1$ m (c), $\ell=10$ m (d), and $\ell=10^2$ m (e). Panel (f) is based on the corresponding relative difference maps for the limiting case $\xi_0=\infty$, in which the comparison is performed over the entire semi-infinite fracture. The regimes correspond to: $\mathcal{R}_0$, negligible poroelastic effects; $\mathcal{R}_1$, dominance of the direct mechanical poroelastic response; $\mathcal{R}_2$, significant influence of the poroelastic backstress; and $\mathcal{R}_3$, the fully coupled regime. The regime classification is based on the combined metric $\Delta$ defined in equation~\eqref{eq:metrics_for_comparison}. The black contour outlines the modelling domain shown in figure~\ref{fig:representative_ranges_governing_parameters}a. The dimensionless poroelastic parameters are fixed at $\beta=0.15$ and $\eta=0.3$.
    }
    \label{fig:Regime_maps}
\end{figure}

Figure~\ref{fig:Regime_maps} presents the resulting regime maps for five finite physical comparison intervals, $\ell=10^{-2}$, $0.1$, $1$, $10$, and $10^2$ m, together with the limiting case $\xi_0=\infty$.

For finite comparison intervals, all four characteristic regimes are realized. By contrast, when the entire semi-infinite fracture is included in the comparison, only Regimes 1--3 are observed for the representative value $\beta=0.15$ considered here. Regime 0 appears at small $\varrho$ and moderate-to-large $\Sigma_0'$. Regime 1 occupies the adjacent low-$\varrho$ region and extends over a broad range of $\Sigma_0'$. Regime 2 covers a large part of the parameter space, especially at larger $\varrho$ and larger $\Sigma_0'$, whereas Regime 3 is localized mainly at large $\varrho$ and small $\Sigma_0'$.

The extent of the characteristic regimes depends on the length of the comparison interval. As $\ell$ increases, the applicability region of Regime 0 progressively shrinks, indicating that neglecting poroelastic effects is acceptable only over sufficiently small distances from the fracture tip. At the same time, the region associated with Regime 3 expands slightly with increasing $\ell$, showing that the mechanical-to-hydraulic coupling becomes more relevant when a larger portion of the semi-infinite fracture solution is included in the comparison. The boundary between Regimes 1 and 2 is less sensitive to $\ell$, suggesting that the importance of the poroelastic backstress is governed primarily by the fluid exchange regime characterized by $(\varrho,\Sigma_0')$, rather than by the physical extent of the hydraulic fracture tip region over which the comparison is performed. Accordingly, shorter comparison intervals representative of laboratory scales allow simpler reduced formulations to remain applicable over a broader part of the parameter space, whereas longer intervals representative of field scales progressively approach the limiting classification obtained for $\xi_0=\infty$.

Finally, we consider the regime map corresponding to the limiting case $\xi_0=\infty$ (figure~\ref{fig:Regime_maps}f). Regime 1 is approximately confined to the domain where $\varrho \leq 0.06$ and $\Sigma_0' < 0.39\,\varrho^{-1}$. This region corresponds to cases in which fluid exchange between the fracture and the host rock remains limited. In dimensional terms, it may include low-permeability sandstone formations with relatively high fracture toughness and moderate elastic stiffness, as well as parts of the granite and shale parameter ranges. For granites and shales, this zone is associated with the more permeable end of their admissible ranges, although these permeabilities may still be low in absolute terms compared with typical sandstones. Under these conditions, the impact of the cross-coupling terms on the fracture fields is weak, and the direct mechanical poroelastic contribution to the normal stress represents the dominant correction to the uncoupled formulation. 

Regime 3 approximately occupies the domain where $\varrho > 0.19$ and $\Sigma_0' < 1.5\,\varrho^{-0.19}$. This region can be associated with rapid fracture propagation in relatively low-toughness formations. For sandstones, it is also associated with overpressured conditions. For granites and shales, which are generally much less permeable, this regime is reached near the upper end of their admissible permeability ranges. In this case, all poroelastic coupling mechanisms become sufficiently important, and the fracture tip formulation must be treated as fully coupled. 

The remaining part of the parameter space, outside Regimes 1 and 3, is assigned to Regime 2, where the poroelastic backstress, together with the direct mechanical poroelastic response, must be retained, while the deformation-induced pore pressure perturbation has only a secondary influence on the fracture fields. The regime classification provides a physical interpretation of the trends observed in figure~\ref{fig:Delta_fully_coupled_vs_uncoupled}: within the Regime 2 domain, the discrepancy between the fully coupled and uncoupled solutions grows with $\varrho$ for $\Sigma_0' > 0.3$, and also tends to increase with $\Sigma_0'$ once $\Sigma_0'$ becomes sufficiently large. This behaviour is primarily caused by the increasing importance of the poroelastic backstress, which is captured by the one-way coupled formulation in Regime 2.

The regime maps presented in this section are based on the combined metric~\eqref{eq:metrics_for_comparison}, which accounts for discrepancies in both the fracture opening and the net pressure. However, in some numerical solvers for hydraulic fracture propagation, the solution for a steadily moving semi-infinite crack is often used primarily through its opening profile. For example, in the ILSA and IMMA schemes, the updated fracture front position over a time step is determined by matching the opening predicted by the tip model with the opening in the survey element, i.e. the cell adjacent to the tip element. In addition, the width assigned to the tip element is obtained by averaging its volume computed from the near-tip opening profile. Thus, from the perspective of such propagation schemes, the accuracy of the fracture opening is of particular importance. For this reason, we have also constructed relative difference and regime maps using the opening-based metric $\Delta_\Omega$ alone. This additional analysis is motivated mainly by numerical implementation considerations rather than by a different physical interpretation of the poroelastic coupling processes near the fracture tip. The resulting maps, together with a brief discussion of their differences from the maps based on the combined metric $\Delta=\max(\Delta_\Omega,\Delta_\Pi)$, are provided in Appendix~\ref{sec:regime_maps_opening}. The region where the fully coupled formulation is required tends to shrink when the regime classification is based only on the opening profile.

\subsection{Solution structure in characteristic coupling regimes}
\label{sec:solution_structure_in_regimes}

We now analyse the structure of the fracture tip solution at representative points selected from the characteristic regimes identified in Section~\ref{sec:regime_maps}. Our objective is to complement the relative difference and regime maps presented in Sections~\ref{sec:rel_dif_maps} and \ref{sec:regime_maps} by showing how the different poroelastic coupling mechanisms modify the fracture opening, net pressure, and fluid displacement function. This analysis also clarifies why a given reduced formulation is sufficient in some regions of the parameter space, whereas the fully coupled formulation is required in others.

We consider three representative points in the $(\varrho, \Sigma_0')$ plane: $(\varrho,\Sigma_0')=(10^{-5}, 10^2)$, $(10^3, 10)$, and $(10^4, 10^{-2})$. The first point is classified as Regime 0 (negligible poroelastic effects; uncoupled approximation) for finite physical comparison lengths $\ell \leq 10$ m. For larger comparison lengths, $\ell>10$ m, and in particular in the limiting case $\xi_0=\infty$, the same point belongs to Regime 1 (dominance of the direct mechanical poroelastic response). The second and third points belong to Regime 2 (significant influence of the poroelastic backstress; one-way coupled approximation) and Regime 3 (fully coupled regime), respectively, for all comparison lengths considered. The dimensionless poroelastic parameters are fixed at the same representative values as in the previous sections, $\beta=0.15$ and $\eta=0.3$.

For each representative point, we compare the fully coupled solution with the uncoupled and one-way coupled formulations in terms of the fracture opening, net pressure, and fluid displacement function profiles. In addition, for the fully coupled model, we decompose the normal stress and pore fluid pressure on the fracture surfaces into their direct and cross-coupling contributions. This decomposition allows us to relate the observed differences between the fully coupled and reduced formulations to the poroelastic coupling mechanisms that control the solution structure in each regime.

\subsubsection{Regimes 0 and 1: negligible poroelastic effects and dominance of the direct mechanical poroelastic response}
\label{sec:solution_structure_in_regime_1}

\begin{figure}[h!]
    \centerline{
    \includegraphics[width=1\textwidth]{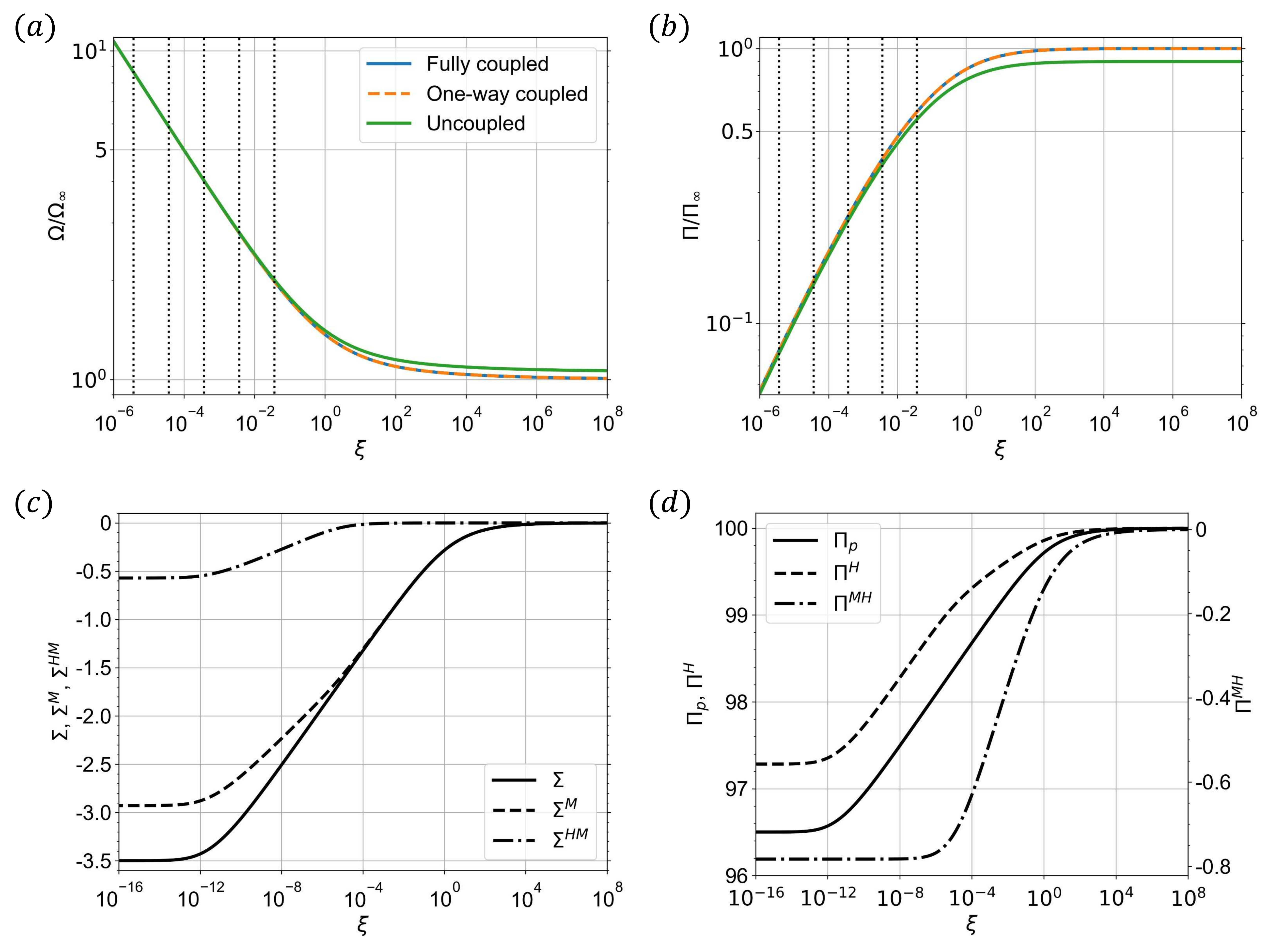}
    }
    \caption{
    Representative solution at $(\varrho,\Sigma_0')=(10^{-5},10^2)$, associated with Regime 0 for finite comparison lengths $\ell \leq 10$ m and with Regime 1 for larger $\ell$ and in the limiting case $\xi_0=\infty$. Panels (a, b) compare the fully coupled solution (blue solid lines), the one-way coupled solution (orange dashed lines), and the uncoupled solution (green solid lines) in terms of (a) the fracture opening and (b) the net pressure, both normalized by their far-field asymptotes as $\Omega/\Omega_\infty$ and $\Pi/\Pi_\infty$, respectively, where $\Omega_\infty$ and $\Pi_\infty$ are given by \eqref{eq:far-field_asymptote_mk_scaling}. The vertical dotted lines mark the endpoints $\xi_0=\ell/\ell_{mk}$ of the comparison intervals, with $\ell=\{10^{-2},0.1,1,10,100\}$ m; the length scale $\ell_{mk}$ is taken as the mean value corresponding to the considered point $(\varrho, \Sigma_0')$ (figure~\ref{fig:l_mk_distribution}b). Panels (c, d) show the decomposition of the normal stress and pore fluid pressure on the fracture surfaces in the fully coupled model into direct and cross-coupling contributions. Panel (c) shows the normal stress $\Sigma$ (black solid line), together with the direct mechanical poroelastic contribution $\Sigma^M$ (black dashed line) and the hydraulic-to-mechanical cross-coupling contribution $\Sigma^{HM}$ (black dash-dotted line). Panel (d) shows the pore fluid pressure $\Pi_p$ (black solid line), together with the direct hydraulic contribution $\Pi^H$ (black dashed line) and the mechanical-to-hydraulic cross-coupling contribution $\Pi^{MH}$ (black dash-dotted line). Because $\Pi^{MH}$ is much smaller in magnitude than $\Pi_p$ and $\Pi^H$, it is plotted against a separate right vertical axis to make its variation visible.} The dimensionless poroelastic parameters are fixed at $\beta=0.15$ and $\eta=0.3$.
    \label{fig:Regime_1}
\end{figure}

We first consider the point $(\varrho, \Sigma_0')=(10^{-5}, 10^2)$, which is representative of Regimes 0 and 1, depending on the finite physical comparison length, as noted above. Panels (a) and (b) in  Figure~\ref{fig:Regime_1} show the fracture opening and net pressure profiles, normalized by their far-field asymptotes \eqref{eq:far-field_asymptote_mk_scaling}. The vertical dotted lines represent the dimensionless endpoints $\xi_0=\ell/\ell_{mk}$ of the comparison intervals used in Sections~\ref{sec:rel_dif_maps} and \ref{sec:regime_maps}, thereby linking the local profile comparison to the relative difference and regime maps constructed for different finite physical lengths $\ell$. 

Over finite comparison lengths $\ell \leq 10$ m, poroelastic effects have a negligible influence on the fracture fields (figures~\ref{fig:Regime_1}a and \ref{fig:Regime_1}b). The fully coupled and uncoupled solutions are therefore in close agreement for both the fracture opening and the net pressure over the corresponding comparison intervals. Farther from the fracture tip, however, the deviation between these two models becomes apparent in both fields, indicating that the direct mechanical poroelastic contribution to the normal stress is no longer negligible. In this region, the fully coupled solution is accurately reproduced by reduced formulations that retain this contribution. In particular, the one-way coupled model provides an accurate approximation of the fracture fields and is practically indistinguishable from the auxiliary model without cross-coupling terms, which was used in Section~\ref{sec:regime_maps} to distinguish between Regime 1 (dominance of the direct mechanical poroelastic response) and Regime 2 (significant influence of the poroelastic backstress). This confirms that the agreement of the one-way coupled model with the fully coupled solution is controlled by the direct mechanical poroelastic response, rather than by the backstress contribution. The uncoupled model, by contrast, is no longer applicable for larger comparison lengths because its far-field behaviour differs from that of the fully coupled model: it overestimates the fracture opening and predicts a less negative net pressure. Thus, the signed net pressure is overestimated, whereas its magnitude is underestimated.

The distributions of the dimensionless normal stress $\Sigma$ and pore fluid pressure $\Pi_p$ on the fracture surfaces in the fully coupled solution are shown in figures~\ref{fig:Regime_1}c and \ref{fig:Regime_1}d, together with their decompositions into direct and cross-coupling contributions. The normal stress and both of its components are negative over the entire semi-infinite fracture. The stress response is controlled primarily by the direct mechanical poroelastic contribution $\Sigma^M$, which nearly coincides with $\Sigma$ beyond a certain distance from the fracture tip. The hydraulic-to-mechanical cross-coupling contribution $\Sigma^{HM}$ (backstress) is much smaller in magnitude. Thus, although the backstress is clearly present in the stress decomposition, it remains a secondary contribution and does not appreciably modify the fracture opening or net pressure profiles.

In contrast, both the pore fluid pressure $\Pi_p$ and the direct hydraulic contribution $\Pi^H$ are positive. They are smaller near the fracture tip and gradually approach the dimensionless effective confining stress $\Sigma_0'$ in the far field. This behaviour follows from the closure relation \eqref{eq:closure_relations_mk_scaling}, together with the vanishing of the net pressure as $\xi \to \infty$. The mechanical-to-hydraulic cross-coupling contribution $\Pi^{MH}$ is negative and remains very small compared to $\Pi^H$. It starts from a finite negative value near the fracture tip and gradually increases towards zero in the far field. Thus, deformation-induced pore pressure perturbations represent only a minor correction to the direct hydraulic response and do not appreciably affect the fracture fields.

The dimensionless fluid displacement function is shown in figure~\ref{fig:Regime_1_3_Y}a in the supplementary material. This quantity is likewise insensitive to the coupling mechanisms in Regimes 0 and 1: the fully coupled, one-way coupled, and uncoupled models give practically identical profiles for this representative point.

\subsubsection{Regime 2: significant influence of poroelastic backstress}
\label{sec:solution_structure_in_regime_2}
\begin{figure}[h!]
    \centerline{
    \includegraphics[width=1\textwidth]{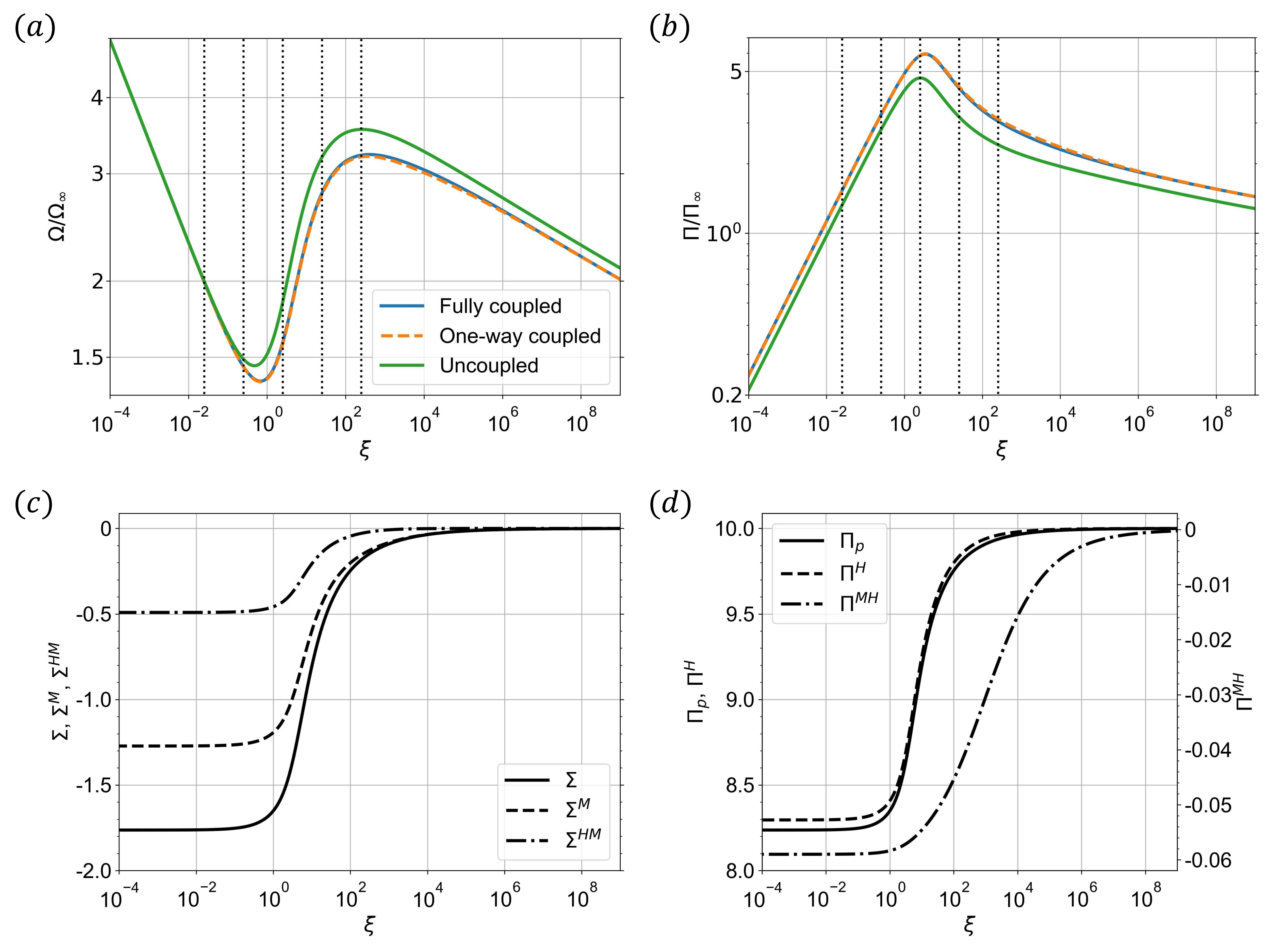}
    }
    \caption{
    Representative solution at $(\varrho, \Sigma_0')=(10^3, 10)$, associated with Regime 2. Panels (a, b) compare the fully coupled, one-way coupled, and uncoupled solutions in terms of (a) the fracture opening and (b) the net pressure, both normalized by their far-field asymptotes. Panels (c, d) show the decomposition of the normal stress and pore fluid pressure on the fracture surfaces in the fully coupled model into direct and cross-coupling contributions. Line styles, normalization, comparison-interval markers, and notation are the same as in figure~\ref{fig:Regime_1}. In panel (d), $\Pi^{MH}$ is likewise plotted against a separate right vertical axis.} The dimensionless poroelastic parameters are fixed at $\beta=0.15$ and $\eta=0.3$.
    \label{fig:Regime_2}
\end{figure}

We next consider the point $(\varrho, \Sigma_0')=(10^3, 10)$, which belongs to Regime 2. Panels (a) and (b) in figure~\ref{fig:Regime_2} show the fracture opening and net pressure profiles, normalized by their far-field asymptotes \eqref{eq:far-field_asymptote_mk_scaling}. Compared with the previous case, this point illustrates a regime in which both the direct and cross-coupling contributions to the normal stress must be retained to reproduce the fracture fields of the fully coupled solution.

In Regime 2, the fully coupled and one-way coupled solutions are practically indistinguishable for both the fracture opening and the net pressure (figures~\ref{fig:Regime_2}a and \ref{fig:Regime_2}b). This agreement indicates that deformation-induced pore pressure perturbations have only a weak influence on the fracture fields. At the same time, the auxiliary model without cross-coupling terms fails to reproduce the fully coupled solution, demonstrating that the poroelastic backstress is an essential contribution in this regime. Thus, both the direct mechanical poroelastic response and the hydraulic-to-mechanical cross-coupling term must be retained. In contrast, the uncoupled model cannot approximate the fully coupled solution, overestimating both the fracture opening and the signed net pressure.

The stress decomposition in figure~\ref{fig:Regime_2}c confirms this interpretation. The normal stress $\Sigma$ and both of its components are negative over the entire semi-infinite fracture. The direct mechanical poroelastic contribution $\Sigma^M$ remains the dominant part of the stress response; however, unlike in Regime 0 (negligible poroelastic effects) and Regime 1 (dominance of the direct mechanical poroelastic response), it approaches the total normal stress $\Sigma$ only at substantially larger distances from the fracture tip. This delayed convergence reflects the broader spatial range over which the hydraulic-to-mechanical cross-coupling contribution $\Sigma^{HM}$ (backstress) remains significant. Although smaller in magnitude than $\Sigma^M$, the backstress provides a substantial additional contribution to $\Sigma$. This explains why both the uncoupled formulation and the auxiliary model without cross-coupling terms fail to reproduce the fully coupled solution, whereas the one-way coupled formulation remains accurate.

The pore fluid pressure distribution and its decomposition in figure~\ref{fig:Regime_2}d show that $\Pi_p$ is controlled almost entirely by the direct hydraulic contribution $\Pi^H$. Both quantities are positive and increase from smaller values near the fracture tip toward $\Sigma_0'$ in the far field. The mechanical-to-hydraulic cross-coupling contribution $\Pi^{MH}$ is negative and remains very small compared with $\Pi^H$. It starts from a finite negative value near the fracture tip and gradually decays in magnitude, tending to zero in the far field. Therefore, deformation-induced pore pressure perturbations represent only a minor correction to the direct hydraulic response and have a negligible impact on the fracture fields.

The dimensionless fluid displacement function is shown in figure~\ref{fig:Regime_1_3_Y}b in the supplementary material. The one-way coupled model reproduces the fully coupled profile. By contrast, the uncoupled model predicts a zero-crossing of $\Upsilon$ closer to the fracture tip and therefore underestimates the spatial extent of the near-tip circulation zone.

\subsubsection{Regime 3: fully coupled regime}
\label{sec:solution_structure_in_regime_3}
\begin{figure}[h!]
    \centerline{
    \includegraphics[width=1\textwidth]{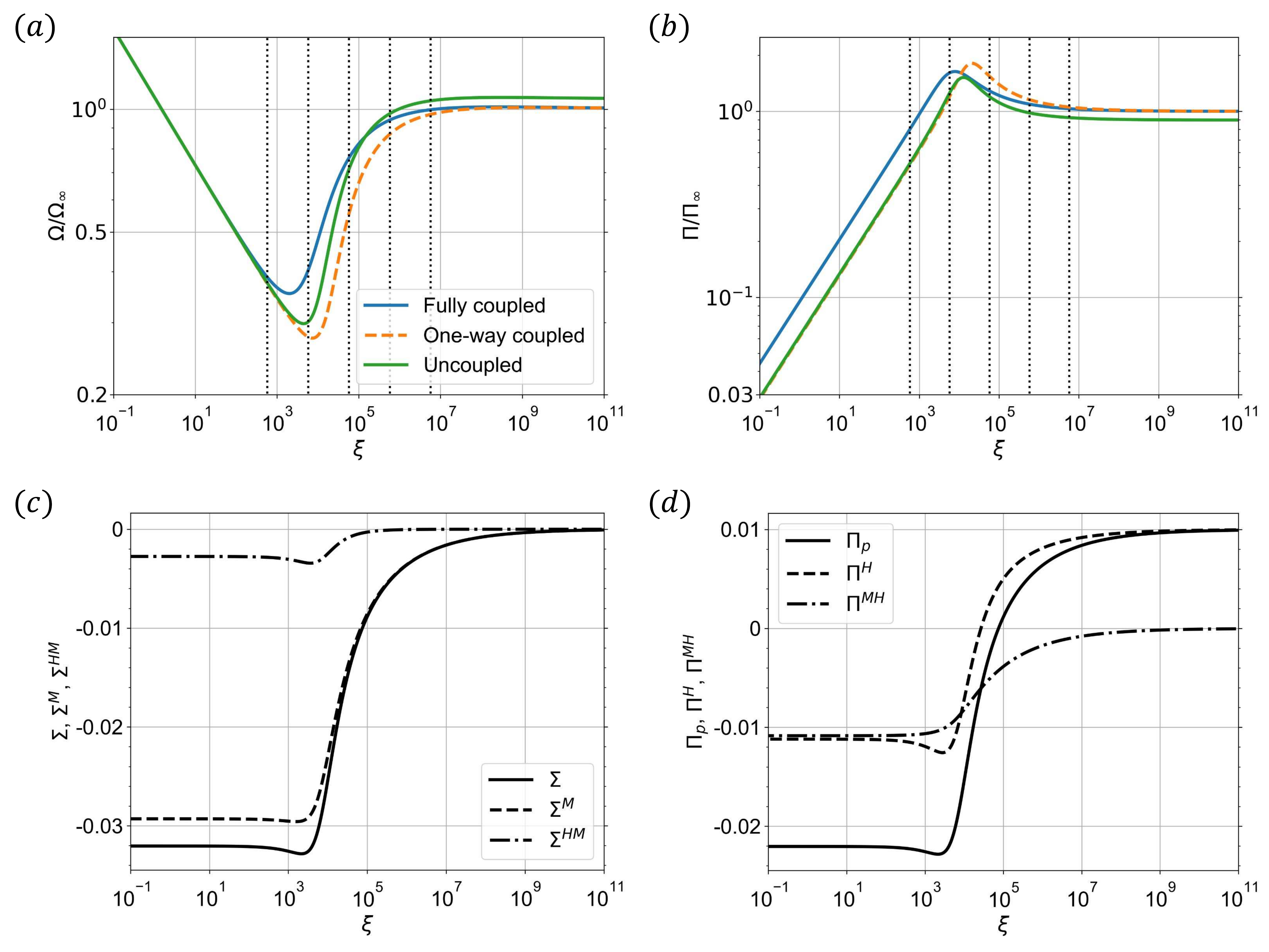}
    }
    \caption{
    Representative solution at $(\varrho, \Sigma_0')=(10^4, 10^{-2})$, associated with Regime 3. Panels (a, b) compare the fully coupled, one-way coupled, and uncoupled solutions in terms of (a) the fracture opening and (b) the net pressure, both normalized by their far-field asymptotes. Panels (c, d) show the decomposition of the normal stress and pore fluid pressure on the fracture surfaces in the fully coupled model into direct and cross-coupling contributions. Line styles, normalization, comparison-interval markers, and notation are the same as in figure~\ref{fig:Regime_1}. Unlike in figures~\ref{fig:Regime_1}(d) and \ref{fig:Regime_2}(d), $\Pi_p$, $\Pi^H$, and $\Pi^{MH}$ are plotted against the same vertical axis in panel (d) because $\Pi^{MH}$ becomes comparable in magnitude to the other two quantities near the fracture tip.} The dimensionless poroelastic parameters are fixed at $\beta=0.15$ and $\eta=0.3$.
    \label{fig:Regime_3}
\end{figure}

We finally consider the point $(\varrho, \Sigma_0')=(10^4, 10^{-2})$, which belongs to Regime 3. Figures~\ref{fig:Regime_3}a and \ref{fig:Regime_3}b show the fracture opening and net pressure profiles, normalized by their far-field asymptotes. In contrast to Regime 2 (significant influence of the poroelastic backstress), the fully coupled formulation is required for this representative point.

The results demonstrate that neither the uncoupled model nor the one-way coupled model reproduces the fracture fields predicted by the fully coupled formulation (figures~\ref{fig:Regime_3}a and \ref{fig:Regime_3}b). This discrepancy is particularly important for the one-way coupled formulation, which already retains both the direct mechanical poroelastic response and the poroelastic backstress, and differs from the fully coupled model only by neglecting deformation-induced pore pressure perturbations. Its failure therefore demonstrates that the mechanical-to-hydraulic cross-coupling contribution to the pore fluid pressure is no longer a secondary effect. This contribution must be retained to accurately reproduce the fracture opening and net pressure profiles, implying that the fracture tip problem must be formulated as fully coupled.

The normal stress on the fracture surfaces and both of its components are negative over the entire semi-infinite fracture (figure~\ref{fig:Regime_3}c). The direct mechanical poroelastic response $\Sigma^M$ is the dominant contribution to the total normal stress. The backstress $\Sigma^{HM}$ is small in magnitude compared with $\Sigma^M$, but, unlike in Regime 0 (negligible poroelastic effects) and Regime 1 (dominance of the direct mechanical poroelastic response), it persists over a much broader spatial interval. Therefore, its relevance cannot be assessed from its pointwise magnitude alone: although it is not the dominant part of $\Sigma$, its spatially extended contribution remains observable in the fracture fields. Nevertheless, the main distinction of Regime 3 is the emergence of significant deformation-induced pore pressure perturbations.

This is confirmed by the pore fluid pressure decomposition in figure~\ref{fig:Regime_3}d. In contrast to Regimes 0--2, the mechanical-to-hydraulic contribution $\Pi^{MH}$ is no longer very small compared to the direct hydraulic contribution $\Pi^H$. Near the fracture tip, both contributions are negative and of comparable order, so that deformation-induced pore pressure perturbations substantially modify the distribution of the pore fluid pressure $\Pi_p$. As a result, $\Pi_p$ becomes comparable in magnitude to the normal stress $\Sigma$ (figure~\ref{fig:Regime_3}c), which is consistent with the closure relations \eqref{eq:closure_relations_mk_scaling} and the small value of $\Sigma_0'$ considered in the present case. The negative value of $\Pi_p$ near the fracture tip indicates that the pore fluid pressure on the fracture surface falls below the initial pore pressure in this region. Farther from the tip, $\Pi^{MH}$ decays in magnitude and tends to zero, while $\Pi_p$ and $\Pi^H$ approach $\Sigma_0'$ in the far field. Thus, the mechanical-to-hydraulic coupling modifies the pore fluid pressure distribution and consequently has a significant effect on the fracture opening and net pressure profiles.

The dimensionless fluid displacement function is shown in figure~\ref{fig:Regime_1_3_Y}c in the supplementary material. The profiles obtained with the uncoupled and one-way coupled models differ noticeably from the fully coupled profile in the transition region, where $\Upsilon$ changes sign. Both reduced models predict the zero-crossing of $\Upsilon$, which marks the outer boundary of the near-tip circulation zone, to be farther from the fracture tip and therefore overestimate the spatial extent of this zone. This confirms that deformation-induced pore pressure perturbations affect not only the fracture opening and net pressure profiles, but also the fluid displacement function profile.

\section{Conclusions and perspectives}
\label{sec:conclusions}

The influence of poroelasticity on hydraulic fracture propagation has been of interest for several decades, starting from the studies of \cite{Ruin78} and \cite{Clea78}. Here, we provide a fully coupled solution for the near-tip fields by solving the problem of a steadily moving semi-infinite hydraulic fracture in a linear isotropic poroelastic solid. For the considered problem setup, the poroelastic response is described by two coupled boundary integral equations derived and validated by \cite{kanin2026steadily}, which relate the normal stress and pore fluid pressure on the fracture surfaces to the fracture opening and fluid displacement function. In the present work, we supplement these equations with appropriate closure relations on the fracture surfaces, the LEFM propagation condition and the non-linear lubrication equation governing fluid flow inside the fracture. The resulting non-linear system is discretized using a Chebyshev spectral collocation framework with precomputed boundary integral operators and solved by root finding.

The asymptotic behaviour in the near field follows the classical LEFM form, $w \sim x^{1/2}$, $p \sim \mathrm{const}$, under drained conditions, and is therefore governed by the drained elastic moduli. In the far field, the well-known storage-viscosity asymptote, $w \sim x^{2/3}$, $p \sim x^{-1/3}$, is recovered under undrained conditions, with the corresponding undrained elastic moduli. Poroelastic coupling and fluid exchange between the fracture and the surrounding porous reservoir control the intermediate behaviour between these near- and far-field asymptotes. In particular, the fluid displacement function reveals a near-tip circulation zone, where fluid leaks into the fracture near the tip and leaks out of the fracture farther behind the tip \citep{detournay2003near, kanin2020near}; beyond this zone, only leak-off from the fracture into the surrounding rock occurs. The extent of the circulation zone is affected by poroelastic coupling. 

In its dimensionless form, written in the viscosity-toughness $mk-$scaling, the problem depends on four dimensionless parameters:
\begin{equation*}
    \varrho=\frac{12 k E'^5 V \mu'}{K_{Ic}^6}, \quad \Sigma_0'=\dfrac{\sigma_0' K_{Ic}^2}{E'^2 V \mu'}, \quad \beta=\frac{\nu_u-\nu}{1-\nu}, \quad \eta =\frac{\alpha (1-2\nu)}{2 (1-\nu)}.
\end{equation*}
The parameters $\beta$, which quantifies the contrast between the undrained and drained Poisson's ratios, and $\eta$, the poroelastic stress coefficient, govern the poroelastic coupling and depend only on the poroelastic material properties. By contrast, the dimensionless permeability $\varrho$, which characterizes the strength of the leak-in contribution, and the dimensionless effective confining stress $\Sigma_0'$, which measures the relative importance of leak-off and leak-in, govern fluid exchange between the fracture and the surrounding porous medium. The parameters $\varrho$ and $\Sigma_0'$, in addition to depending on the poroelastic material properties, incorporate the hydraulic and fracture properties of the rock, the fluid viscosity, the fracture propagation velocity, and the in-situ effective confining stress. Variations in $\varrho$ and $\Sigma_0'$ strongly affect the solution, namely the fracture opening, net pressure, and fluid displacement function, in the intermediate region between the near- and far-field asymptotes.

We have systematically explored the parameter space by basing the ranges of variation on available measurements of poroelastic properties of several rock types, including sandstones, granites, and shales, and by using representative hydraulic and fracture properties, fluid viscosities, in-situ conditions, and fracture propagation velocities. To assess the influence of poroelastic coupling, we compared the fully coupled poroelastic model with two reduced formulations: an uncoupled model, in which poroelastic coupling is neglected while two-dimensional fluid exchange is retained, and a one-way coupled model, in which the effect of fracture opening, or mechanical deformation, on pore pressure is neglected. We used a relative difference metric to compare the opening and net pressure profiles of the fully coupled and reduced solutions over spatial intervals extending from the fracture tip to a physical distance $\ell$. The length $\ell$ is interpreted as the physical extent of the near-tip region and is taken to scale with the characteristic size of a finite hydraulic fracture \citep{garagash2011multiscale}. Since the influence of near-tip processes on finite fracture propagation depends on how their characteristic length scales compare with the fracture length, we considered values of $\ell$ ranging from 1 cm to 100 m, thereby spanning laboratory-to-field scales. We also examined the limiting case, in which the comparison is performed over the entire semi-infinite fracture.

Over the range of realistic values of $(\varrho, \Sigma_0')$, four distinct regimes emerge. In Regime 0, poroelastic effects are negligible. This regime is confined to small $\varrho$ and moderate-to-large $\Sigma_0'$ for finite spatial intervals close to the tip. It disappears when the entire semi-infinite fracture is considered, because the far-field asymptote of the fully coupled model is governed by the undrained response and therefore cannot be captured by the uncoupled formulation, whose far field is governed by the drained response. Regime 1 occupies the low-to-moderate $\varrho$ region. Here, the direct mechanical poroelastic contribution to the normal stress must be retained, while the cross-coupling terms remain secondary. In Regime 2, which covers a broad part of the $(\varrho, \Sigma_0')$ parameter space, the pore pressure-induced poroelastic backstress becomes essential. This effect is accurately captured by the one-way coupled model. Finally, in Regime 3, located mainly at large $\varrho$ and low $\Sigma_0'$, all poroelastic coupling mechanisms become important and the problem must be treated in a fully coupled manner. In particular, the mechanical-to-hydraulic cross-coupling contribution to pore pressure becomes comparable to the direct hydraulic contribution. This regime can be realized under overpressured reservoir conditions and rapid fracture propagation. The applicability domains of these characteristic regimes vary with $\ell$, reflecting changes in the relative importance of the underlying near-tip processes across laboratory-to-field scales.

Our results clarify when poroelastic coupling matters in the near-tip region of a propagating hydraulic fracture and should therefore be explicitly accounted for. They also provide a basis for selecting the simplest sufficiently accurate near-tip model for use in finite fracture simulators based on ILSA or IMMA schemes. For a homogeneous rock, the dimensionless poroelastic parameters $\beta$ and $\eta$ are fixed, so efficient evaluation of the near-tip opening and fluid displacement function profiles could rely on precomputed solutions tabulated over the $(\varrho, \Sigma_0')$ parameter space and interpolation in these two parameters. If the initial effective confining stress is also uniform, changes in the local propagation velocity trace a path with $\varrho\Sigma_0'=\mathrm{const}$ in this plane. Moreover, introducing poroelasticity and two-dimensional fluid exchange only through the near-tip model would not provide a consistent finite fracture description; the corresponding physical processes must also be represented in the global finite fracture problem. The present investigation has been restricted to a steadily moving semi-infinite crack. For finite hydraulic fracture geometries, such as plane strain, penny-shaped, or blade-like fractures propagating under a prescribed injection rate, the fracture velocity is time dependent. As a result, multiple time scales associated with poroelasticity may emerge. A complete and detailed investigation is still lacking, especially for quantifying the accuracy of simplified models, such as the recently proposed one-way coupled approach with one-dimensional pressure-dependent leak-off \citep{dontsov2021efficient}. 

Several extensions of the present model are worth investigating in the future. First, we have assumed perfect hydraulic communication between the fracture and the surrounding porous medium, with identical properties of the fracturing fluid and the pore fluid and no additional hydraulic resistance at the fracture surfaces. Extending the formulation to two distinct fluids would allow contrasts in their properties to be accounted for. Filter cake formation could also be included to represent time-dependent sealing of the fracture surfaces and the associated modification of the fluid exchange process. These processes likely have a first-order impact in practical reservoir applications. In addition, replacing the LEFM propagation condition with a cohesive zone description would enable analysis of the interaction between poroelastic coupling and the finite spatial extent of the fracture process zone.

\section*{Data availability statement}
The numerical solver and notebooks used to reproduce the principal results and figures are available in the GitHub repository: \url{https://github.com/evgenii-kanin/poroelastic-hydraulic-fracture-tip}.

\section*{Funding Statement}
This work was partly funded by the Swiss National Science Foundation, Switzerland under grant \# 192237 to BL.

\section*{Declaration of Interests. }
The authors report no conflict of interest.

\section*{CRediT author statement}
\textbf{Evgenii Kanin:} Conceptualization, Formal analysis, Investigation, Methodology, Software, Validation, Visualization, Writing - original draft. \textbf{Andreas Möri:} Conceptualization, Formal analysis, Investigation, Methodology, Software, Validation, Visualization, Writing - original draft. \textbf{Dmitry Garagash:} Conceptualization, Investigation, Methodology, Supervision, Writing - review \& editing. \textbf{Brice Lecampion:} Conceptualization, Funding acquisition, Investigation, Methodology, Project administration, Supervision, Writing - review \& editing.

\begin{appen}

\setcounter{figure}{0}
\renewcommand{\thefigure}{A\arabic{figure}}

\section{Governing equations in the diffusion scaling}
\label{sec:diffusion_scaling}

This appendix presents the governing equations in the diffusion scaling \citep{Kovalyshen10}. In the $mk-$scaling (Section~\ref{sec:normalized_equations_mk_scaling}), the equations for the normal stress \eqref{eq:normal_stress_dimensionless_mk_scaling} and the pore fluid pressure \eqref{eq:pore_fluid_pressure_dimensionless_mk_scaling} on the fracture surfaces contain the ratio $\mathcal{S}/\varrho$ within the integral kernels. To avoid this parametric dependence, the general numerical solution is computed in the diffusion scaling using the algorithm described in Appendix~\ref{sec:modeling_approach}, where the kernels are non-parametric; the solution is then mapped back to the $mk-$scaling.

\subsection{Governing equations in the diffusion scaling}
\label{sec:normalized_equations_diffusion_scaling}

In the diffusion scaling, the characteristic length $\ell_{dif}$, opening $w_{dif}$, fluid displacement function $v_{dif}$, and pressure $p_{dif}$ are defined as follows:
\begin{equation}
    \ell_{dif} = \frac{2c}{V}, \quad w_{dif} = v_{dif} = \frac{\sigma_0'\ell_{dif}}{E'} = \frac{2c\sigma_0'}{E'V}, \quad p_{dif} = \sigma_0'.
    \label{eq:diffusion_scaling}
\end{equation}

The normalized distance from the fracture tip $\xi$, the fracture opening $\Omega$, the net pressure $\Pi$, the fluid displacement function $\Upsilon$, the normal stress perturbation $\Sigma$, and the pore pressure perturbation $\Pi_p$ are defined as follows:
\begin{equation}
    \xi = \frac{x}{\ell_{dif}}, \quad \Omega = \frac{w}{w_{dif}}, \quad \Pi = \frac{p}{p_{dif}}, \quad \Upsilon = \frac{v}{v_{dif}}, \quad \Sigma = \frac{\sigma_{yy} - \sigma_0}{p_{dif}}, \quad \Pi_p = \frac{p_p - p_0}{p_{dif}}.
    \label{eq:dimensionless_variables_diffusion_scaling}
\end{equation}

Substituting these dimensionless variables into equations~\eqref{eq:normal_stress}, \eqref{eq:pore_fluid_pressure}, \eqref{eq:closure_relations}, \eqref{eq:lubrication_equation}, and \eqref{eq:propagation_criterion} yields their dimensionless form in the diffusion scaling:
\begin{itemize}    
    \item poroelastic boundary integral equations:
    \begin{itemize}[label=$\circ$]
    \item normal stress equation:
    \begin{flalign}
        & \Sigma(\xi) = \Sigma^M(\xi) + \Sigma^{HM}(\xi), \nonumber \\
        & \Sigma^M(\xi) = \frac{1}{4\pi (1-\beta)} \int_0^{\infty} \frac{\diff \Omega}{\diff s}  \frac{\mathcal{S}^e \left(\xi-s\right)}{\xi-s} \diff s, \quad \Sigma^{HM}(\xi) = - \frac{\eta}{2\pi \mathcal{S}} \int_0^{\infty} \frac{\diff \Upsilon}{\diff s} \mathcal{S}^s \left(\xi -s\right) \diff s;
        \label{eq:normal_stress_dimensionless_diffusion_scaling}
    \end{flalign}
    \item pore fluid pressure equation:
    \begin{flalign}
        & \Pi_p(\xi) = \Pi^H(\xi) + \Pi^{MH}(\xi), \nonumber \\
        & \Pi^H(\xi) = \frac{1}{2\pi \mathcal{S}} \int_{0}^{\infty} \frac{\diff \Upsilon}{\diff s} \mathcal{P}^s \left(\xi -s\right)\diff s, \quad \Pi^{MH}(\xi) = - \frac{\eta}{2\pi \mathcal{S}}\int_{0}^{\infty} \frac{\diff \Omega}{\diff s} \mathcal{P}^e \left(\xi-s\right) \diff s;
        \label{eq:pore_fluid_pressure_dimensionless_diffusion_scaling}
    \end{flalign}
    \end{itemize}
    \item closure relations:
    \begin{equation}
        \Sigma(\xi) = \Pi(\xi), \quad \Pi_p(\xi) = \Pi(\xi) + 1;
        \label{eq:closure_relations_diffusion_scaling}
    \end{equation}
    
    \item lubrication equation:
    \begin{equation}
        \frac{\Omega^3}{\mathcal{M}} \frac{\diff \Pi}{\diff \xi} = \Omega + \Upsilon;
        \label{eq:lubrication_equation_dimensionless_diffusion_scaling}
    \end{equation}
    
    \item propagation criterion:
    \begin{equation}
        \Omega = \mathcal{K} \sqrt{\xi}, \quad \xi \rightarrow 0.
        \label{eq:propagation_criterion_dimensionless_diffusion_scaling}
    \end{equation}
\end{itemize}

In the normalized system, we introduce the following dimensionless parameters:
\begin{equation}
    \mathcal{K} = \sqrt{\frac{32}{\pi}}\frac{K_{Ic}}{\sigma_0'\sqrt{\ell_{dif}}} =  \sqrt{\frac{32}{\pi}}\frac{K_{Ic}}{\sigma_0'} \sqrt{\frac{V}{2c}}, \quad \mathcal{M} = \frac{E'^2 V \mu'}{\ell_{dif} \sigma_0'^3} = \frac{E'^2 V^2 \mu'^2 S}{24 k \sigma_0'^3},
    \label{eq:dimensionless_governing_parameters_diffusion_sclaing}
\end{equation}
where $\mathcal{K}$ is the dimensionless material toughness, and $\mathcal{M}$ is the dimensionless fluid viscosity.

In the diffusion scaling, the solution is governed by four independent dimensionless parameters, $\mathcal{K}$, $\mathcal{M}$, $\beta$, and $\eta$, while the dimensionless storage coefficient $\mathcal{S}$ is expressed in terms of $\beta$ and $\eta$ via equation~\eqref{eq:dimensionless_storage}.

Finally, based on the asymptotic expressions derived in the $mk-$scaling in Sections~\ref{sec:near_field} and~\ref{sec:far_field}, we provide the corresponding near-field and far-field limits in the diffusion scaling:
\begin{flalign}
    & \Omega_0(\xi) = \mathcal{K}\sqrt{\xi}, \quad \Pi_0(\xi) = \mathrm{const}, \quad \Upsilon_0(\xi) = -\mathcal{K}\sqrt{\xi};
    \label{eq:near-field_asymptote_diffusion_scaling} \\
    & \Omega_\infty(\xi) = 2^{1/3}3^{5/6} \mathcal{M}^{1/3} \left(1-\beta\right)^{1/3} \xi^{2/3}, \quad \Pi_\infty(\xi) = -6^{-2/3} \mathcal{M}^{1/3} \left(1-\beta\right)^{-2/3}\xi^{-1/3}, \quad \Upsilon_\infty(\xi) = 4\sqrt{\frac{2}{\pi}}\mathcal{S}\sqrt{\xi}.
    \label{eq:far-field_asymptote_diffusion_scaling}
\end{flalign}

\subsection{Relation between the $mk-$ and diffusion scalings}
\label{sec:relation_between_scalings}

To obtain the solution in the $mk-$scaling, we first specify the governing parameters $\varrho$, $\Sigma_0'$, $\beta$, and $\eta$, which determine the value of $\mathcal{S}$ \eqref{eq:dimensionless_storage}. The corresponding parameters in the diffusion scaling, $\mathcal{K}$ and $\mathcal{M}$, are then computed using:
\begin{equation}
    \mathcal{K} = \sqrt{\frac{32}{\pi}}\frac{1}{\Sigma_0'}\sqrt{\frac{\mathcal{S}}{\varrho}}, \quad \mathcal{M} = \frac{\mathcal{S}}{\varrho \Sigma_0'^3}. 
    \label{eq:relation_parameters_1}
\end{equation}

The problem is subsequently solved in the diffusion scaling for the parameter set $\mathcal{K}$, $\mathcal{M}$, $\beta$, $\eta$, and $\mathcal{S}$, using the numerical method described in Appendix~\ref{sec:modeling_approach}, and the resulting solution is then mapped to the $mk-$scaling using:
\begin{equation}
    \xi = \tilde{\xi} \frac{\mathcal{M}^2}{\mathcal{K}^6}\left(\frac{32}{\pi}\right)^3, \quad \Omega = \tilde{\Omega} \frac{\mathcal{M}}{\mathcal{K}^4}\left(\frac{32}{\pi}\right)^2, \quad \Pi = \tilde{\Pi} \frac{\mathcal{K}^2}{\mathcal{M}}\frac{\pi}{32}, \quad \Upsilon = \tilde{\Upsilon} \frac{\mathcal{M}}{\mathcal{K}^4}\left(\frac{32}{\pi}\right)^2,
    \label{eq:relation_parameters_2}
\end{equation}
where tildes denote quantities defined in the diffusion scaling, while quantities without tildes correspond to the $mk-$scaling.


\section{Numerical solver}
\label{sec:modeling_approach}

This appendix outlines the numerical approach used to solve the governing system and to compute the fracture opening, net pressure, and fluid displacement function for prescribed values of the governing parameters $\mathcal{K}$, $\mathcal{M}$, $\beta$, $\eta$, and $\mathcal{S}$. We perform all computations in the diffusion scaling (Appendix~\ref{sec:normalized_equations_diffusion_scaling}), in which all equations in this appendix are expressed. This choice is motivated by the fact that, in this scaling, the integral kernels in the equations for the normal stress \eqref{eq:normal_stress_dimensionless_diffusion_scaling} and the pore fluid pressure \eqref{eq:pore_fluid_pressure_dimensionless_diffusion_scaling} on the fracture surfaces are non-parametric. The relations between the governing parameters in the $mk-$ and diffusion scalings, as well as the transformations of the dimensionless distance, opening, net pressure, and fluid displacement function between the two scalings, are provided in Appendix~\ref{sec:relation_between_scalings}.

The method is based on a spectral collocation framework adapted from \citep{garagash2000tip, garagash2011multiscale, viesca2018numerical, garagash2019cohesive}. An earlier version of this algorithm, which did not account for lubrication flow within the fracture and relied on a logarithmic computational mesh, was presented in \citep{kanin2026steadily}.

We begin by introducing a cubic mapping that transforms the semi-infinite crack, $\xi \in (0, \infty)$, onto the finite interval $z \in (-1, 1)$ \citep{garagash2019cohesive}:
\begin{equation}
    \xi(z) = \left(\frac{1 + z}{1 - z}\right)^3.
    \label{eq:cubic_mapping}
\end{equation}

The mapped domain $z \in (-1, 1)$ is discretized using two sets of nodes \citep{viesca2018numerical}:
\begin{itemize}
    \item primary nodes (roots of Chebyshev polynomials of the first kind):  
    \begin{equation}
        \mathbf{s} = \{s_j\}_{j=1}^{N_p} = \left\{\cos{\left[\frac{\pi(j-1/2)}{N_p}  \right]}\right\}_{j=1}^{N_p};
        \label{eq:primary_nodes}
    \end{equation}
    \item complementary nodes (roots of Chebyshev polynomials of the second kind):
    \begin{equation}
        \mathbf{z} = \{z_i\}_{i=1}^{N_c} = \left\{\cos{\left(\frac{\pi i}{N_p} \right)}\right\}_{i=1}^{N_c}, ~  N_c = N_p - 1.
        \label{eq:complementary_nodes}
    \end{equation}
\end{itemize}
By construction, larger values of $s_j$ and $z_i$ correspond to smaller indices $j$ and $i$. Bold symbols denote vectors, and $N_p$ and $N_c$ represent the numbers of primary and complementary nodes.

The mapped gradients of the fracture opening, $\diff \Omega / \diff z$, and the fluid displacement function, $\diff \Upsilon / \diff z$, can be written as \citep{viesca2018numerical}:
\begin{equation}
    \frac{\diff \Omega}{\diff z} = W_{\Omega}(z) F_{\Omega}(z), \quad \frac{\diff \Upsilon}{\diff z} = W_{\Upsilon}(z) F_{\Upsilon}(z),
    \label{eq:w_v_mapped_gradients}
\end{equation}
where the weight functions $W_{\Omega}(z)$ and $W_{\Upsilon}(z)$ incorporate the near-field \eqref{eq:near-field_asymptote_diffusion_scaling} and far-field \eqref{eq:far-field_asymptote_diffusion_scaling} asymptotic behaviour of the fracture opening and fluid displacement function, respectively:
\begin{equation*}
    W_{\Omega}(z) = \frac{\sqrt{1+z}}{\left(1-z\right)^3}, \quad W_{\Upsilon}(z) = \frac{\sqrt{1+z}}{\left(1-z\right)^{5/2}}.
\end{equation*}
The functions $F_{\Omega}(z)$ and $F_{\Upsilon}(z)$ are regular (bounded). The weight functions therefore ensure that the correct asymptotic behaviour is captured:
\begin{flalign*}
    & \frac{\diff \Omega}{\diff z} \sim \sqrt{1+z} \quad \text{as } z \to -1, 
    \quad 
    \frac{\diff \Omega}{\diff z} \sim (1 - z)^{-3} \quad \text{as } z \to 1, \\
    & \frac{\diff \Upsilon}{\diff z} \sim \sqrt{1+z} \quad \text{as } z \to -1,
    \quad
    \frac{\diff \Upsilon}{\diff z} \sim (1-z)^{-5/2} \quad \text{as } z \to 1.
\end{flalign*}

The functions $F_{\Omega}(z)$ and $F_{\Upsilon}(z)$ are discretized at the primary nodes, $\mathbf{F}_{\Omega} = \{F_{\Omega}(s_j)\}_{j=1}^{N_p}$ and $\mathbf{F}_{\Upsilon} = \{F_{\Upsilon}(s_j)\}_{j=1}^{N_p}$, while the fracture opening, fluid displacement function, and net pressure are evaluated at the complementary nodes, $\boldsymbol{\Omega} = \{\Omega(\zeta_i)\}_{i=1}^{N_c}$, $\boldsymbol{\Upsilon} = \{\Upsilon(\zeta_i)\}_{i=1}^{N_c}$, $\boldsymbol{\Pi} = \{\Pi(\zeta_i)\}_{i = 1}^{N_c}$, where $\zeta_i = \xi(z_i)$.

Before discretizing the governing equations, we introduce a simplification of equations~\eqref{eq:normal_stress_dimensionless_diffusion_scaling} and \eqref{eq:pore_fluid_pressure_dimensionless_diffusion_scaling}, together with the corresponding closure relations~\eqref{eq:closure_relations_diffusion_scaling}, to facilitate the numerical implementation. Specifically, the fluid displacement function is decomposed into its far-field behaviour \eqref{eq:far-field_asymptote_diffusion_scaling} and a residual part, $\Upsilon(\xi) = \Upsilon_{\infty}(\xi) + \Upsilon_r(\xi)$, which yields:
\begin{flalign}
    & \Pi(\xi) = \frac{1}{4\pi (1-\beta)} \int_0^{\infty} \frac{\diff \Omega}{\diff s} \frac{\mathcal{S}^e \left(\xi-s\right)}{\xi-s} \diff s - \frac{\eta}{2\pi \mathcal{S}} \int_0^{\infty} \frac{\diff \Upsilon_r}{\diff s} \mathcal{S}^s \left(\xi-s\right) \diff s; 
    \label{eq:normal_stress_dimensionless_simplified} \\
    & \Pi(\xi) = \frac{1}{2\pi \mathcal{S}} \int_{0}^{\infty} \frac{\diff \Upsilon_r}{\diff s} \mathcal{P}^s \left(\xi-s\right)\diff s - \frac{\eta}{2\pi \mathcal{S}}\int_{0}^{\infty} \frac{\diff \Omega}{\diff s} \mathcal{P}^e \left(\xi-s\right) \diff s,
    \label{eq:pore_fluid_pressure_dimensionless_simplified}
\end{flalign}
where we used the following identities:
\begin{equation*}
    \int_0^{\infty} \frac{\diff \Upsilon_\infty}{\diff s} \mathcal{S}^s \left(\xi-s\right) \diff s = 0, \quad \frac{1}{2\pi \mathcal{S}} \int_{0}^{\infty} \frac{\diff \Upsilon_\infty}{\diff s} \mathcal{P}^s \left(\xi-s\right)\diff s = 1.
\end{equation*}
The second identity cancels the unit contribution arising from the normalized effective confining stress in the closure relation~\eqref{eq:closure_relations_diffusion_scaling}.

In the present computational framework, the integrals in equations~\eqref{eq:normal_stress_dimensionless_simplified} and \eqref{eq:pore_fluid_pressure_dimensionless_simplified} are evaluated in the original semi-infinite domain. The relations between the gradients of the fracture opening, $\omega(\xi) = \diff \Omega / \diff \xi$, and the residual fluid displacement function, $\upsilon(\xi) = \diff \Upsilon_r / \diff \xi$, in the original domain and the corresponding functions $F_\Omega(z)$ and $F_\Upsilon(z)$ are given by:
\begin{equation}
    \omega_j = \omega \left[\xi(s_j)\right] = \frac{W_{\Omega}(s_j) F_{\Omega}(s_j)}{\xi'(s_j)}, \quad \upsilon_j = \upsilon \left[\xi(s_j)\right] = \frac{W_{\Upsilon}(s_j) F_{\Upsilon}(s_j)}{\xi'(s_j)} - \frac{\diff \Upsilon_\infty}{\diff \xi}\left[\xi(s_j)\right],
    \label{eq:gradients_original_mapped}
\end{equation}
where $\xi'(z) = \diff \xi / \diff z$ is the derivative of the mapping \eqref{eq:cubic_mapping}. 

We next describe the procedure for evaluating the integrals in equations~\eqref{eq:normal_stress_dimensionless_simplified} and \eqref{eq:pore_fluid_pressure_dimensionless_simplified}, which follows the approach outlined in \citep{kanin2026steadily}. For brevity, the functions $\omega(\xi)$ and $\upsilon(\xi)$ are collectively denoted by $f(\xi)$, i.e. $f \in \{\omega, \upsilon\}$. The variation of $f(\xi)$ along the semi-infinite domain is then approximated as:
\begin{equation}
    f(\xi) = 
    \begin{cases}
        f_0^*(\xi), & \xi \leq \xi_{N_p}, \\
        A_j f_0(\xi) + B_j f_\infty(\xi), & \xi \in [\xi_j, \xi_{j-1}], ~ j = N_p, \ldots, 2, \\
        f_\infty^*(\xi), & \xi \geq \xi_1,
    \end{cases}
    \label{eq:omega_v_approximation}
\end{equation}
where $\xi_j = \xi(s_j)$, $j = 1, \ldots, N_p$. The functions $f_0^*(\xi)$ and $f_\infty^*(\xi)$ represent the near-field and far-field asymptotic behaviour of $f(\xi)$, respectively. The functions $f_0(\xi) = \xi^a$ and $f_\infty(\xi) = \xi^b$ serve as power-law basis functions, where the exponents $a$ and $b$ are chosen to match the corresponding asymptotic behaviour of $f(\xi)$, i.e. $f_0(\xi) \sim f_0^*(\xi)$ and $f_\infty(\xi) \sim f_\infty^*(\xi)$.

By enforcing continuity of the approximation \eqref{eq:omega_v_approximation} at the internal nodes $\xi_j$, $j = N_p - 1, \ldots, 2$, the coefficients $A_j$ and $B_j$ can be expressed in terms of the values of the field $f(\xi)$ at the endpoints of the $j$th subinterval, $f_j = f(\xi_j)$:
\begin{flalign}
    & A_j=\frac{f_\infty\left(\xi_j\right)f_{j-1}-f_\infty\left(\xi_{j-1}\right)f_j}{f_\infty\left(\xi_j\right)f_0(\xi_{j-1})-f_\infty\left(\xi_{j-1}\right)f_0(\xi_j)}, \quad B_j=-\frac{f_0\left(\xi_j\right)f_{j-1}-f_0\left(\xi_{j-1}\right)f_j}{f_\infty\left(\xi_j\right)f_0(\xi_{j-1})-f_\infty\left(\xi_{j-1}\right)f_0(\xi_j)}.
    \label{eq:coefficients_approximation}
\end{flalign}

The continuity of the field $f(\xi)$ at the endpoints would require $f_{N_p} = f_0^*(\xi_{N_p})$ and $f_1 = f_\infty^*(\xi_1)$. Instead, we impose relations that ensure the correct limiting behaviour of the unknown functions in the near- and far-field regions (see below).

The numerical evaluation of each integral in equations~\eqref{eq:normal_stress_dimensionless_simplified} and \eqref{eq:pore_fluid_pressure_dimensionless_simplified} at the complementary nodes, $\zeta_i = \xi(z_i)$, can be expressed as:
\begin{align}
    & \mathcal{J}(\zeta_i) = \int_0^\infty f(s) \mathcal{F}(\zeta_i - s) \diff s = \underbrace{\int_0^{\xi_{N_p}}f_0^*(s) \mathcal{F}(\zeta_i - s) \diff s}_{\mathcal{J}_1} + \nonumber \\
    & + \sum_{j = N_p}^{2}\Bigg[A_j \underbrace{\int_{\xi_j}^{\xi_{j-1}} f_0(s) \mathcal{F}(\zeta_i - s) \diff s}_{\mathcal{J}_2} + B_j \underbrace{\int_{\xi_j}^{\xi_{j-1}} f_\infty(s) \mathcal{F}(\zeta_i - s) \diff s}_{\mathcal{J}_3}\Bigg] + \underbrace{\int_{\xi_1}^{\infty}f_\infty^*(s) \mathcal{F}(\zeta_i - s) \diff s}_{\mathcal{J}_4},
    \label{eq:calculation_integral}
\end{align}
where the integrals are precomputed numerically once, prior to solving the problem, highlighting the advantage of the diffusion scaling.

According to equations~\eqref{eq:near-field_asymptote_diffusion_scaling} and \eqref{eq:far-field_asymptote_diffusion_scaling}, the functions introduced in \eqref{eq:omega_v_approximation} can be written as follows:
\begin{flalign}
    & \omega_0^*(\xi) = \frac{\diff \Omega_0}{\diff \xi}, \quad \omega_0(\xi) = \xi^{-1/2}, \quad \omega_\infty^*(\xi) = \frac{\diff \Omega_\infty}{\diff \xi}, \quad \omega_\infty(\xi) = \xi^{-1/3}; \nonumber \\ 
    & \upsilon_0^*(\xi) = \frac{\diff \Upsilon_0}{\diff \xi} - \frac{\diff \Upsilon_\infty}{\diff \xi}, \quad \upsilon_0(\xi) = \xi^{-1/2}, \quad \upsilon_\infty^*(\xi) = 0, \quad \upsilon_\infty(\xi) = \xi^{-1/3}.
    \label{eq:omega_v_approximation_function}
\end{flalign}
Since $\upsilon(\xi)$ vanishes in the far-field by definition, the power-law form of $\upsilon_{\infty}(\xi)$ may be chosen arbitrarily; here, we adopt the same exponent as for the opening gradient.

Using equations~\eqref{eq:coefficients_approximation}, \eqref{eq:calculation_integral}, and \eqref{eq:omega_v_approximation_function}, we rewrite equations~\eqref{eq:normal_stress_dimensionless_simplified} and \eqref{eq:pore_fluid_pressure_dimensionless_simplified} in matrix form:
\begin{equation}
    \begin{pmatrix}
    \boldsymbol{\Pi} \\ \boldsymbol{\Pi}
    \end{pmatrix}
    = \begin{pmatrix}
    \frac{1}{4\pi (1-\beta)} \mathbb{S}^e & - \frac{\eta}{2 \pi \mathcal{S}}\mathbb{S}^s \\
     - \frac{\eta}{2 \pi \mathcal{S}}\mathbb{P}^e &  \frac{1}{2\pi\mathcal{S}} \mathbb{P}^s
    \end{pmatrix}
    \begin{pmatrix}
    \boldsymbol{\omega} \\ 
    \boldsymbol{\upsilon}
    \end{pmatrix} + \mathbf{L},
    \label{eq:linear_system}
\end{equation}
where $\boldsymbol{\omega} = \{\omega_j\}_{j=1}^{N_p}$, $\boldsymbol{\upsilon} = \{\upsilon_j\}_{j=1}^{N_p}$ are related to $\boldsymbol{F}_{\Omega}$ and $\boldsymbol{F}_\Upsilon$ through equation~\eqref{eq:gradients_original_mapped}. The matrices $\mathbb{S}^e$, $\mathbb{S}^s$, $\mathbb{P}^e$, $\mathbb{P}^s$ contain the integrals $\mathcal{J}_2$ and $\mathcal{J}_3$ \eqref{eq:calculation_integral}, combined with the coefficients \eqref{eq:coefficients_approximation}, where the function $\mathcal{F}(\xi)$ corresponds, respectively, to $\mathcal{S}^e(\xi)/\xi$, $\mathcal{S}^s(\xi)$, $\mathcal{P}^s(\xi)$, and $\mathcal{P}^e(\xi)$. The vector $\mathbf{L}$ contains the constant values resulting from integrals $\mathcal{J}_1$ and $\mathcal{J}_4$ \eqref{eq:calculation_integral}.

The lubrication equation \eqref{eq:lubrication_equation_dimensionless_diffusion_scaling} in discretized form reads:
\begin{equation}
    \frac{\Omega_i^3}{\mathcal{M}} \frac{1}{\xi'(z_i)} \left(\frac{\diff \Pi}{\diff z}\right)_i  = \Omega_i + \Upsilon_i, \quad i = 1, \ldots, N_c,
\end{equation}
where $\Omega_i = \Omega(\zeta_i)$, $\Upsilon_i = \Upsilon(\zeta_i)$, and the mapped pressure gradient is evaluated using the barycentric Lagrange differentiation matrix \citep{viesca2018numerical}:
\begin{flalign*}
    & \left(\frac{\diff \Pi}{\diff z}\right)_i = \sum_{i' = 1}^{N_c}\mathcal{D}_{ii'}\Pi_{i'}.
\end{flalign*}
The entries of the differentiation matrix are given by:
\begin{equation*}
    \mathcal{D}_{ii' (i \neq i')} = \frac{\omega_{i'}/\omega_i}{z_i - z_{i'}}, \quad \mathcal{D}_{ii} = -\sum_{i'=1, i' \neq i}^{N_c} \mathcal{D}_{ii'},
\end{equation*}
where $\omega_i = (-1)^i \sin^2{(\pi i / N_p)}$ are the barycentric weights associated with the complementary Chebyshev nodes.

The fracture opening and fluid displacement function are computed by numerically integrating their gradients \eqref{eq:w_v_mapped_gradients} \citep{viesca2018numerical, garagash2019cohesive}:
\begin{equation*}
     \Omega_i = \sum_{j=1}^{N_p} (\mathcal{S}_\Omega^*)_{ij} F_{\Omega}(s_j), \quad \Upsilon_i = \sum_{j=1}^{N_p} (\mathcal{S}_\Upsilon^*)_{ij} F_{\Upsilon}(s_j).
\end{equation*}
In the above equation, the integration matrices are given by:
\begin{equation*}
    (\mathcal{S}_\Omega^*)_{ij} = \sum_{k = 0}^{N_c} B_{kj} \Phi_k^\Omega(z_i), \quad (\mathcal{S}_\Upsilon^*)_{ij} = \sum_{k = 0}^{N_c} B_{kj} \Phi_k^\Upsilon(z_i),
\end{equation*}
with
\begin{flalign*}
     & B_{kj} = 
     \begin{cases}
        1 / N_p, & \quad k = 0, \\
        \frac{2}{N_p} \cos{\left[\frac{\pi k (j - 1/2)}{N_p}\right]}, & \quad k > 0;
     \end{cases} \\
     & \Phi_k^{\Omega}(z) = \int_{-1}^{z} W_{\Omega}(s) T_k(s) \diff s, \quad \Phi_k^{\Upsilon}(z) = \int_{-1}^{z} W_{\Upsilon}(s) T_k(s) \diff s,
\end{flalign*}
where $T_k(s)$ denotes Chebyshev polynomials of the first kind, and the integrals are to be evaluated numerically.

The problem involves $3N_p - 1$ unknowns, $\mathbf{F}_{\Omega}$, $\mathbf{F}_{\Upsilon}$, and $\boldsymbol{\Pi}$, and $3N_p - 3$ relations arising from the equations for the normal stress and pore fluid pressure on the fracture surfaces, together with the lubrication equation. The system is completed with additional conditions, namely the propagation criterion \eqref{eq:near-field_asymptote_diffusion_scaling}, the far-field asymptote of the fluid displacement function \eqref{eq:far-field_asymptote_diffusion_scaling}, and the requirement of vanishing net pressure at infinity \citep{viesca2018numerical, garagash2019cohesive}:
\begin{equation*}
    F_\Omega(-1) = \sum_{j=1}^{N_p} \mathcal{P}_j F_\Omega(s_j) = 3 \sqrt{2}\mathcal{K}, \quad F_\Upsilon(1) = \sum_{j=1}^{N_p} \mathcal{Q}_j^s F_\Upsilon(s_j) = 12 \mathcal{S} \sqrt{\frac{2}{\pi}}, \quad \Pi(1) = \sum_{i=1}^{N_c} \mathcal{Q}_i^z \Pi(z_i) = 0,
\end{equation*}
where the interpolation matrices are defined as:
\begin{equation*}
    \mathcal{P}_j = \frac{(-1)^j}{N_p} \tan\left(\frac{\pi(j-1/2)}{2 N_p}\right), \quad \mathcal{Q}_j^s = \frac{(-1)^{j + 1}}{N_p} \cot\left(\frac{\pi(j-1/2)}{2 N_p}\right), \quad \mathcal{Q}_i^z = (-1)^{i + 1} (1 + z_i).
\end{equation*}

To balance the number of equations and unknowns, the fluid pressure equation at the complementary node nearest the fracture tip is omitted. The resulting non-linear system is solved by root finding using the \texttt{root} routine from the \texttt{scipy.optimize} module of the SciPy library \citep{2020SciPy-NMeth}. For the simulations presented here, we set $N_p = 300$. A resolution study for the fully coupled solutions at two representative points in Regimes 2 and 3 is presented in Appendix~\ref{sec:numerical_convergence}.

\section{Numerical convergence of the fully coupled solution}
\label{sec:numerical_convergence}

We examine the resolution dependence of the fully coupled solution for two representative parameter combinations, $(\varrho,\Sigma_0')=(10^3,10)$ and $(10^4,10^{-2})$, associated with Regimes 2 and 3, respectively (Sections~\ref{sec:solution_structure_in_regime_2}
and~\ref{sec:solution_structure_in_regime_3}). For both cases, the dimensionless poroelastic parameters are fixed at $\beta=0.15$ and $\eta=0.3$. Following the methodology described in Appendix~\ref{sec:modeling_approach}, we compute solutions with $N_p=100$, $200$, $300$, $400$, $500$, and $600$ primary nodes \eqref{eq:primary_nodes}.

All fracture fields are evaluated at the complementary nodes \eqref{eq:complementary_nodes}. Because each tested $N_p$ is a multiple of 100, all six grids share the 99 complementary nodes $z_k=\cos(\pi k/100)$, $k=1,\ldots,99$. For each representative case, these nodes map to the same $mk-$scaled coordinates $\xi_k$ at every resolution, so no interpolation is required.

For $Q\in\{\Omega,\Pi,\Upsilon\}$, we define the pairwise root-mean-square difference between successive resolutions as
\begin{equation}
    D_Q(N_p)=
    \sqrt{\frac{1}{99}\sum_{k=1}^{99}\left(
    \frac{Q_{N_p}(\xi_k)-Q_{N_p+100}(\xi_k)}{S_Q(\xi_k)}\right)^2},
    \label{eq:convergence_metric}
\end{equation}
where $S_\Omega(\xi)=\Omega_\infty(\xi)$, $S_\Pi(\xi)=1$, and $S_\Upsilon(\xi)=\Upsilon_\infty(\xi)$. The far-field asymptotes \eqref{eq:far-field_asymptote_mk_scaling} for opening and fluid displacement are used to reduce the influence of large profile magnitudes far from the tip. For pressure, we take $S_\Pi=1$: since $\Pi_\infty(\xi)$ tends to zero, using it as a denominator would give increasing weight to distant nodes.

For each fracture field and representative case, we fit the five values of $D_Q(N_p)$ to
\begin{equation}
    D_Q(N_p)\simeq C_Q \left[N_p^{-r_Q}-(N_p+100)^{-r_Q}\right],
    \label{eq:convergence_rate_fit}
\end{equation}
with $C_Q$ and $r_Q$ determined by least squares in the logarithm of
$D_Q$. The fitted exponent $r_Q$ is an observed convergence order, under the assumption that the leading discretization error varies approximately as $N_p^{-r_Q}$ over the resolutions considered.

Figure~\ref{fig:numerical_convergence} shows the measured pairwise differences and their fits.
\begin{figure}[h!]
    \centerline{
    \includegraphics[width=1\textwidth]{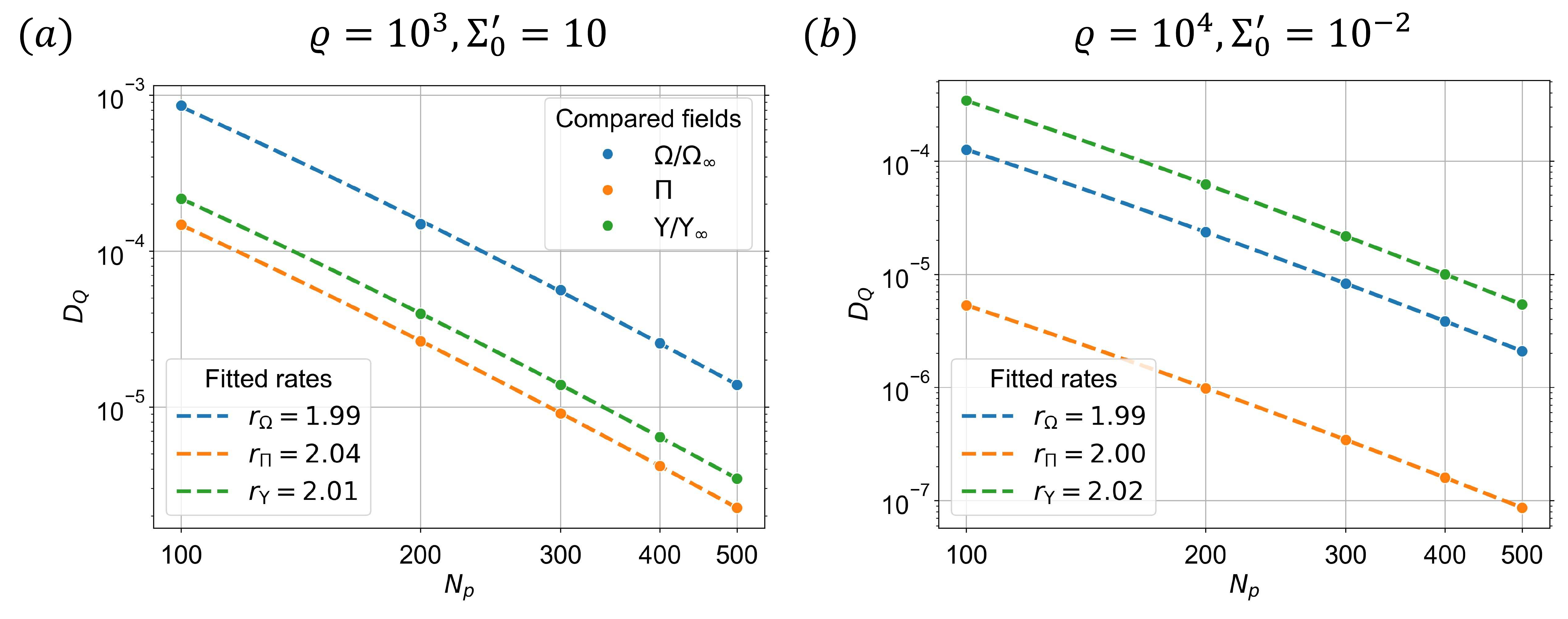}
    }
    \caption{
    Panel (a) shows the resolution dependence of the fully coupled solution for $(\varrho,\Sigma_0')=(10^3,10)$ (Regime 2), and panel (b) shows the corresponding results for $(\varrho,\Sigma_0')=(10^4,10^{-2})$ (Regime 3). In both cases, $\beta=0.15$ and $\eta=0.3$. Circles show the pairwise root-mean-square differences  \eqref{eq:convergence_metric} for the fracture opening, net pressure,
    and fluid displacement function at 99 common complementary nodes.
    Dashed curves show the fits \eqref{eq:convergence_rate_fit}.
    Each value of $N_p$ on the horizontal axis denotes a comparison
    between solutions with $N_p$ and $N_p+100$ primary nodes.
    }
    \label{fig:numerical_convergence}
\end{figure}
For all three fields in both representative cases, $D_Q(N_p)$ decreases as $N_p$ increases. All fitted exponents $r_Q$ are close to $2$. At the resolution used for the reported results, $N_p=300$, each $D_Q(300)$, comparing the $N_p=300$ and $400$ solutions, is below $6\times10^{-5}$, and the differences decrease further upon refinement. For these two representative cases, in which poroelastic effects substantially influence the near-tip fields, the small differences and their continued decrease support the use of $N_p=300$ for the reported calculations. The values of $D_Q$ quantify differences between successive resolutions rather than errors relative to an exact solution.

\section{Effects of the dimensionless poroelastic parameters $\beta$ and $\eta$}
\label{sec:beta_eta_variation}

In Section~\ref{sec:general_solution}, we analysed the influence of the dimensionless permeability $\varrho$ and the dimensionless effective confining stress $\Sigma_0'$ on the fracture opening, net pressure, and fluid displacement function profiles in the fully coupled fracture tip solution, while keeping the dimensionless poroelastic parameters fixed at the representative values $\beta=0.15$ and $\eta=0.3$. This appendix examines how variations in $\beta$ and $\eta$ affect the near-tip fracture fields predicted by the fully coupled formulation at the three representative points considered in Section~\ref{sec:solution_structure_in_regimes}: $(\varrho, \Sigma_0')=(10^{-5}, 10^2)$, $(10^3, 10)$, and $(10^4, 10^{-2})$. The first point is associated with Regimes 0 and 1, depending on the comparison interval, whereas the second and third points correspond to Regimes 2 and 3, respectively, as defined in Section~\ref{sec:regime_maps}. Both dimensionless poroelastic parameters are varied within the admissible ranges identified in Section~\ref{sec:representative_values_governing_parameters}.

The parameter $\beta$ controls the contrast between the drained and undrained elastic responses and therefore directly affects the far-field behaviour of the hydraulic fracture, as given by \eqref{eq:far-field_asymptote_mk_scaling}. In turn, the poroelastic stress coefficient $\eta$ controls the strength of the cross-coupling contributions to the normal stress and pore fluid pressure. When one parameter is varied, the other is kept fixed at its representative value used in the main text: $\eta=0.3$ for variations of $\beta$, and $\beta=0.15$ for variations of $\eta$.

\subsection{Effect of the dimensionless parameter $\beta$}

We first vary the dimensionless parameter $\beta$. The resulting fracture opening and net pressure profiles are shown in figure~\ref{fig:variation_beta}. Since the far-field behaviour of these fracture fields depends on $\beta$, the profiles are normalized by the corresponding far-field asymptotes in the uncoupled limit, i.e. by \eqref{eq:far-field_asymptote_mk_scaling} evaluated at $\beta=0$.
\begin{figure}[h!]
    \centerline{
    \includegraphics[width=1\textwidth]{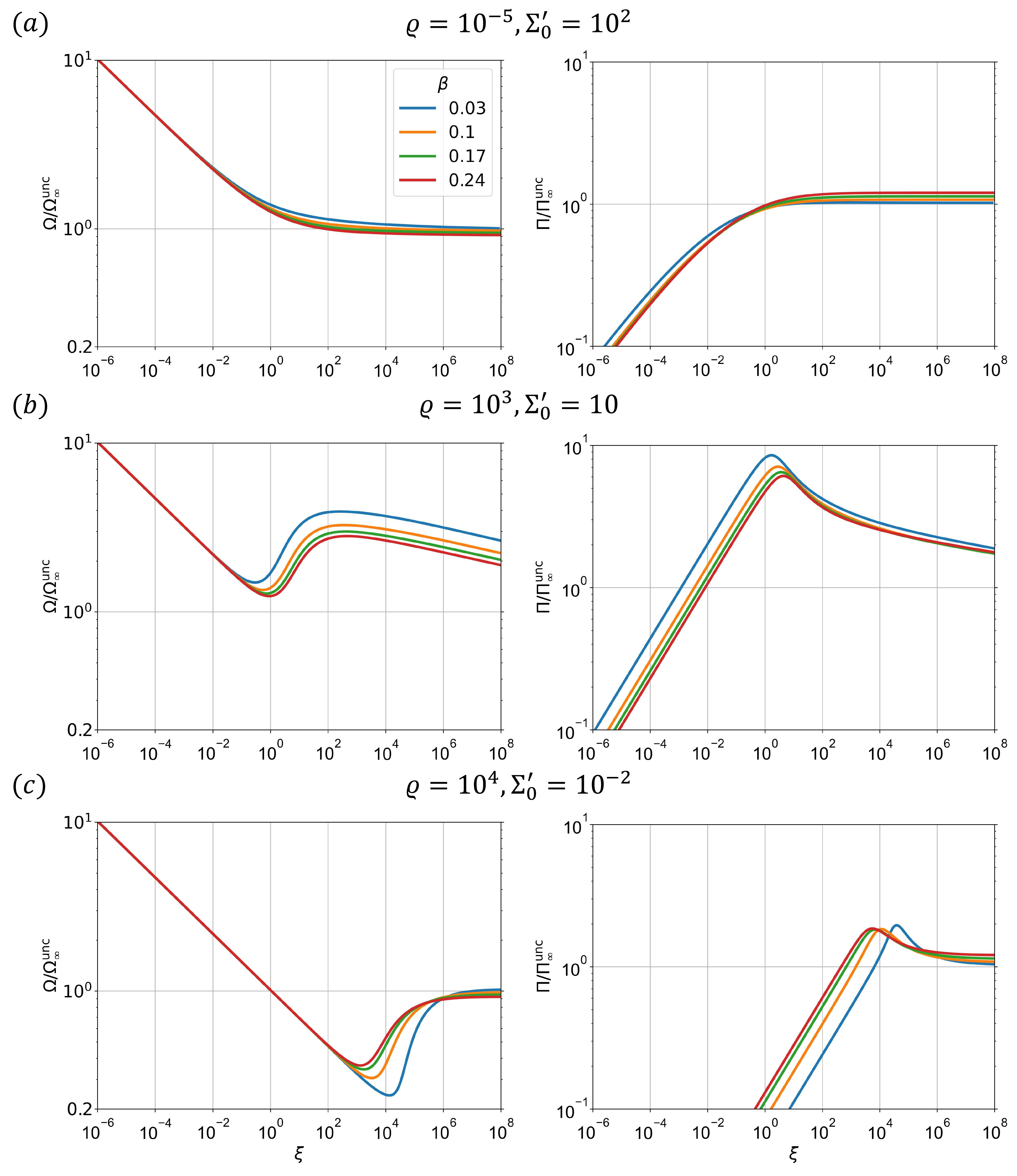}
    }
    \caption{
    Fracture opening $\Omega(\xi)$ (left column) and net pressure $\Pi(\xi)$ (right column) profiles for different values of the dimensionless parameter $\beta$ at three representative points in the parameter space: $(\varrho, \Sigma_0')=(10^{-5}, 10^2)$ (a), $(10^3, 10)$ (b), and $(10^4, 10^{-2})$ (c). The poroelastic stress coefficient is fixed at $\eta=0.3$. The profiles are shown in normalized form as $\Omega/\Omega_\infty^{\mathrm{unc}}$ and $\Pi/\Pi_\infty^{\mathrm{unc}}$, where $\Omega_\infty^{\mathrm{unc}}$ and $\Pi_\infty^{\mathrm{unc}}$ denote the far-field asymptotes in the uncoupled limit, obtained from \eqref{eq:far-field_asymptote_mk_scaling} by setting $\beta=0$. 
    }
    \label{fig:variation_beta}
\end{figure}

Larger values of $\beta$ reduce the far-field fracture opening and increase the absolute value of the far-field net pressure; equivalently, since the net pressure is negative, the signed far-field net pressure becomes more negative. This trend is visible through the limiting far-field levels of $\Omega/\Omega_\infty^{\mathrm{unc}}$ and $\Pi/\Pi_\infty^{\mathrm{unc}}$ in figure~\ref{fig:variation_beta}, and is consistent with \eqref{eq:far-field_asymptote_mk_scaling}.

The influence of $\beta$ is also reflected in the near-tip net pressure plateau. For the first and second representative points, associated with Regimes 0/1 and 2, respectively, larger $\beta$ leads to a smaller absolute value of the near-tip net pressure. By contrast, at the Regime 3 point, the near-tip pressure plateau becomes larger in magnitude.

To quantify how the influence of poroelastic effects changes with $\beta$, we evaluate the combined metric $\Delta$, defined in \eqref{eq:metrics_for_comparison}, between the fully coupled and uncoupled solutions. For the first representative point, $\Delta$ rises from approximately $2\%$ to $16.7\%$ over the range of $\beta$ considered. This growth is mainly associated with the different far-field asymptotic behaviours of the fully coupled and uncoupled solutions. For the Regime 2 point, the metric changes only weakly, varying from about $25\%$ to $26.4\%$. For the Regime 3 point, the change is more pronounced, with $\Delta$ rising from approximately $17.2\%$ to $36.6\%$. This demonstrates that the impact of poroelastic effects on the fracture opening and net pressure profiles becomes stronger as the drained--undrained contrast becomes larger.

The effect of $\beta$ on the spatial transition between the near- and far-field regimes is less systematic. The distance from the fracture tip at which the profiles approach their far-field asymptotes changes only weakly with $\beta$. However, larger $\beta$ slightly extends the range over which the near-field asymptote remains applicable for the first and second representative points, whereas this range shrinks for the Regime 3 point. Overall, variations in $\beta$ mainly affect the intermediate region, where the solution transitions from the drained near-tip response to the undrained far-field response. This is also the region where the non-monotonic features of the opening and net pressure profiles are most sensitive to $\beta$.

The fluid displacement function profiles for varying $\beta$ are shown in figure~\ref{fig:variation_beta_Y} in the supplementary material, demonstrating that $\Upsilon$ is also affected by changes in the drained--undrained contrast. For the first and second representative points, the zero-crossing of $\Upsilon$ moves farther from the fracture tip as $\beta$ increases, indicating an enlargement of the near-tip circulation zone. By contrast, for the Regime 3 point, the circulation zone shrinks with increasing $\beta$. In all three cases, the far-field magnitude of $\Upsilon$ decreases with $\beta$, consistently with the far-field asymptote \eqref{eq:far-field_asymptote_mk_scaling} through the dependence of the dimensionless storage coefficient $\mathcal{S}$ on $\beta$.

\subsection{Effect of the poroelastic stress coefficient $\eta$}

Next, we vary the poroelastic stress coefficient $\eta$. The resulting fracture opening and net pressure profiles, normalized by their far-field asymptotes \eqref{eq:far-field_asymptote_mk_scaling}, are shown in figure~\ref{fig:variation_eta}.
\begin{figure}[h!]
    \centerline{
    \includegraphics[width=1\textwidth]{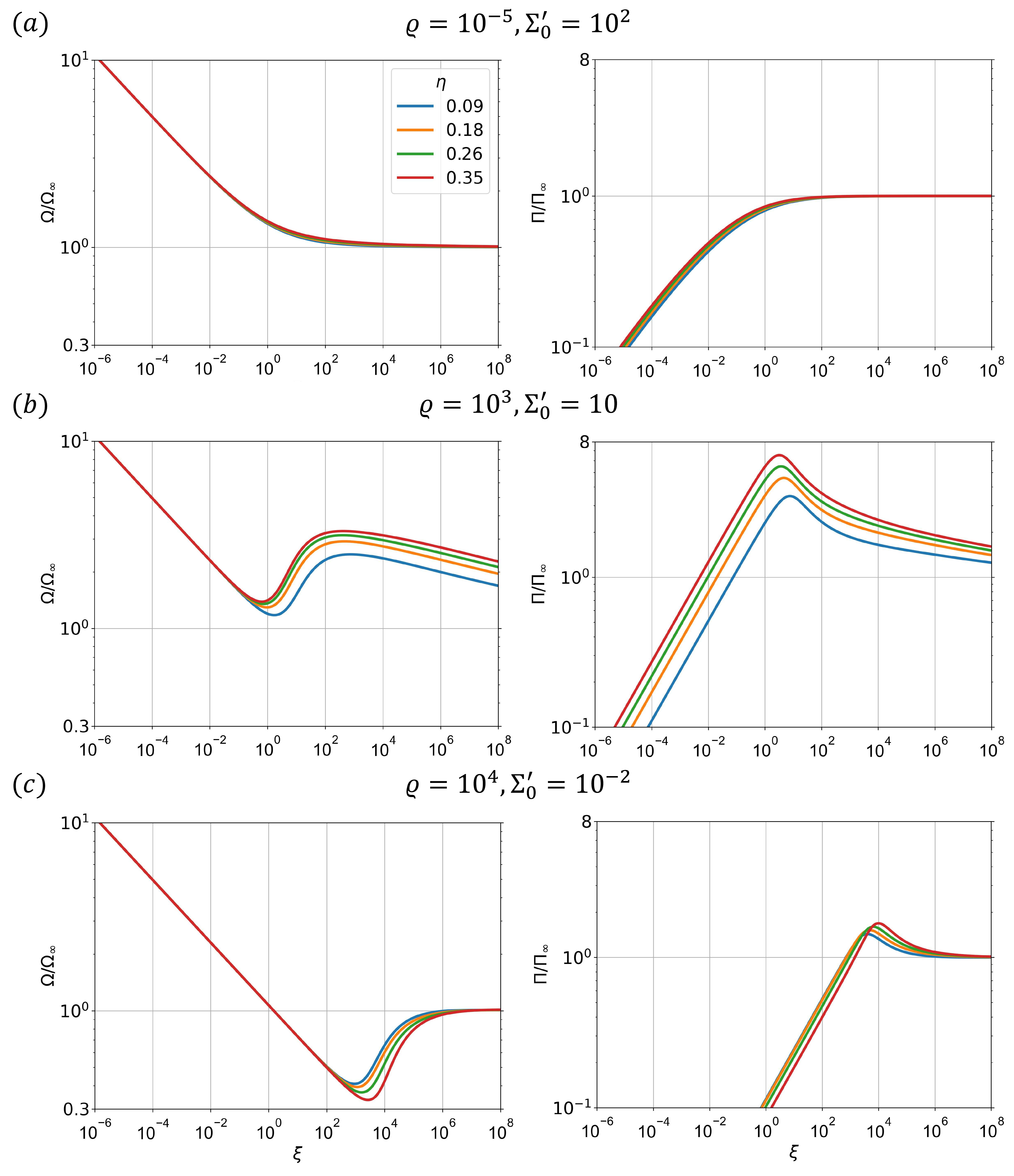}
    }
    \caption{
    Fracture opening $\Omega(\xi)$ (left column) and net pressure $\Pi(\xi)$ (right column) profiles for different values of the poroelastic stress coefficient $\eta$ at three representative points in the parameter space: $(\varrho, \Sigma_0')=(10^{-5}, 10^2)$ (a), $(10^3, 10)$ (b), and $(10^4, 10^{-2})$ (c). The dimensionless poroelastic parameter $\beta$ is fixed at $\beta=0.15$. The profiles are shown in normalized form as $\Omega/\Omega_\infty$ and $\Pi/\Pi_\infty$, where $\Omega_\infty$ and $\Pi_\infty$ are the far-field asymptotes given by \eqref{eq:far-field_asymptote_mk_scaling}.
    }
    \label{fig:variation_eta}
\end{figure}

Since $\eta$ controls the strength of the cross-coupling contributions, its influence is weak at the first representative point, associated with Regimes 0/1. The fracture opening and net pressure profiles remain almost unchanged over the range of $\eta$ considered, consistent with the negligible role of cross-coupling mechanisms in this part of the parameter space.

At the Regime 2 and Regime 3 points, the effect of $\eta$ becomes more pronounced. In Regime 2, larger values of $\eta$ enhance the poroelastic backstress. In Regime 3, variations in $\eta$ affect not only the backstress but also the mechanical-to-hydraulic coupling, so that deformation-induced pore pressure perturbations are modified as well. Consequently, at both points, the opening and net pressure profiles change noticeably: the opening mainly in the intermediate field, and the net pressure in both the near-tip and intermediate regions.

The near-tip pressure plateau also depends on $\eta$. For the first and second representative points, associated with Regimes 0/1 and 2, respectively, the absolute value of the near-tip net pressure increases with $\eta$. By contrast, at the Regime 3 point, the near-tip pressure magnitude decreases as $\eta$ increases.

To quantify this sensitivity, we evaluate the combined metric $\Delta$ \eqref{eq:metrics_for_comparison} between the fully coupled and uncoupled solutions. For the first representative point, $\Delta$ remains almost unchanged at approximately $10.3\%$ over the considered range of $\eta$. By contrast, for the Regime 2 point, it increases from about $10.8\%$ to $30.2\%$, while for the Regime 3 point it rises from approximately $19.6\%$ to $34.9\%$. Thus, increasing $\eta$ enhances the influence of poroelastic effects on the fracture opening and net pressure profiles, except in the Regimes 0/1 case, where the cross-coupling mechanisms remain weak.

The corresponding fluid displacement function profiles are shown in figure~\ref{fig:variation_eta_Y} in the supplementary material. For the first and second representative points, the zero crossing of $\Upsilon$ moves closer to the fracture tip as $\eta$ increases, indicating a reduction of the near-tip circulation zone. By contrast, for the Regime 3 point, the circulation zone expands with increasing $\eta$. In all cases, the far-field magnitude of $\Upsilon$ increases with $\eta$, consistently with the far-field asymptote \eqref{eq:far-field_asymptote_mk_scaling} through the dependence of the dimensionless storage coefficient $\mathcal{S}$ on $\eta$.

\section{Regime maps based on the opening profiles}
\label{sec:regime_maps_opening}

This appendix presents the relative difference and regime maps obtained using the opening-based metric $\Delta_\Omega$~\eqref{eq:metrics_for_comparison_2} and discusses their main differences from the corresponding maps based on the combined metric $\Delta$~\eqref{eq:metrics_for_comparison}, presented in Sections~\ref{sec:rel_dif_maps} and~\ref{sec:regime_maps}. This analysis is motivated by hydraulic fracture propagation algorithms in which the fracture tip solution enters primarily through the opening profile. 

The validity domains of the reduced models, $\mathcal{D}_0$, $\mathcal{D}_1$, and $\mathcal{D}_2$, are determined following the procedure described in Section~\ref{sec:regime_maps}, with the opening-based metric $\Delta_\Omega$ used instead of the combined metric $\Delta$. Unlike the domains obtained using the combined metric, however, the opening-based domains are not strictly nested. This occurs because the effects of different poroelastic coupling mechanisms on the fracture opening can partially compensate for one another. Consequently, a simpler reduced model may satisfy the criterion $\Delta_\Omega < 5\%$ at a point where a more complete reduced model does not. Such agreement  should therefore not be interpreted as evidence that the mechanisms neglected by the simpler model are physically negligible.

To account for this lack of nesting while preserving a hierarchical classification based on model complexity, the applicability regions of the characteristic coupling regimes are defined as: $\mathcal{R}_0 = \mathcal{D}_0 \cap \mathcal{D}_1 \cap \mathcal{D}_2$, $\mathcal{R}_1 = (\mathcal{D}_1 \cap \mathcal{D}_2) \setminus \mathcal{D}_0$, $\mathcal{R}_2 = \mathcal{D}_2 \setminus \mathcal{D}_1$, and  $\mathcal{R}_3 = \mathcal{D}\setminus\mathcal{D}_2$. These definitions ensure that the four applicability regions are mutually disjoint and together cover the entire modelling domain $\mathcal{D}$. A point is assigned to Regime 0 only if all three reduced models reproduce the opening profile within the prescribed accuracy. Regime 1 corresponds to the region where the auxiliary model without cross-coupling terms and the one-way coupled model are accurate, whereas the uncoupled model is not. Regime 2 applies where the one-way coupled model is accurate but the auxiliary model is not. Finally, Regime 3 comprises the part of the modelling domain where the one-way coupled model is insufficient.

The relative difference maps comparing the fracture opening profiles obtained with the fully coupled model and with the uncoupled and one-way coupled models are provided in figures~\ref{fig:Delta_fully_coupled_vs_uncoupled_Delta_Omega} and \ref{fig:Delta_fully_coupled_vs_one_way_coupled_Delta_Omega} in the supplementary material, respectively. For the comparison with the uncoupled model, the region where $\Delta_\Omega<5\%$ is substantially larger than the corresponding region obtained using the combined metric. In particular, for the physical comparison interval $\ell=10^{-2}$ m, this region occupies almost the entire modelling domain. As $\ell$ increases, it progressively shrinks and eventually disappears in the limiting case $\xi_0=\infty$. For the representative value $\beta=0.15$ considered here, this occurs because the opening profiles in the fully coupled and uncoupled models still have different far-field asymptotic behaviours \eqref{eq:far-field_asymptote_mk_scaling}, leading to a relative difference exceeding $5\%$. The maximum opening discrepancy in this limiting case reaches approximately $58\%$ in the region with large $\varrho$ and small $\Sigma_0'$.

For the comparison with the one-way coupled model, the region where $\Delta_\Omega < 5\%$ also occupies almost the entire modelling domain for $\ell=10^{-2}$ m and shrinks only moderately as $\ell$ increases. The largest opening discrepancy occurs at large $\varrho$ and small $\Sigma_0'$, reaching approximately $66\%$ in the limiting case $\xi_0=\infty$. Thus, even when only the fracture opening is used to compare the fully coupled and reduced models, the mechanical-to-hydraulic coupling remains important in the same part of the parameter space as identified using the combined metric.

Figure~\ref{fig:Regime_maps_Delta_Omega} shows the regime maps based on $\Delta_\Omega$ for five finite physical comparison intervals, $\ell=10^{-2}$, $0.1$, $1$, $10$, and $10^2$ m, together with the limiting case $\xi_0=\infty$. 
\begin{figure}[h!]
    \centerline{
    \includegraphics[width=1\textwidth]{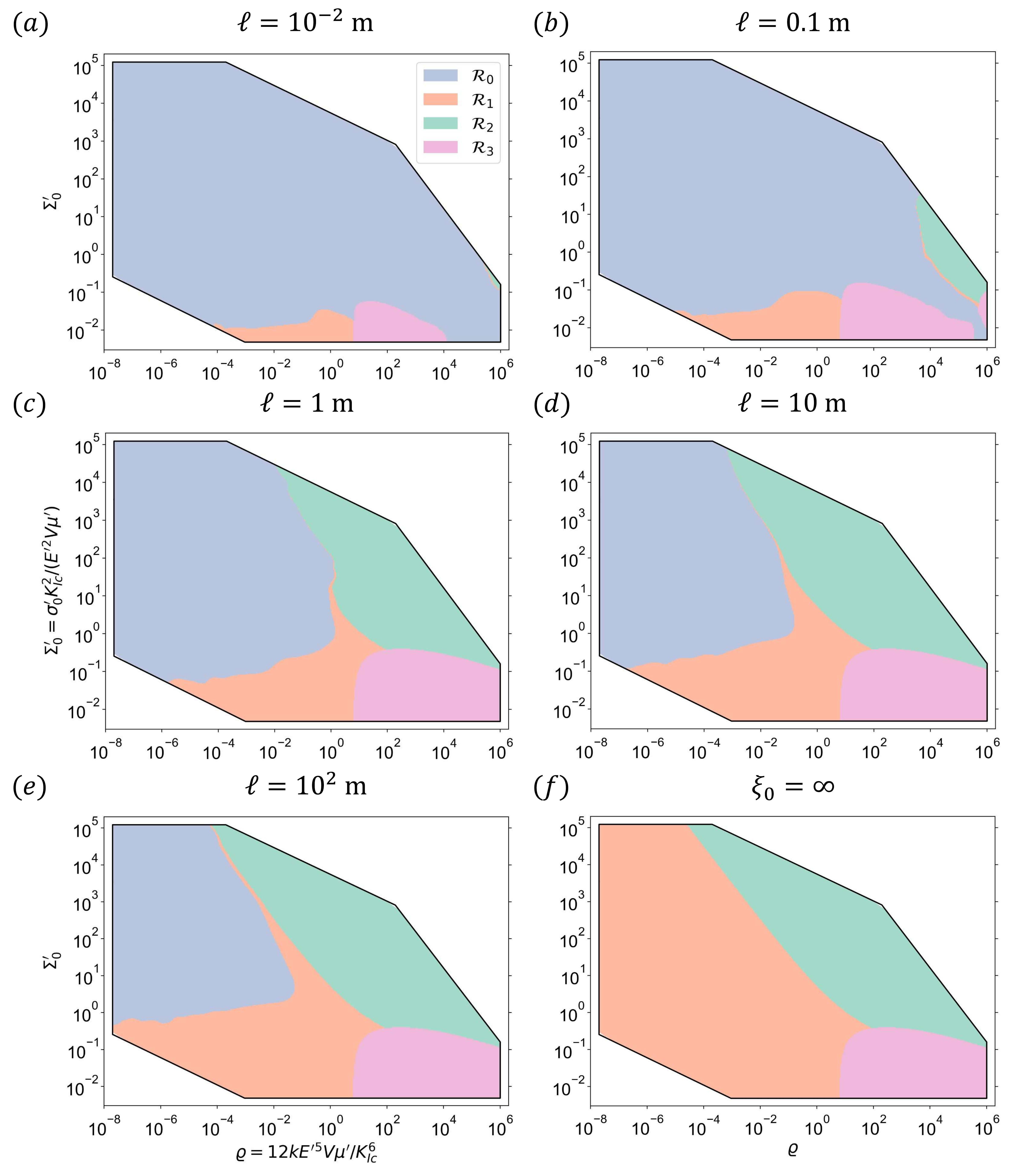}
    }
    \caption{
    Regime maps identifying the dominant poroelastic coupling mechanisms in the $(\varrho,\Sigma_0')$ parameter space. Panels (a)--(e) are constructed from relative difference maps comparing the fully coupled and reduced models over the finite physical comparison intervals: $\ell=10^{-2}$ m (a), $\ell=0.1$ m (b), $\ell=1$ m (c), $\ell=10$ m (d), and $\ell=10^2$ m (e). Panel (f) is based on the corresponding relative difference maps for the limiting case $\xi_0=\infty$, in which the comparison is performed over the entire semi-infinite fracture. The regimes correspond to: $\mathcal{R}_0$, negligible poroelastic effects; $\mathcal{R}_1$, dominance of the direct mechanical poroelastic response; $\mathcal{R}_2$, significant influence of the poroelastic backstress; and $\mathcal{R}_3$, the fully coupled regime. The regime classification is based on the opening-based metric $\Delta_\Omega$ defined in equation \eqref{eq:metrics_for_comparison_2}. The black contour outlines the modelling domain shown in figure~\ref{fig:representative_ranges_governing_parameters}a. The dimensionless poroelastic parameters are fixed at $\beta=0.15$ and $\eta=0.3$.
    }
    \label{fig:Regime_maps_Delta_Omega}
\end{figure}
Comparing the regime maps based on $\Delta_\Omega$ (figure~\ref{fig:Regime_maps_Delta_Omega}) with those based on the combined metric $\Delta$ (figure~\ref{fig:Regime_maps}), we observe that, for finite comparison intervals, Regime 0 occupies a significantly larger region when the opening-based metric is used for classification. In particular, for $\ell=10^{-2}$ m, Regime 0 covers almost the entire modelling domain, indicating that the uncoupled formulation can reproduce the near-tip opening profile with good accuracy over this short physical interval. This occurs because, for sufficiently small $\ell$, the comparison is performed mainly within the range over which the near-field LEFM asymptote~\eqref{eq:near-field_asymptote_mk_scaling} applies; this asymptote is identical in the fully coupled and uncoupled models. As $\ell$ increases, the applicability region of Regime 0 decreases, and it disappears when the entire semi-infinite fracture is included in the comparison. Correspondingly, the applicability regions of Regimes 1--3 gradually expand, reflecting the increasing influence of poroelastic effects on the fracture opening profile as a larger portion of the semi-infinite fracture solution is taken into account.

In the limiting case $\xi_0=\infty$, the regime map based on the opening-based metric preserves the same qualitative structure as the map obtained with the combined metric, although the regime boundaries are shifted. In particular, the applicability region of Regime 1 becomes larger, whereas that of Regime 3 becomes smaller. This indicates that the fracture opening is generally less sensitive than the net pressure to the mechanical-to-hydraulic coupling. Nevertheless, the fully coupled formulation remains necessary at large $\varrho$ and small $\Sigma_0'$.

\end{appen}

\clearpage



\section*{Nomenclature}

\begin{longtable}{@{}p{0.24\linewidth}p{0.72\linewidth}@{}}
\multicolumn{2}{@{}l}{\textit{Fracture geometry and fields}}\\[2pt]
$(X,Y)$ & Coordinates in the fixed frame \\
$(x,y)$ & Coordinates in the frame moving with the fracture tip \\
$t$, $V$ & Time; fracture propagation velocity \\
$w$, $q$ & Fracture opening; fluid flux within the fracture \\
$g$, $v$ & Fluid exchange rate; fluid displacement function \\
$p_f$, $p$, $p_p$ & Fracturing fluid pressure; net pressure; pore fluid pressure on the fracture surfaces \\
$p_0$, $\sigma_0$, $\sigma_0'$ & Initial pore pressure; confining stress; effective confining stress \\
$\sigma_{yy}$ & Normal stress on the fracture surfaces \\
$\sigma_{yy}^{M}$, $\sigma_{yy}^{HM}$ & Direct mechanical and hydraulic-to-mechanical contributions to $\sigma_{yy}$ \\
$\sigma_{yy}^{M,e}$ & Purely elastic approximation of $\sigma_{yy}^{M}$ \\
$p^{H}$, $p^{MH}$ & Direct hydraulic and mechanical-to-hydraulic contributions to $p_p$ \\
$p_{1\mathrm{D}}^{H}$ & Direct hydraulic contribution to $p_p$ in the one-dimensional pressure-dependent fluid exchange model \\
$v_{1\mathrm{D}}$, $v_{\mathrm C}$ & Fluid displacement functions for one-dimensional pressure-dependent fluid exchange and Carter's leak-off \\[5pt]

\multicolumn{2}{@{}l}{\textit{Rock and fluid properties}}\\[2pt]
$E$, $E'$ & Drained Young's modulus; drained plane strain modulus \\
$G$, $K$ & Shear modulus; drained bulk modulus \\
$\nu$, $\nu_u$ & Drained and undrained Poisson's ratios \\
$\alpha$, $M$, $\phi$ & Biot coefficient; Biot modulus; porosity \\
$K_f$, $K_s$ & Bulk moduli of the pore fluid and solid grains \\
$K_I$, $K_{Ic}$ & Mode-I stress intensity factor; fracture toughness \\
$k$, $\kappa$, $c$ & Intrinsic permeability; fluid mobility; hydraulic diffusivity \\
$\mu$, $\mu'$ & Fluid viscosity; viscosity parameter \\
$S$ & Dimensional storage coefficient \\
$\beta$, $\eta$ & Drained-undrained Poisson's ratio contrast; poroelastic stress coefficient \\[5pt]

\multicolumn{2}{@{}l}{\textit{Boundary integral kernels}}\\[2pt]
$\sigma_{yy}^{e}$, $\sigma_{yy}^{s}$ & Normal stress responses to a unit normal edge dislocation and a unit fluid source \\
$p^{e}$, $p^{s}$ & Pore pressure responses to a unit normal edge dislocation and a unit fluid source \\
$S^{e}$, $S^{s}$ & Dimensionless normal stress kernel functions \\
$P^{e}$, $P^{s}$ & Dimensionless pore pressure kernel functions \\[5pt]

\multicolumn{2}{@{}l}{\textit{Dimensionless fracture fields and parameters}}\\[2pt]
$\xi$, $\Omega$, $\Pi$, $\Upsilon$ & Distance from the tip; fracture opening; net pressure; fluid displacement function \\
$\Sigma$, $\Pi_p$ & Normal stress and pore pressure perturbations on the fracture surfaces \\
$\Sigma^{M}$, $\Sigma^{HM}$ & Direct mechanical and hydraulic-to-mechanical contributions to $\Sigma$ \\
$\Sigma^{M,e}$ & Purely elastic approximation of $\Sigma^{M}$ \\
$\Pi^{H}$, $\Pi^{MH}$ & Direct hydraulic and mechanical-to-hydraulic contributions to $\Pi_p$ \\
$\Pi_{1\mathrm{D}}^{H}$ & Direct hydraulic contribution to $\Pi_p$ in the one-dimensional pressure-dependent fluid exchange model \\
$\Upsilon_{1\mathrm{D}}$, $\Upsilon_{\mathrm C}$ & Fluid displacement functions for one-dimensional pressure-dependent fluid exchange and Carter's leak-off \\
$\varrho$, $\Sigma_0'$ & Dimensionless permeability; dimensionless effective confining stress \\
$\mathcal{S}$ & Dimensionless storage coefficient \\
$\chi$, $\zeta$ & Dimensionless leak-off and leak-in coefficients \\[5pt]

\multicolumn{2}{@{}l}{\textit{Scales, asymptotes and model comparison}}\\[2pt]
$\ell_{mk}$, $w_{mk}$, $v_{mk}$, $p_{mk}$ & Length, opening, fluid displacement and pressure scales in the $mk$-scaling \\
$\ell_{\mathrm{dif}}$ & Diffusion length scale \\
$\ell$, $\xi_0$ & Physical comparison length; corresponding dimensionless interval endpoint \\
$\Delta_\Omega$, $\Delta_\Pi$, $\Delta$ & Opening, pressure and combined relative difference measures \\
$\mathcal{D}$, $\mathcal{D}_0$, $\mathcal{D}_1$, $\mathcal{D}_2$ & Modelling domain and reduced model validity domains \\
$\mathcal{R}_0$, $\mathcal{R}_1$, $\mathcal{R}_2$, $\mathcal{R}_3$ & Characteristic poroelastic coupling regimes \\[5pt]
\end{longtable}

\clearpage
\setcounter{figure}{0}
\renewcommand{\thefigure}{S\arabic{figure}}

\section*{Supplementary figures}

\begin{figure}[!ht]
    \centerline{
    \includegraphics[width=1\textwidth]{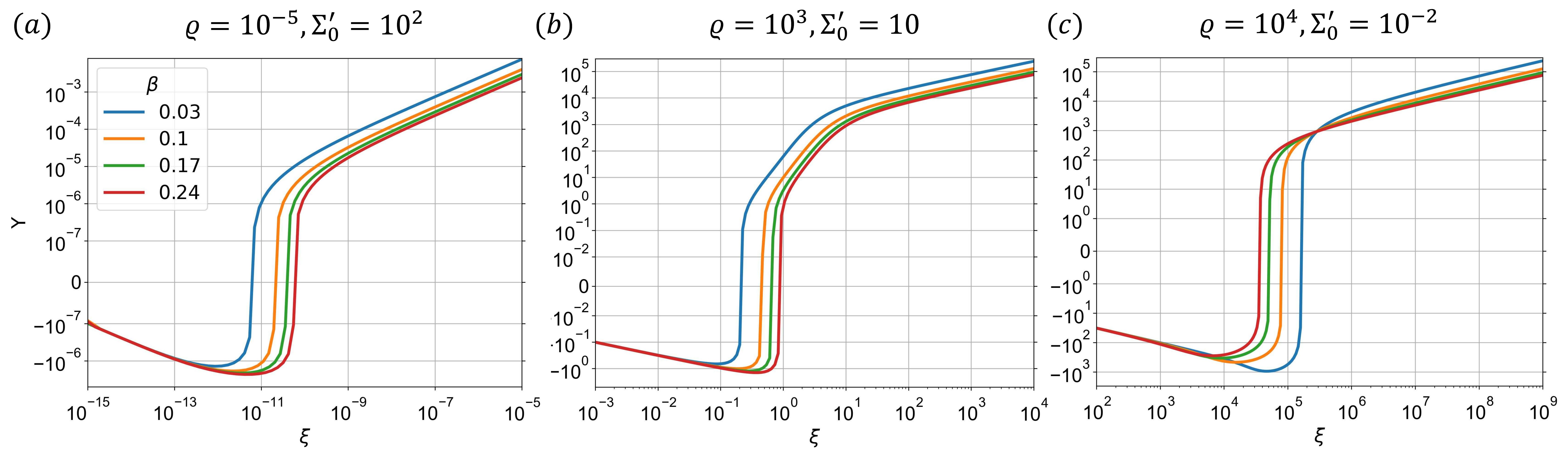}
    }
    \caption{
    The fluid displacement function $\Upsilon(\xi)$ for different values of the dimensionless parameter $\beta$ at three representative points in the parameter space: $(\varrho, \Sigma_0')=(10^{-5}, 10^2)$ (a), $(10^3, 10)$ (b), and $(10^4, 10^{-2})$ (c). The poroelastic stress coefficient is fixed at $\eta=0.3$. A symmetric logarithmic scale is used on the vertical axis to show both negative and positive values of $\Upsilon$.    
    }
    \label{fig:variation_beta_Y}
\end{figure}

\begin{figure}[!ht]
    \centerline{
    \includegraphics[width=1\textwidth]{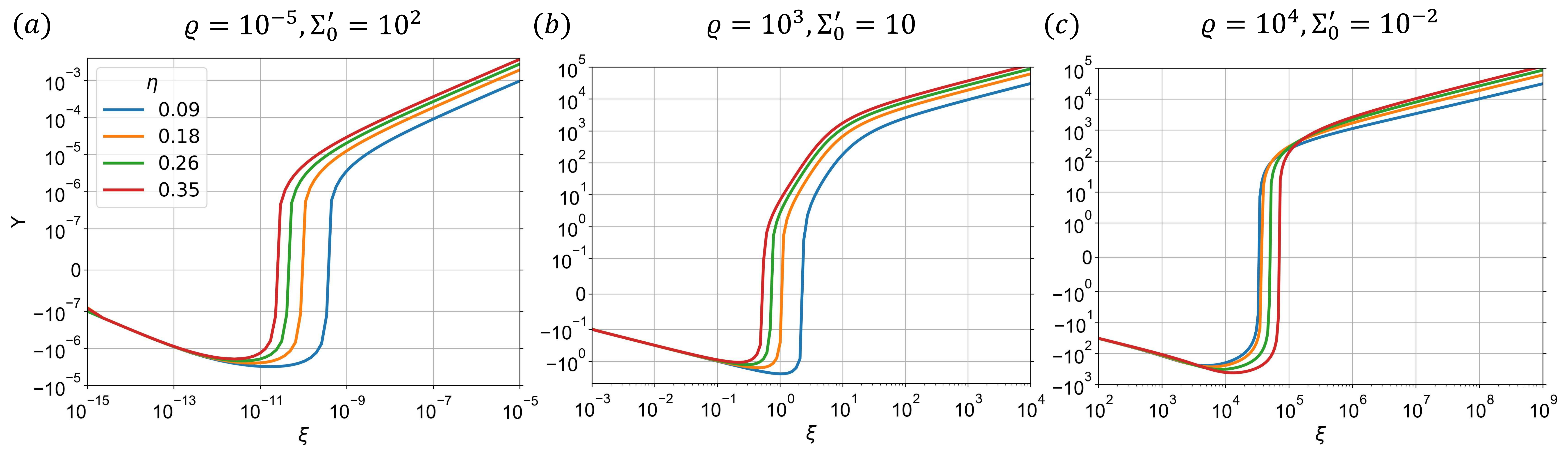}
    }
    \caption{
    The fluid displacement function $\Upsilon(\xi)$ for different values of the poroelastic stress coefficient $\eta$ at three representative points in the parameter space: $(\varrho, \Sigma_0')=(10^{-5}, 10^2)$ (a), $(10^3, 10)$ (b), and $(10^4, 10^{-2})$ (c). The dimensionless poroelastic parameter $\beta$ is fixed at $\beta=0.15$. A symmetric logarithmic scale is used on the vertical axis to show both negative and positive values of $\Upsilon$.   
    }
    \label{fig:variation_eta_Y}
\end{figure}

\begin{figure}[!ht]
    \centering
    \includegraphics[width=1\textwidth]{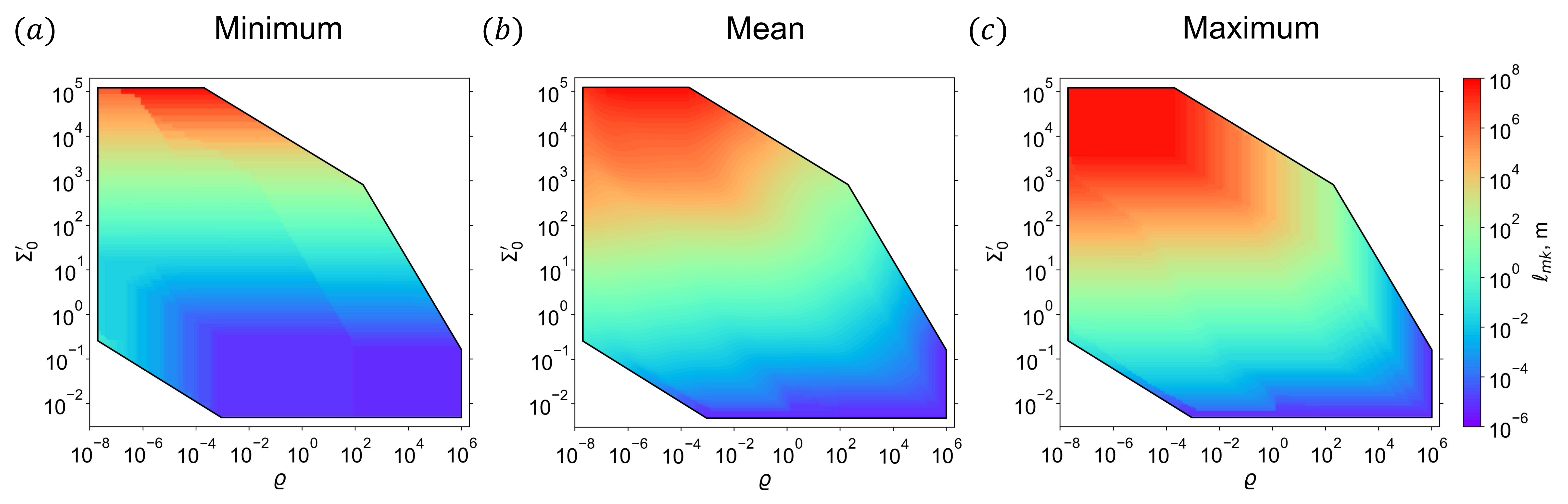}
    \caption{
    Distribution of the minimum (a), mean (b), and maximum (c) values of the length scale $\ell_{mk}$, in meters, over the $(\varrho,\Sigma_0')$ parameter space. The admissible range of $\ell_{mk}$ at each point $(\varrho,\Sigma_0')$ is evaluated from the dimensional parameter ranges discussed in  Section~\ref{sec:representative_values_governing_parameters}. The black contour outlines the modelling domain shown in figure~\ref{fig:representative_ranges_governing_parameters}a.
    }
    \label{fig:l_mk_distribution}
\end{figure}

\begin{figure}[!ht]
    \centerline{
    \includegraphics[width=1\textwidth]{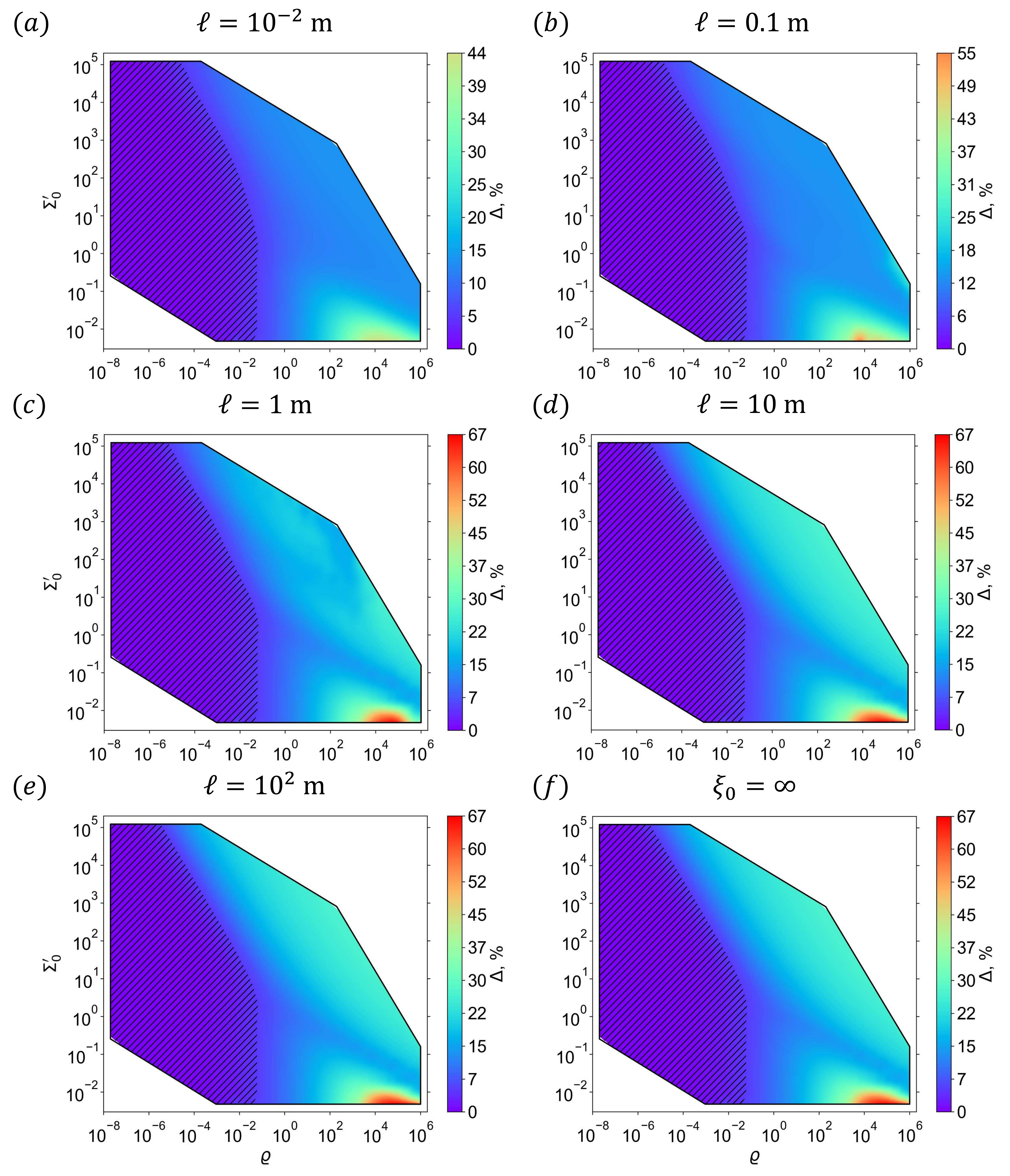}
    }
    \caption{
    Relative difference $\Delta$~\eqref{eq:metrics_for_comparison} between the fully coupled model and the auxiliary model without cross-coupling terms, evaluated over the $(\varrho,\Sigma_0')$ parameter space. Panels show the results for finite physical comparison intervals $\ell=10^{-2}$ m (a), $\ell=0.1$ m (b), $\ell=1$ m (c), $\ell=10$ m (d), and $\ell=10^2$ m (e), together with the limiting case $\xi_0=\infty$ (f). The colour scale is common to all panels, so that identical colours correspond to the same values of $\Delta$. The black contour outlines the modelling domain shown in figure~\ref{fig:representative_ranges_governing_parameters}a. Hatched regions indicate $\Delta<5\%$, where the model without cross-coupling terms approximates the fully coupled solution within the prescribed accuracy. The dimensionless poroelastic parameters are fixed at $\beta=0.15$ and $\eta=0.3$.
    }
    \label{fig:Delta_fully_coupled_vs_no_cross_couplings}
\end{figure}

\begin{figure}[!ht]
    \centerline{
    \includegraphics[width=1\textwidth]{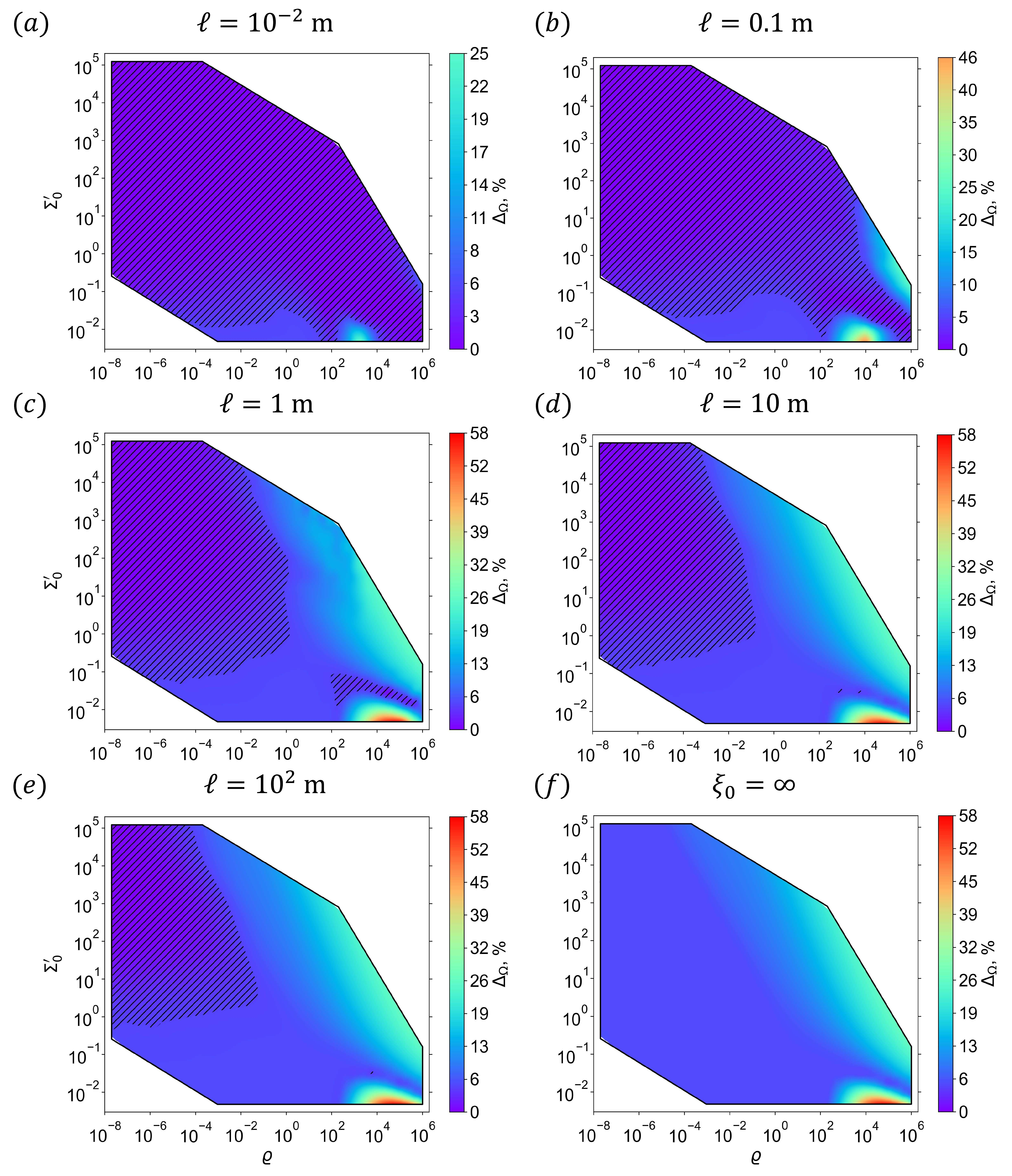}
    }
    \caption{
    Relative difference $\Delta_\Omega$~\eqref{eq:metrics_for_comparison_2} between the fracture opening profiles obtained with the fully coupled and uncoupled models, evaluated over the $(\varrho,\Sigma_0')$ parameter space. Panels show the results for finite physical comparison intervals $\ell=10^{-2}$ m (a), $\ell=0.1$ m (b), $\ell=1$ m (c), $\ell=10$ m (d), and $\ell=10^2$ m (e), together with the limiting case $\xi_0=\infty$ (f). The colour scale is common to all panels, so that identical colours correspond to the same values of $\Delta_\Omega$. The black contour outlines the modelling domain shown in figure~\ref{fig:representative_ranges_governing_parameters}a. Hatched regions indicate $\Delta_\Omega<5\%$, where the uncoupled model reproduces the fracture opening profile of the fully coupled solution within the prescribed accuracy. The dimensionless poroelastic parameters are fixed at $\beta=0.15$ and $\eta=0.3$.
    }
    \label{fig:Delta_fully_coupled_vs_uncoupled_Delta_Omega}
\end{figure}

\begin{figure}[!ht]
    \centerline{
    \includegraphics[width=1\textwidth]{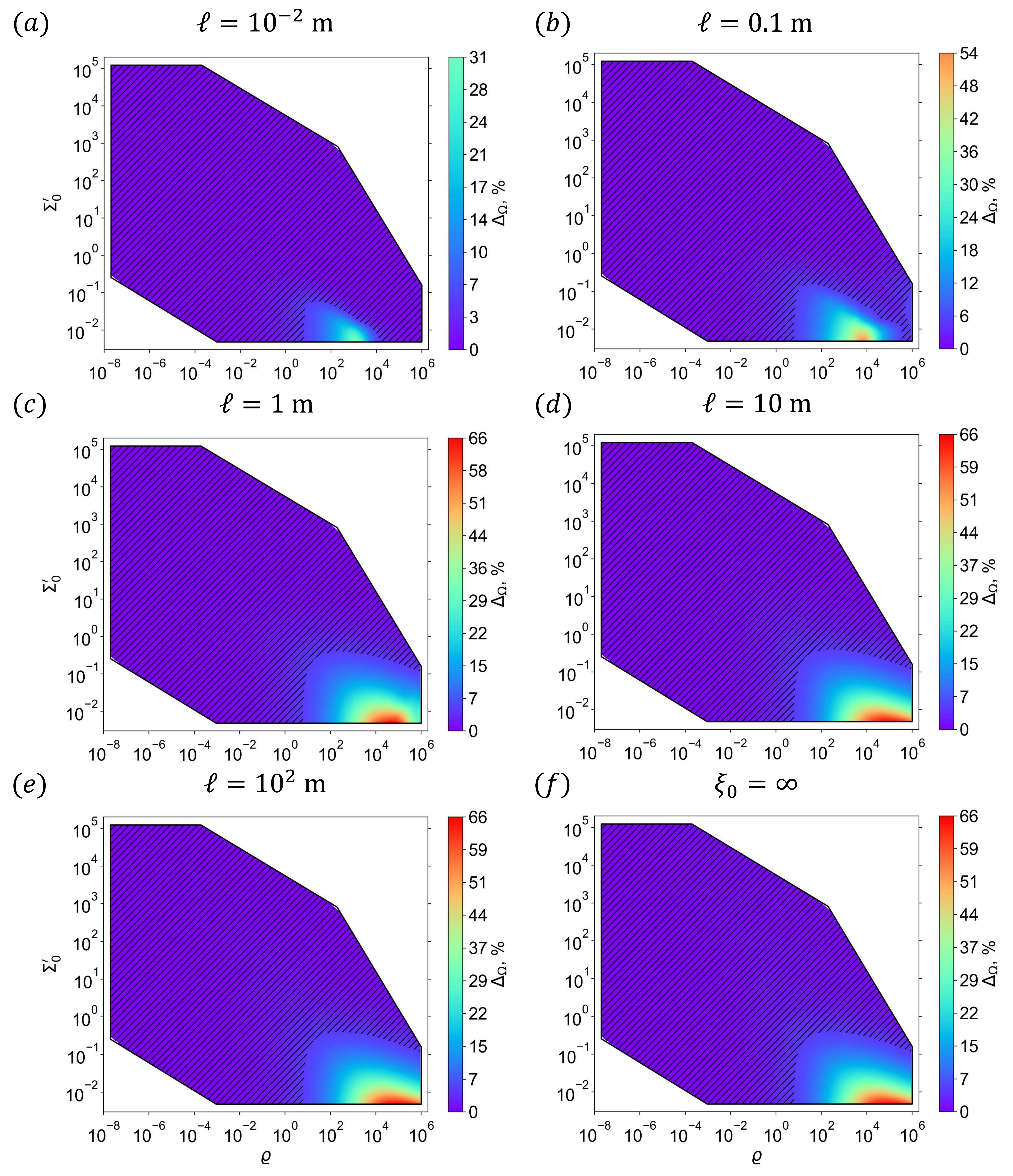}
    }
    \caption{
    Relative difference $\Delta_\Omega$~\eqref{eq:metrics_for_comparison_2} between the fracture opening profiles obtained with the fully coupled and one-way coupled models, evaluated over the $(\varrho,\Sigma_0')$ parameter space. Panels show the results for finite physical comparison intervals $\ell=10^{-2}$ m (a), $\ell=0.1$ m (b), $\ell=1$ m (c), $\ell=10$ m (d), and $\ell=10^2$ m (e), together with the limiting case $\xi_0=\infty$ (f). The colour scale is common to all panels, so that identical colours correspond to the same values of $\Delta_\Omega$. The black contour outlines the modelling domain shown in figure~\ref{fig:representative_ranges_governing_parameters}a. Hatched regions indicate $\Delta_\Omega<5\%$, where the one-way coupled model reproduces the fracture opening profile of the fully coupled solution within the prescribed accuracy. The dimensionless poroelastic parameters are fixed at $\beta=0.15$ and $\eta=0.3$.
    }
    \label{fig:Delta_fully_coupled_vs_one_way_coupled_Delta_Omega}
\end{figure}

\begin{figure}[!ht]
    \centerline{
    \includegraphics[width=1\textwidth]{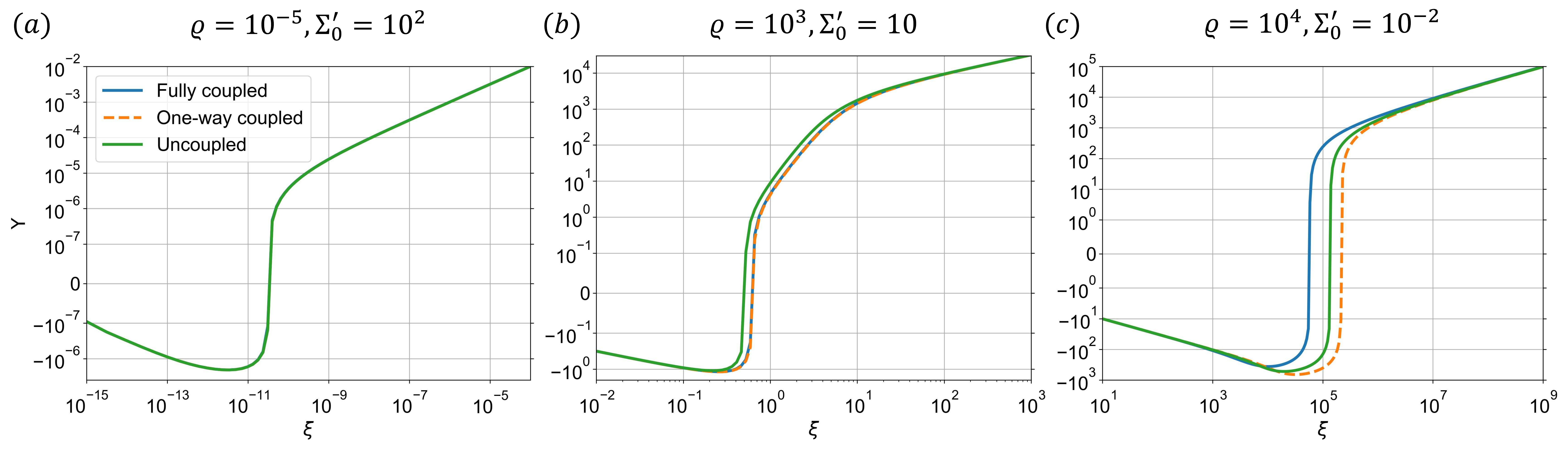}
    }
    \caption{
    The dimensionless fluid displacement function $\Upsilon(\xi)$ for representative solutions in the characteristic regimes. Panel (a) corresponds to $(\varrho, \Sigma_0')=(10^{-5}, 10^2)$, which is associated with Regime 0 for finite comparison lengths $\ell \leq 10$ m and with Regime 1 for larger $\ell$ and in the limiting case $\xi_0=\infty$. Panel (b) corresponds to Regime 2 at $(\varrho, \Sigma_0')=(10^3, 10)$, and panel (c) to Regime 3 at $(\varrho, \Sigma_0')=(10^4, 10^{-2})$. In each panel, the fully coupled solution (blue solid line), the one-way coupled solution (orange dashed line), and the uncoupled solution (green solid line) are compared. The dimensionless poroelastic parameters are fixed at $\beta=0.15$ and $\eta=0.3$. A symmetric logarithmic scale is used on the vertical axis to show both negative and positive values of $\Upsilon$.
    }
    \label{fig:Regime_1_3_Y}
\end{figure}


\clearpage

\bibliographystyle{jfm}
\bibliography{jfm}

\end{document}